\documentclass[%
 reprint,
 amsmath,
 amssymb,
 aps,
pra,
]{revtex4-1}

\usepackage{graphicx}% Include figure files
\usepackage{dcolumn}% Align table columns on decimal point
\usepackage{bm}% bold math
\usepackage{color}

\newcommand{\bD}{\mathbf{D}}

\newcommand{\bH}{\mathbf{H}}
\newcommand{\bI}{\mathbf{I}}
\newcommand{\bM}{\mathbf{M}}
\newcommand{\bQ}{\mathbf{Q}}
\newcommand{\bS}{\mathbf{S}}

\renewcommand{\Re}{\operatorname{Re}}
\renewcommand{\Im}{\operatorname{Im}}

\newcommand{\bLam}{\boldsymbol{\Lambda}}
\newcommand{\diag}{\mathrm{diag}}

\begin{document}

% \preprint{APS/123-QED}

\title{Determinantal polynomial wave functions induced by random matrices}% Force line breaks with \\

\author{Anthony Mays}
 \email{Anthony.Mays@unimelb.edu.au}
 \author{Anita K. Ponsaing}
\affiliation{%
School of Mathematics and Statistics, ARC Centre of Excellence for Mathematical and Statistical Frontiers, University of Melbourne, Victoria 3010, Australia
}%

\author{David M. Paganin}
\affiliation{
 School of Physics and Astronomy, Monash University, Victoria 3800, Australia
}%

\date{\today}% It is always \today, today,

\begin{abstract}
Random-matrix eigenvalues have a well-known interpretation as a gas of like-charge particles. We make use of this to introduce a model of vortex dynamics by defining a time-dependent wave function as the characteristic polynomial of a random matrix with a parameterized deformation, the zeros of which form a gas of interacting vortices in the phase. By the introduction of a quaternionic structure, these systems are generalized to include anti-vortices and non-vortical topological defects: phase maxima, phase minima and phase saddles. The commutative group structure for complexes of such defects generates a hierarchy, which undergo topologically-allowed reactions. Several special cases, including defect-line bubbles and knots, are discussed from both an analytical and  computational perspective. Finally, we return to the quaternion structures to provide an interpretation of two-vortex fundamental processes as states in a quaternionic space, where annihilation corresponds to scattering out of real space, and identify a time--energy uncertainty principle.
\end{abstract}

\pacs{Valid PACS appear here}% PACS, the Physics and Astronomy
\maketitle

\section{Introduction}\label{s:Intro}

The emergent phenomenon of the quasi particle is a profoundly useful concept  that pervades much of physics \cite{Annett2004,Volovik2003}.  The zoo of quasi-particles includes phonons, surface plasmons, polaritons, oscillons, solitons and excitons \cite{AshcroftMermin,Oscillons1996,Drazin1989}.  Topological defects \cite{VilenkinShellard1994} may also be regarded as quasi-particles in the broad sense of the term.  Such defects may be in a complex scalar optical field (e.g. phase vortices \cite{Dirac1931}, phase maxima and mimima, and phase saddles \cite{Freund1999b}), real vectorial optical fields (e.g. C-lines, disclinations, skyrmions \cite{Nye1999}) and tensorial optical fields (e.g. homotopy-group classification of tensor defects \cite{VilenkinShellard1994}).  

The topological defects of optical fields in particular \cite{Nye1999}, and classical fields more generally \cite{VilenkinShellard1994}, are well known to exhibit phenomena that have direct analogs in the behavior of genuine particles.  Examples include the obvious parallel between the fundamental electromagnetic process of pair production (e.g. $\gamma  \rightarrow e^+ + e^-$) and the spontaneous formation of a paired phase vortex and phase anti-vortex \cite{BerryLesHouches}, and the parallel between the decay of unstable particles and the decay of higher-order phase vortices into a set of lower-order phase vortices \cite{NyeBerry1974} (the ``critical point explosions'' \cite{Freund1999a}).

This paper is devoted to defect-line dynamics and topological reactions in the phase of classical complex scalar optical fields. Such interacting defects include local maxima and minima in the wave-function phase, together with phase saddles, phase vortices and phase anti-vortices. We approach this topic from the perspective of polynomial wave functions \cite{NyeBerry1974,BerryLesHouches,Nye1999,Dennis2011} generated by determinantal polynomials of random matrices.  This approach is motivated by: (i) the utility and mathematical simplicity of low-order polynomial functions as local descriptors of topologically non-trivial fields \cite{NyeBerry1974,Nye1999,BerryLesHouches,Dennis2011,PagaBeltPete2018}; (ii) the ensemble of such fields that can be generated by suitable ensembles of random matrices; (iii) the opportunity to explore, at length, a non-standard physical application of the eigenvalue dynamics of random matrices \cite{BohiDeCaPato2012}, in which individual random matrices are put into a correspondence with polynomial wave functions, with a further correspondence being developed between the evolution law for the random-matrix ensemble and the physical law governing the spatio-temporal evolution of the associated determinantal wave function.  

Random matrices have found several applications in physics, beginning with Wigner in 1955, who hypothesized that the eigenvalue statistics of some ensembles of Hermitian matrices ``may reproduce some features of the ... behavior of atomic nuclei'' \cite{Wigner1955}. Wigner's conjecture and preliminary work was followed-up by Dyson, with papers in 1962 \cite{Dyso1962I,Dyso1962II,Dyso1962III,Dyso1962ThreeFold} that laid the foundation for modern random-matrix theory. Part of his work (building on that of Wigner) established fundamental differences between random matrices with real, complex and quaternionic entries, and found deep connections between the symmetry classes of random matrices and those of various algebraic structures.  This is known as Dyson's Threefold Way \cite{Dyso1962ThreeFold}. 

Another application is quantum chaos, which, broadly speaking, is the study of quantum systems whose classical analog exhibits chaotic behavior \cite{Berr1987}. Canonical examples are the Sinai and stadium billiards as studied in \citeauthor{BohiGianSchm1984} \cite{BohiGianSchm1984}. It has been found \cite{Berr1987} that the statistics of these classically chaotic systems match those of the eigenvalues of random matrices (the energy levels are strongly correlated and repel), while the statistics of the classically non-chaotic systems are Poissonian (the energy levels are essentially independent). This has become known as the \textit{quantum chaos conjecture} \cite{KosLjubPros2017}, and has enabled the description of quantum chaotic systems which do not have a classical analog: through an appeal to universality, such systems are identified as quantum chaotic if their statistics match those of the corresponding random matrix ensembles. While the quantum chaos conjecture is well established by statistical data, a detailed theoretical understanding of the connection between the quantum systems and random matrices is still lacking. A recent work \cite{KosLjubPros2017} makes progress in this direction by calculating the form factor (Fourier transform of the two-point correlation function) for an Ising model in a periodically kicking transverse field, showing that it agrees in the two leading orders to the corresponding random matrix form factor, that of the circular orthogonal ensemble.

A particularly important and widely studied class of random matrices is the statistical ensemble of $N \times N$ matrices of the form
\begin{align}
\label{d:Ginmat} \bM = \left[ \begin{array}{cccc}
a_{1,1} +i b_{1,1}& a_{1,2} +i b_{1,2}& \dots & a_{1,N} +i b_{1,N}\\
a_{2,1} +i b_{2,1}& a_{2,2} +i b_{2,2}& \dots & a_{2,N} +i b_{1,N}\\
\vdots& \vdots& \ddots& \\
a_{N,1} +i b_{N,1}& a_{N,2} +i b_{N,2}& \dots & a_{N,N} +i b_{N,N}
\end{array}\right].
\end{align}
Here, each entry of $\bM$ is a complex random variable $a+i b$, in which the real and imaginary parts are independently and identically distributed (iid) as Gaussians with mean zero and variance $(2N)^{-1/2}$. These matrices are known as Ginibre matrices \cite{Gini1965}. Although eigenvalues can be degenerate in principle, for Ginibre matrices there is a vanishing probability of eigenvalues coinciding.

It is well known that the eigenvalues of Hermitian operators $\bM= \bM^{\dagger}$ are strictly real numbers \cite{Krey1978}, however when the operators are non-Hermitian then the eigenvalues are generic complex numbers. In a series of work beginning in 1984 \cite{Girk1984, TaoVuKris2010} it has been shown that in the limit of large matrix dimension the eigenvalues of independent and identically distributed (non-Hermitian) matrices are supported only on the unit disk, centred at the origin, on which the eigenvalues are uniformly distributed.  This has become known as the ``circular law''. The fact that the eigenvalues of non-Hermitian operators can be non-real presents many technical problems, yet several techniques have been developed to deal with these. One approach is to ``Hermitize'' the non-Hermitian matrices, which is the approach used to establish the circular law. The basic idea of Hermitization is to create a four dimensional quaternionic space to perform calculations over $\mathbb{C}$ \cite{JaniNowaPappZahe1997a, JaniNowaPappZahe2002, FeinZee1997}.  This is analogous to the use of complex variables allowing one to analytically define the Stieltjes transform for Hermitian problems \cite{FeinZee1997}. We will discuss further connections to the theory of quaternions in the present work.

The 2-dimensional (complex) eigenvalue distributions of non-Hermitian operators can be studied as stochastic point processes, under the category of ``determinantal point processes'', see \citeauthor{HougKrisPereVira2009} \cite{HougKrisPereVira2009} for some examples. The joint probability density function (jpdf) for the (complex) eigenvalues of matrices of the form in Eq.~(\ref{d:Ginmat}) is proportional to \cite{Gini1965}
\begin{align}
\label{e:evjpdf} \prod_{j=1}^N e^{- |\lambda|^2} \prod_{1\leq j< k \leq N} |\lambda_j - \lambda_k|^2,
\end{align}
which has a natural interpretation as a Coulomb gas with logarithmic intra-particle electrostatic repulsion, confined within a Gaussian potential (see e.g. \citeauthor{Forr2010} \cite{Forr2010}). It is this interpretation that inspires us to introduce dynamical behavior to the eigenvalue ``particle gas''. Note that the vanishing probability of eigenvalue degeneracy for a Ginibre matrix is evident in Eq.~(\ref{e:evjpdf}). 

The precedent for this approach (in the case of Hermitian matrices) is the work by Dyson \cite{Dyso1962BrownianMotion, Dyso1972}, pointing out that Eq.~(\ref{e:evjpdf}) is the stationary solution to the equations of motion of a Coulomb gas undergoing Brownian motion. The conclusion is that when the matrix executes Brownian motion (according to a specified law) then the eigenvalues of the matrix also undergo Brownian motion. This approach has been continued (e.g. \cite{SmolSimo2003a,SmolSimo2003b}), and very recently with investigations into adapting the results to non-Hermitian ensembles \cite{BlaiGrelNowaTarnWarc2016}.

In contrast to these approaches, we would like to use the random matrix to generate a distribution of particles, which then undergo deterministic evolution. On this theme, the dynamics of eigenvalues in the Hermitian--non-Hermitian cross-over regime of \cite{FyodKhorSomm1996, FyodSommKhor1998} (the elliptic ensembles) have been studied using matrices of the form
\begin{align}
\bH(t) = \frac{1}{2} (\bM+ \bM^{\dagger}) + \frac{t}{2} (\bM- \bM^{\dagger}),
\end{align}
where $\bM$ is a Ginibre matrix as in Eq.~(\ref{d:Ginmat}) and $t\in [0,1]$ is a dimensionless time parameter. The matrix interpolates between the matrix $\bH(0)$, which is Hermitian, and $\bH(1)= \bM$, which has no Hermitivity constraint. By calculating the time derivative of the diagonalized matrix $\bD = \bQ^{-1} \bH(t) \bQ$ a set of $2N(N +1)$ coupled first order differential equations can be found, which determine the velocities of the eigenvalues \cite{BohiDeCaPato2012}. The initial velocities are in the purely imaginary direction, implying that the first motion of the eigenvalues as they begin to explore the complex plane is perpendicularly away from the real line.

Section \ref{s:Evs} interprets the eigenvalues of a random Hamiltonian (which depends on a time parameter) as an evolving system of vortices or like-charge particles. This is in the context of earlier work along the same lines (see above), although we discuss how this interpretation can be applied to a wide class of evolution equations. By defining our wave function to be the characteristic polynomial of the matrix, the zeros of the wave function are the eigenvalues $\lambda= x+iy \in\mathbb{C}$ which thereby map to locations $(x,y)$ in two spatial dimensions. The characteristic polynomial is a polynomial in the single variable $\lambda$, with complex coefficients, and so in general it has complex solutions. By interpreting the real and imaginary parts of the zeros as the coordinates of a vortex we create an interacting gas of these zeros. By analysing the phase of the resulting wave function we see that every zero has positive winding number, and so eigenvalues must always be interpreted as phase vortices \cite{Dirac1931,Nye1999} (as opposed to anti-vortices, which have negative winding number). Additionally, we present some data from simulations using a specific Hamiltonian, and then present some exact calculations for small matrix size using this Hamiltonian. While much of this material is of course well known, it forms a foundation, as well as establishing notation, for the results that are subsequently developed. It can also be seen that this is unrelated to other determinantal constructions of wave functions, such as the Slater determinant \cite{Slater1929}.

In Sec.~\ref{s:QEvs} we discuss a different determinantal polynomial expression for the wave function, where half of the polynomial variables are replaced by the complex conjugate. This creates a two-variable polynomial with complex coefficients, imposing the condition that half of the zeros are now anti-vortices with the remainder being vortices. This opens up the possibility of vortex--anti-vortex annihilation and creation; indeed we find through simulation that it is very easy to generate scenarios with a rich structure of behavior. While this former point is well known, we use it to establish connections to the theory of quaternionic random matrix ensembles. Then in Sec.~\ref{s:GenEvs} we generalize the system by allowing for any number of vortices and anti-vortices.  Defects in the phase of the relevant wave functions are seen in Sec.~\ref{s:Top} to obey topological rules for creation and annihilation events associated with phase vortices, phase anti-vortices, phase saddles and phase extrema (i.e. phase maxima and phase minima).  Examples are given of defect-line reactions, which are both richer and more general than those that only involve vortices and anti-vortices, including the construction of defect-line knots and bubbles. The defect-line knots, while closed when considered only at the level of nodal lines, are seen to be not closed when considered at the more general level of defect lines.  Similarly, the possible closed-defect-line structures are richer than those merely associated with nodal lines. The countably infinite set of all possible defect complexes is considered in Sec.~\ref{s:AlgebraicStructureAndSupermultiplets}, as generated by the underlying algebraic structure of the possible defect complexes.  These are arranged into multiplets and super multiplets, which are somewhat analogous to the corresponding constructs in the quark model of hadrons.  Section~\ref{s:fundproc} considers transient quaternionic solutions that comprise unstable intermediate states associated with a certain vortex--anti-vortex annihilation event and a delayed but nevertheless associated subsequent vortex--anti-vortex creation.  For the system studied, the quaternionic states obey an energy--lifetime uncertainty principle. The role of scattering into quaternionic degrees of freedom arises naturally, a connection which is considered in some detail. We then discuss broader implications and avenues for future work, in Sec.~\ref{s:Discussion}. We conclude with Sec.~\ref{Sec:Conclusion}.

\section{Evolving-matrix model for vortex gas}\label{s:Evs}

We consider a direct correspondence between the characteristic polynomial of a square matrix, and an associated complex wave function that may in turn correspond to a specified physical system.  The continuous temporal evolution of any one matrix ${\bf M}$ induces a time-varying determinantal wave function $\Psi$, with an associated governing equation for the latter that may be chosen to coincide with a particular physical law.  We pay particular attention to the vortical character of the time-dependent wave functions induced by random matrices, setting up a formalism and establishing a notation that is subsequently generalized to the more general topic of defect-lines.  As we shall see, many aspects of the associated nodal-line networks, and more generally of the defect-line networks, may be {\em locally} described by considering the evolution of a $2\times 2$ matrix and its associated determinantal wave function.  

\subsection{Characteristic-polynomial wave function}

Consider an $N \times N$ complex matrix $\bf{M}$, the eigenvalues $\lambda_j$ of which obey the usual characteristic equation
\begin{align}
\label{e:charpoly0}\chi(\lambda)= \det ({\bf M}- \bLam_N)= 0.
\end{align}
\noindent Here, $\bLam_N= \lambda {\bf I}$, ${\bf I}$ is the identity matrix and $\det$ denotes matrix determinant.  In general one has $N$ eigenvalues $\lambda_1,\lambda_2,\cdots \lambda_N$ in the complex plane.  These eigenvalues may be degenerate, but (as discussed in Sec.~\ref{s:Intro} for the specific case of the Ginibre ensemble) if one considers matrix deviates drawn from an ensemble with specified continuous probability distribution independently governing each element, the likelihood of degeneracy is infinitesimally small.

By making the identification 
\begin{align}
\lambda = x + iy,
\end{align}
where the real numbers $x$ and $y$ are spatial coordinates in two transverse dimensions, one can consider the characteristic polynomial $\chi$ to be a complex wave function $\Psi$:
\begin{align}
\label{e:wavfn} \chi(\lambda = x + iy; t)= \Psi_{N, N} (x,y; t).
\end{align}
\noindent We have added the time label $t$ above, to allow the characteristic polynomial and the associated wave function to evolve with time in an as-yet unspecified manner. This label may be continuous or discrete. The subscripts on $\Psi_{N,w}$ denote that the wave function originates from an $N\times N$ matrix and the winding number of the wave-function phase is $w$ when traversing a contour enclosing all zeros, which, in this section, is also equal to $N$ (we will discuss phase winding numbers in more detail below). When using the above wave function to model a given (2+1)-dimensional physical system, the evolution of $\chi(x,y;t)$ should be such that $\Psi(x,y;t)$ evolves in time in a manner consistent with the relevant physical law governing $\Psi(x,y;t)$.  We shall return later in the present section to the choice of temporal evolution law for $\chi(x,y;t)$, but for the moment we leave this unspecified.

The wave function defined by Eqs.~(\ref{e:charpoly0}) and (\ref{e:wavfn}) will be a polynomial of order $N$ in $x$ and $y$.  While finite-order polynomials are guaranteed to diverge for large $x$ and $y$, there are many contexts in which they have proved to be a powerful approximation for the local behavior of complex scalar wave fields obeying a rich variety of evolution equations \cite{NyeBerry1974, BerryLesHouches, Nye1999, Dennis2011, PagaBeltPete2018}.  These include, but are not limited to, exact polynomial solutions to the (2+1)-dimensional Schr\"{o}dinger equation, the d'Alembert wave equation and the wide class of evolution equations associated with linear shift-invariant coherent imaging systems \cite{NyeBerry1974,BerryLesHouches, Nye1988,Nye1999,PagaBeltPete2018,Paganin2018b}.

There is a close correspondence between choices for the matrix ${\bf M}(t=0)$ from a suitable ensemble of complex random matrices, and wave functions describing a random arrangement of vortices over a disc. If each element of the $N\times N$ matrix ${\bf M}$ is a complex random variable with zero mean and variance $(2N)^{-1/2}$, then for $N\gg 1$ the eigenvalues are uniformly and randomly distributed within the unit disc in the complex plane (the ``circular law'' mentioned in Sec.~\ref{s:Intro}).  This corresponds to a characteristic-polynomial wave function $\Psi(x,y;t=0)$, within which is embedded a random gas of like-charge vortices.  As mentioned earlier (see Eq.~(\ref{e:evjpdf})), this vortex gas may under certain circumstances behave as a Coulomb gas.  Our core focus, however, is on application to a much broader class of system. 

Whatever the structure of ${\bf M}$, the associated finite-order polynomial wave functions can describe vortical structures, for which the wave-function phase exhibits a screw-type phase dislocation with integer winding.  This has been well studied e.g. in the context of finite-order polynomial approximations to vortical coherent scalar electromagnetic fields \cite{NyeBerry1974,BerryLesHouches,Nye1999}.  The factorisability of a polynomial of order $N$ in $\lambda = x + iy \in \mathbb{C}$ implies that
\begin{align}
\label{e:wavfact} \Psi_{N,N} (x,y;t)=\prod_{j=1}^{N}\{[x-\Re(\lambda_j)]+i [y- \Im(\lambda_j)]\},
\end{align}
which is manifestly vortical in the sense described below.

Recall that typically the random-matrix eigenvalues at $(x,y)=(\Re(\lambda),\Im(\lambda))$ will be non-degenerate. Each eigenvalue will then be an isolated zero of the characteristic polynomial.  Further, each wave-function zero will be a branch point for the phase  
\begin{align}
\Phi(x,y;t)\equiv \arg[\Psi (x,y;t)] 
\end{align}
of the associated wave function, with unit phase winding.  This corresponds to the $m=1$ case of the more general expression for admissible phase windings \cite{Dirac1931,NyeBerry1974}:
\begin{align}
\frac{1}{2\pi}\oint_{\Gamma}d\Phi = m,
\label{e:PhaseWinding}
\end{align}
\noindent where $\Gamma$ is a simple anticlockwise-traversed smooth closed contour in the $x{\text{--}}y$ plane and $m$ is the \textit{winding number}.  The integer $m$ is often called the net topological charge of the vortex or vortices enclosed within $\Gamma$. See Sec.~\ref{s:Top} for more discussion of the winding number.

Since a characteristic polynomial is a polynomial in the complex variable $\lambda=x+iy$, the associated wave function can only support vortices with $m \ge 1$ (see Appendix \ref{a:vorPf}). Physical systems such as Bose--Einstein condensates (BECs) in a sufficiently rapidly rotating trap \cite{Pitaevskii2003} or Abrikosov vortices in the order-parameter field of a Type-II superconductor \cite{Wells2015}, naturally form wave functions in which all topological charges have the same sign and magnitude.  One final example of quantum systems described by wave functions containing vortices, all of which have the same topological charge, is the interior of one lobe of an Onsager vortex cluster formed in a turbulent vortical cold quantum gas \cite{Groszek2016}.

\subsection{Toy model for topology of defect-line collisions}

\begin{figure}[h]
\includegraphics[width=7cm]{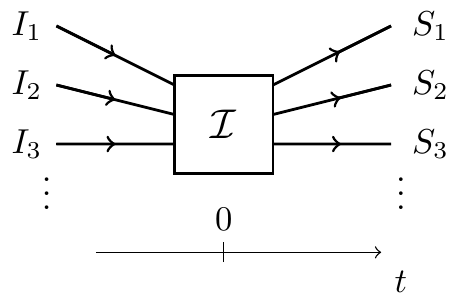}
\caption{\label{f:GenericScatteringScenario} Generic scattering scenario for phase-defect collisions.  Time $t$ runs from left to right.  For $t$ large and negative, incident free phase defects $I_1,I_2,\cdots$ converge with purely-radial free motion towards an interaction region $\mathcal{I}$.  Within the space-time volume $\mathcal{I}$, for which the internal lines are not shown, various topological reactions of the phase defects may occur.  After the interaction, when $t$ is large and positive, one has a series of scattered free phase defects $S_1,S_2,\cdots$ diverging with purely-radial free motion away from $\mathcal{I}$.}
\end{figure}

We introduce a simple model for the defect-line topology of multi-vortex collisions.  While trivial in mathematical form, it will later be apparent that this model generates topological dynamics for a range of phase defects beyond merely multiple-vortex collisions.  However, in the present sub-section we restrict attention to the application of the toy model to multiple-vortex collisions. See Fig.~\ref{f:GenericScatteringScenario}.  

The explicit model that we consider is:
\begin{align}
\label{e:Hmat} \bM(t)= \bM_0 + t\bS
\end{align}
where $\bM_0$ is a (fixed) random complex Ginibre matrix, 
\begin{align}
\bS= \mathrm{diag} (\underbrace{s, \dots, s}_{\footnotesize \mbox{$\frac{N}{2}$ copies}},\underbrace{-s, \dots, -s}_{\footnotesize \mbox{$\frac{N}{2}$ copies}}),
\end{align}
with $s\in \mathbb{C}$ a deformation parameter and $t\in \mathbb{R}$ a dimensionless time parameter. Despite this rather simple form, we find that the evolution of the vortex systems exhibits quite non-trivial behavior. Unless otherwise specified, we restrict ourselves to the cases where $N$ is even.

To explore the utility of this toy model in more detail, first note that when $\bM_0= 0$ then the wave function for ${\bf M}(t)$ is $(x + iy - st)^{N/2} (x + iy + st)^{N/2}$, corresponding to $N/2$ vortices at position $(-st, 0)$ and another $N/2$ vortices at $(st,0)$, however any small perturbation from $\bM_0\neq 0$ explodes the degeneracy of these vortices. In that case, for large $|t|$ the Ginibre eigenvalue jpdf Eq.~(\ref{e:evjpdf}) decomposes into two non-interacting factors
\begin{align}
\label{e:limjpdf}\prod_{j=1}^{N/2} e^{-|\sigma_j|^2} e^{-|\tau_j|^2} \prod_{1\leq j< k \leq N/2} |\sigma_j-\sigma_k|^2 |\tau_j-\tau_k|^2,
\end{align}
up to a proportionality factor which depends on $t$, where (without loss of generality) $\sigma_j= \lambda_j+ts$ and $\tau_j= \lambda_j -ts$. We interpret Eq.~\ref{e:limjpdf} as our system splitting into two disjoint Ginibre eigenvalue distributions for large $|t|$ (where the two pieces are consequently separated by a large distance)---this scenario is represented by the right- and left-hand sides of Fig.~\ref{f:GenericScatteringScenario}. Taking the time evolution chronologically, as $t$ increases, the vortices converge towards the $(x,y)$ origin and enter the interaction region $\mathcal{I}$. Then as $t$ becomes increasingly positive the vortices exit the interaction region and diverge from the $(x,y)$ origin when $t$ is large and positive. Hence our association of the determinantal wave function induced by Eq.~(\ref{e:Hmat}) with a quasi-particle phase-defect collision problem (elastic scattering): 
\begin{align}
v+v+\cdots\rightarrow v+v+\cdots, 
\end{align}
where $v$ denotes a vortex. In this context, the matrix ${\bf M}_0$ may be viewed as perturbing the coalescence of the converging vortex trajectories, introducing non-trivial interaction dynamics.

We are primarily interested in the topologically distinct reactions of phase defects in this toy model, in the ``black box'' interaction region $\mathcal{I}$, hence it is not unphysical to fix the asymptotic behavior in this toy model in the manner that has been adopted.  Indeed, this choice amounts to continuous deformation of the space-time trajectories associated with the phase defects, a process which does not alter their topological structure (i.e. the topological classification of the defect-line graph that is generated).  Interestingly, as shall be seen throughout the paper, while the trivial  temporal evolution defined by Eq.~(\ref{e:Hmat}) is linear in $t$, the induced defect-line trajectories exhibit a behavior that is both highly non-linear and remarkably rich. All defect-line plots in the remainder of the paper can be viewed as special cases of possible topological reactions in the interaction region $\mathcal{I}$ of  Fig.~\ref{f:GenericScatteringScenario}.

\subsection{Time evolution}

We now consider time evolution explicitly. We assume the $N \times N$ complex matrix ${\bf M}(t)$ to evolve as a continuous function of time $t$.  The evolution may be deterministic or stochastic.  For deterministic evolution laws we are particularly interested in matrices ${\bf M}(t)$ for which the associated polynomial wave function obeys a specified physical law, although the formalism explored here permits arbitrary smooth evolution laws to govern ${\bf M}(t)$.  This point will be explored in further detail later.  An example of stochastic evolution for ${\bf M}(t)$ is given by complex random matrices whose eigenvalues undergo continuous Brownian motion in the complex plane \cite{Dyso1962III, BlaiGrelNowaTarnWarc2016}.   

We place relatively little emphasis on the particular means for evolving ${\bf M}(t)$ in time, since, as already emphasized,  we are principally concerned with topological aspects of the phase of the induced wave function in $2+1$ dimensions.  As pointed out by Dirac, these topological aspects of the wave-function phase arise solely from the continuity and single-valuedness of the said complex wave functions, independent of the particular field equation governing their spatio-temporal evolution \cite{Dirac1931}.  

The evolution law governing ${\bf M}(t)$ may be viewed as inducing an associated evolution law for the corresponding polynomial wave function.  

More interestingly, in the context of setting up a correspondence between evolving matrices and evolving wave functions, a given evolution law for a wave function may be considered to induce an associated evolution law for ${\bf M}(t)$. Suppose, in this latter context, that one is given a physical law of evolution for a specified (2+1)-dimensional wave function.  Restrict consideration to partial differential equations of first order with respect to time, hence the physical law may be written as
\begin{align}
i \frac{\partial \Psi(x,y;t)}{dt}=H(x,y;t)\Psi(x,y;t),
\label{e:SchrodingerEquation}
\end{align}
where $H(x,y;t)$ is the Hamiltonian operator (infinitesimal generator of time evolution).  If the wave function is specified at a time $t=t_0$, evolution through a subsequent infinitesimal time $\delta t > 0$ gives
\begin{align}
\nonumber \Psi(x,y; &t_0+\delta t) \\ &= \Psi(x,y;t_0)-i\delta t H(x,y;t_0)\Psi(x,y;t_0).
\end{align}

\begin{figure}
\includegraphics[scale=0.6]{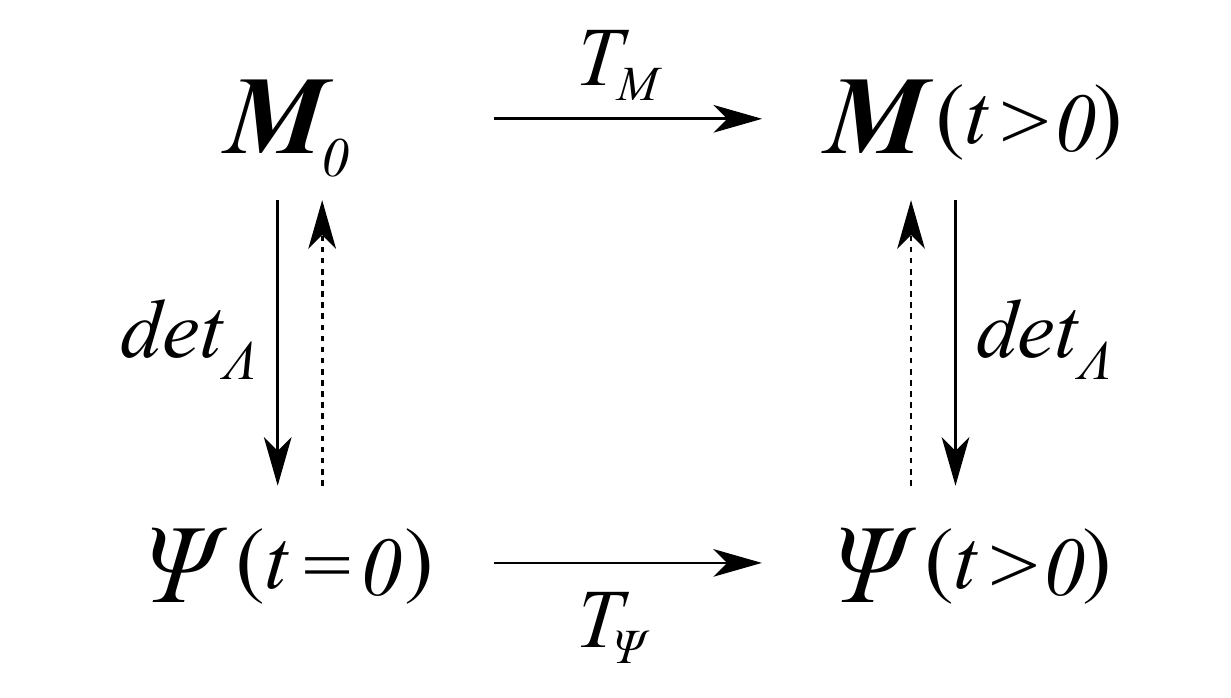}
\caption{Chart outlining correspondence between evolving matrices ${\bf M}(t)$ (or ensembles thereof) and associated evolving determinantal wave functions (or wave-function ensembles).   \label{f:commdiag}}
\end{figure}

One then has the following iterated cycle of steps to evolve ${\bf M}$ in time so that the associated characteristic polynomial obeys Eq.~(\ref{e:SchrodingerEquation}) (see Fig.~\ref{f:commdiag}): (i) Begin with the $N \times N$ matrix ${\bf M}(t=t_0)$ and determine the associated wave function $\Psi_{N,N} (x,y;t=t_0)$ by applying the operator $\det_{\Lambda} \bM = \det \Big( \bM- \lambda \bI \Big)$; (ii) Use the Hamiltonian operator to determine $\Psi_{N,N} (x,y;t=t_0+\delta t)$, using a suitable finite difference scheme such as that given above, or more sophisticated schemes such as the Cayley form \cite{Goldberg1967}; (iii) Solve the updated characteristic equation $\Psi_{N,N} (x,y;t=t_0+\delta t)=0$ for all eigenvalues of the as-yet-unknown $N \times N$ matrix ${\bf M}(t=t_0+\delta t)$; (iv) Choose the updated matrix ${\bf M}(t=t_0+\delta t)$ to be any member of the class of matrices that has the desired eigenvalues at $t=t_0+\delta t$, such that each element of the updated matrix differs at most by a term of order $\delta t$ from the corresponding matrix element at time $t= t_0$.  

Typically, the class of matrices in (iv) will have a continuous infinity of members at each $t$, and so there is no inverse of $\det_{\Lambda}$ (hence the dashed vertical arrows in Fig.~\ref{f:commdiag}).  One may then apply any convenient  auxiliary condition to render the particular choice of updated matrix to be unique. The freedom in the choice of auxiliary condition, corresponding to the class of matrices which generate the same eigenvalues, is somewhat analogous to gauge freedom. 

The above is summarised via the chart in Fig.~\ref{f:commdiag}.  The upper row corresponds to the evolving matrix field ${\bf M}(t)$, with the lower row  corresponding to the associated determinantal wave function $\Psi(x,y;t)$.  Here, ${\bf M}_0 \equiv {\bf M}(t=0)$ denotes the initial matrix, which has time-evolution operator $T_{\bM}$, such that 
\begin{align}
{\bf M}(t)=T_{\bM} \, {\bf M}_0 = e^{-it H_{\bM}} {\bf M}_0.
\label{e:ArbitraryMatrixEvolutionLaw}
\end{align}
In Eq.~(\ref{e:ArbitraryMatrixEvolutionLaw}), we have used the matrix exponential, and $H_{\bM}$ is some matrix operator. The wave function $\Psi(x,y;t=0)\equiv\Psi(t=0)$ corresponding to ${\bf M}_0$ is evolved in time via:
\begin{align}
\Psi(t)=T_{\Psi} \Psi(t=0) = e^{-it H} \Psi(t=0), 
\label{e:ArbitraryWavefunctionEvolutionLaw}
\end{align}
where $H$ is given by Eq.~(\ref{e:SchrodingerEquation}) (to first order in $t$) and $T_{\Psi}$ is the induced operator
\begin{align}
T_{\Psi}={\text{det}}_{\Lambda} \, T_{\bM} \, {\text {det}}_{\Lambda}^{-1},
\end{align}
where we again stress that $\det^{-1}_{\Lambda}$ is the operation of mapping a monic polynomial onto any matrix in the equivalence class of matrices that yield the same characteristic polynomial. (This is not a bijective operation and so, strictly speaking, there is no inverse, however we would like to avoid the distracting complications of defining maps on equivalence classes.) There are many ways of defining this operator, and we will use some specific examples, but at an ensemble level a general way to create a random matrix having a specific set of eigenvalues is via conjugation. If $\{\lambda_1, \dots, \lambda_N \}$ are the (complex) zeros of a polynomial wave function $\Psi$ then let $\bD= \diag (\lambda_1, \dots, \lambda_N)$ and let $\bQ$ be an $N\times N$ random matrix (perhaps Ginibre or unitary; as long as the inverse exists).  Then define
\begin{align}
\label{e:matconj}{\text{det}}_{\Lambda}^{-1} \bD = \bQ \bD \bQ^{-1}.
\end{align}

The determinantal formalism maps matrices to wave functions (downward-pointing arrows in Fig.~\ref{f:commdiag}), with the corresponding ``inverse'' operation denoted by upward-pointing arrows. Both upward- and downward-pointing arrows, which correspond to a change of representation, have the previously mentioned gauge-like freedom.  The equation of motion corresponding to the top row of the chart is the matrix-evolution law in Eq.~(\ref{e:ArbitraryMatrixEvolutionLaw}), with the corresponding wave-function evolution law given by Eq.~(\ref{e:ArbitraryWavefunctionEvolutionLaw}).  If one wishes to work with statistical mixtures rather than pure states, each matrix $\bM$ and associated wave function $\Psi$ can be assigned a real non-negative statistical weight, with each member of the ensemble then being evolved in the manner described above.

When a given matrix is evolved through time, the associated eigenvalues (vortex cores) in the characteristic polynomial of the matrix (the wave function induced by the matrix), will in general trace out a complex nodal-line network in space-time. An indication of the level of complexity that is possible is given by the fractal nodal-line networks associated with visible-light vortical speckle fields  \cite{OHolleran2008} and the tangled nodal-line networks associated with turbulent quantum gases \cite{Ruben2008}.  

While the nodal-line network may be rather complicated, it is natural to consider a {\em local} analysis of topological reactions exhibited by a small number of nodal lines.  Indeed, many two-vortex topological reactions (and topological reactions involving phase maxima, phase minima and phase saddle points; see below) may be locally described by a $2\times 2$ matrix. Hence the significant attention paid, in this paper, to the topological dynamics in the nodal-line evolution (and, more generally, the defect-line evolution) associated with characteristic-polynomial wave functions induced by a $2\times 2$ matrix.  

For a $2\times 2$ matrix
\begin{align}
\bM_0= \left[ \begin{array}{cc}
a_0& b_0\\
c_0& d_0
\end{array}
\right], \quad a_0,b_0,c_0,d_0 \in \mathbb{C}
\end{align}
at time $t= t_0$, the characteristic polynomial gives
\begin{align}
\chi(\lambda;t_0)= \lambda^2 - (a_0+d_0) \lambda + a_0 d_0 - b_0c_0,
\end{align}
which we then evolve in time to some new polynomial
\begin{align}
\chi(\lambda;t_0+ \delta t)= \lambda^2 - k_1 \lambda + k_2.
\end{align}
Note that the determinantal nature of this polynomial forces it to be monic, that is it forces the coefficient of $\lambda^2$ to be unity.  This may be viewed as permitting a time-varying normalization in the induced wave functions $\chi$, which can be accounted for in the usual way e.g. by calculating expectation values of operators $\hat{A}$ via $\langle\chi | \hat{A} \chi\rangle/\langle\chi | \chi \rangle$, the integration being over a specified finite-volume region since finite-order polynomial wave functions are not square integrable. For this reason, for the remainder of the paper we shall work with non-normalized wave functions. 

To construct the matrix $\bM_1$ at time $t_0 +\delta t$ we solve the pair of equations
\begin{align}
k_1 = a_1 + d_1, \qquad k_2 = a_1 d_1 - b_1 c_1.
\end{align}
We can choose $d_1 = d_0 \pm \mathrm{const.} \delta t$ and $c_1 = c_0 \pm \mathrm{const.} \delta t$, (these choices then fix the other parameters $a_1, b_1$) which gives us a continuous set of possible matrices $\bM_1$, each of which is no more than $O(\delta t)$ away from $\bM_0$ in some suitable metric.

As a simple example of this means for evolving a $2\times 2$ matrix so as to conform with a specified Hamiltonian, consider 
\begin{align}
T_{\Psi}= e^{- it H}, \quad H= \frac{\partial}{\partial x}+ \frac{\partial}{\partial y}. 
\end{align}
Then
\begin{align}
T_{\Psi} \Psi(0) = \Psi(t)
\end{align}
and for the degree-two polynomial we have (to first order in $t$)
{\small \begin{align}
&T_{\Psi} \Psi(0) =\left( 1- it \frac{\partial}{\partial x}- it \frac{\partial}{\partial y}\right) \big[ \lambda^2 - (a+ d) \lambda + (ad- bc) \big]\\
&= \lambda^2 - \big[ a+d -2t(1 -i) \big] \lambda +(ab-bc) -t(1-i) (a+d).
\end{align}}
This corresponds to (up to leading order in $t$)
\begin{align}
\bM(t)= T_{\bM} \bM_0
\end{align}
with
\begin{align}
T_{\bM} \bM= \bM - (1- i) \left[ \begin{array}{cc}
t& 0\\
0& t
\end{array}\right].
\end{align}

Conversely, if we assume the simple matrix evolution of Eq.~(\ref{e:Hmat}), with $s= 1$, then a corresponding operator on the wave function is given by
\begin{align}
T_{\Psi} \Psi = e^{-it H} \Psi, \qquad H= i (2d_0- [\lambda]),
\end{align}
where $d_0$ is the lower right element of the matrix $\bM_0$ and $[x^k]$ is the operator that returns the coefficient of $x^k$. 

\subsection{Simulations}\label{s:Evsims}

\begin{figure}[h]
\includegraphics[scale=0.44]{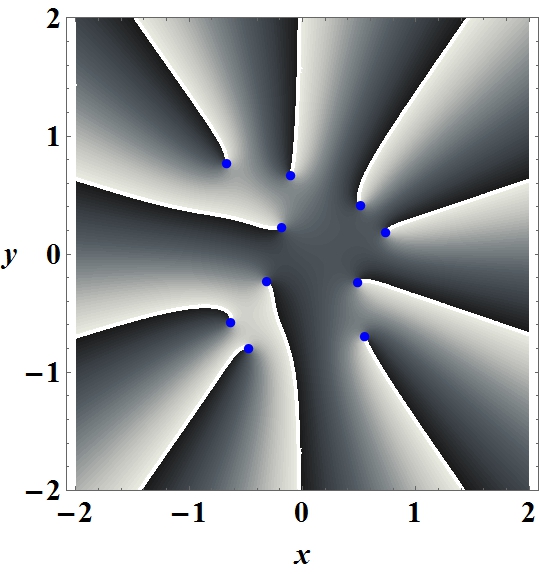}
\caption{\label{f:PhaseEv1}(Color online) Phase of $\Psi_{10, 10} (x,y;0)$ for a random $10\times 10$ matrix as in Eq.~(\ref{d:Ginmat}).  All zeros---which in this case are eigenvalues of $\bM$---are vortices, and are marked as blue dots. In this and all subsequent phase plots, phase $\Phi$ is given modulo $2\pi$, with a linear grayscale between black ($\Phi~\textrm{mod}~2\pi=-\pi$) and white ($\Phi~\textrm{mod}~2\pi=\pi$). }
\end{figure}

As a first numerical example, we generate a random matrix $\bM_0$ as in Eq.~(\ref{d:Ginmat}) with $N=10$ and then evolve in $t$ using the toy Hamiltonian Eq.~(\ref{e:Hmat}), with $s=1$ and $0\leq t \leq 1$. Recall that for these Ginibre matrices, there is a vanishing probability of eigenvalue degeneracy, so the wave function $\Psi_{10, 10} (x,y;t)$ of Eq.~(\ref{e:wavfn}) will have $10$ isolated zeros for almost all $t$.  These must all be vortices of the same helicity (winding number, which is defined in Eq.~(\ref{e:PhaseWinding})). Figure~\ref{f:PhaseEv1} plots the phase $\Phi_{10,10} (x,y;0)=\arg[\Psi_{10,10} (x,y;0)]$ of the initial polynomial wave function, with $t=0$. The vortex cores are labelled with a blue dot, each serving as a branch point for the Riemann sheets of the multi-valued phase.  Note that while the branch-point locations have the physical meaning that they correspond to vortex cores, the branch-line locations have no physical meaning. Indeed, the branch lines move if one alters the wave function via a meaningless global phase factor corresponding to multiplication by any complex constant with modulus unity. 

\begin{figure}[h]
\includegraphics[scale=0.4]{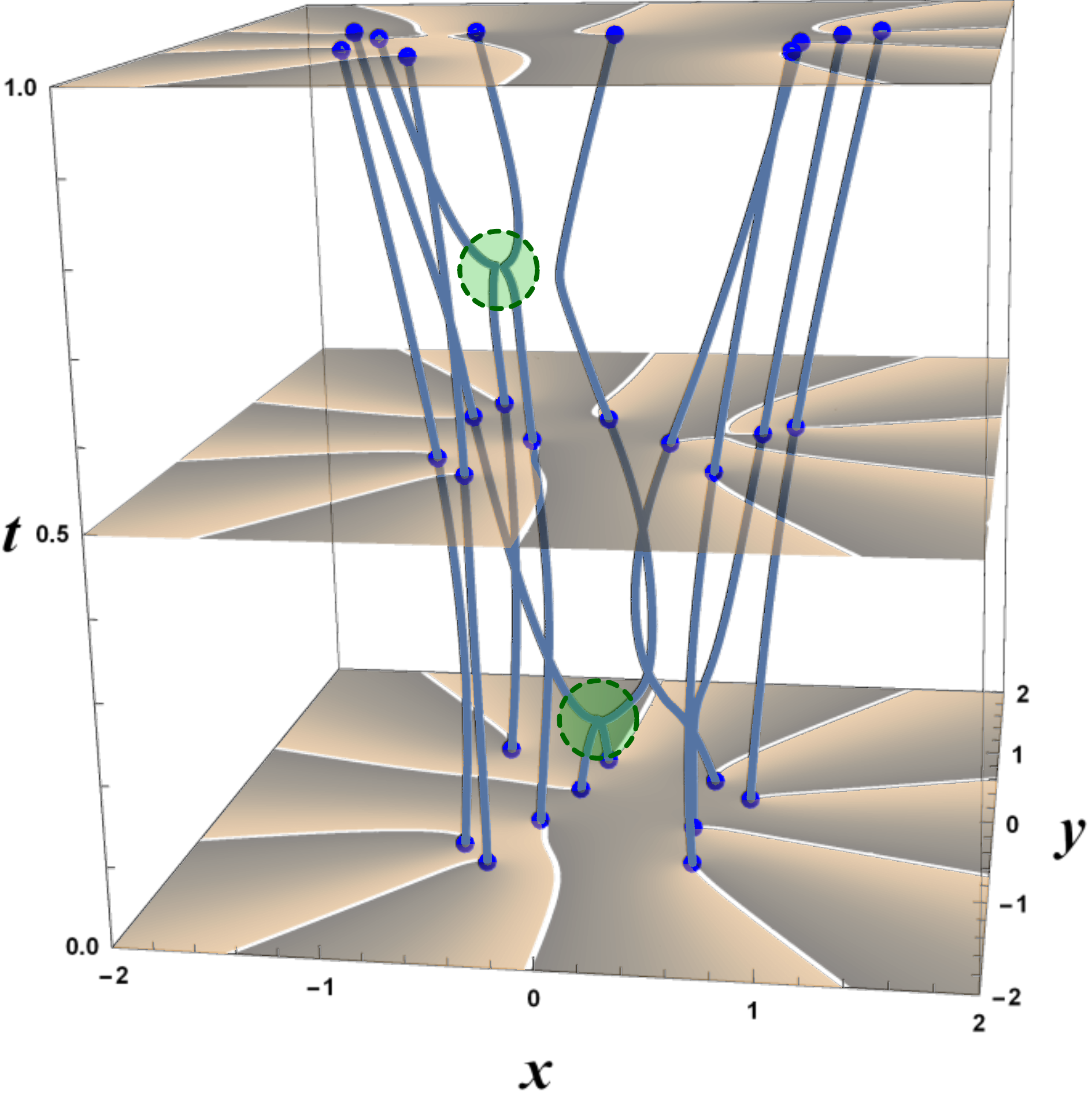}
\caption{\label{f:NodalEv1}(Color online) Trajectories of the vortices of $\Psi_{10, 10} (x,y;t)$ from Fig.~ \ref{f:PhaseEv1}. Figure~\ref{f:PhaseEv1} is the bottom layer of this diagram ($t=0$), with additional phase maps corresponding to $t=0.5$ and $t=1$ also  shown. Although there are some close collisions (highlighted by the dashed green circle) there are no annihilation events, consistent with the vanishing probability of eigenvalue degeneracy (instability of vortices with $m>1$). See Supplemental Material at [psi1010\_500frame50co100dpi.avi] for a video of this system.}
\end{figure}

Evolving the system through time $t\in[0,1]$ (according to Eq.~(\ref{e:Hmat})) we obtain the (2+1)-dimensional representation of the trajectories of the zeros (eigenvalues) of the polynomial wave function---see Fig.~\ref{f:NodalEv1}. The topological conservation laws governing phase vortices and phase anti-vortices are seen to apply: (i) The total topological charge (total winding number) is conserved over time; (ii) the nodal lines (wave-function zeros, eigenvalue trajectories) threading the vortex cores are continuous one-dimensional manifolds that may neither begin nor end at any point within the volume \cite{Paganin2006}. Recall the observation of Dirac \cite{Dirac1931}, that the conservation of topological charge for the phase of the evolving polynomial wave function is independent of the particular equation of motion governing the spatio-temporal evolution of the wave function. We will discuss these topological aspects of phase defects in more depth in Sec.~\ref{s:Top}.

\subsection{Special cases of characteristic equation}

As emphasized by several workers, a particular utility of finite-order polynomial wave functions is their giving a convenient {\em local} description of a wide variety of complex fields \cite{NyeBerry1974,Nye1999}.  While the order-$N$ polynomial wave functions induced by $N \times N$ matrices may indeed provide such a local description, leading to space-time networks of $N$ nodal lines such as that in Fig.~\ref{f:NodalEv1}, one may seek a more localized analysis still, in which only a small number of nodal lines feature.  In particular, one may be inspired by an evident analogy with the ``elementary processes'' of quantum electrodynamics and its generalisations, in which Feynman diagrams of arbitrary complexity may be assembled by constructing all topologically-distinct concatenations of a relatively small number of processes (e.g. the electron--photon vertex, the quark--gluon vertices etc.) \cite{Maggiore2005}.  Similarly, we may examine the space-time nodal-line networks such as that given in Fig.~\ref{f:NodalEv1}---together with the more complicated networks that shall arise later in the paper---and seek to describe the corresponding ``elementary topological processes'' by considering the temporal evolution of characteristic-polynomial wave functions of very low order.  While an $N=1$ characteristic-polynomial wave function would suffice to locally describe the trivial topological dynamics evident in Fig.~\ref{f:NodalEv1}, we shall see that the $N=2$ and $N=3$ cases suffice to cover many of the topological dynamics considered in the present paper.  This motivates consideration of exact formulae for the nodal-line dynamics corresponding to the $N=2$ and $N=3$ cases, a topic to which we now turn.

We consider the Hamiltonian in $\bM(t)$ in Eq.~(\ref{e:Hmat}), but emphasize that the logic below may be applied more generally. Expand each eigenvalue $\lambda_j$ to first order in $t$:
\begin{align}
\label{e:linev} \lambda_j (t) = \lambda_{j, 0} + \lambda_{j, 1} t + O(t^2).
\end{align}
Hence the eigenvalue velocity at position $\lambda_{j, 0}$, together with the velocity of the associated vortex core in the characteristic-polynomial wave function, is given by $\lambda_{j, 1}$.  We now calculate this velocity for the cases $N=2$ and $N=3$, before generalizing to the case of arbitrary $N \ge 1$.

\subsubsection{Vortex velocity for N=2}

Consider the $2\times 2$ matrix
\begin{align}
\bM_0= \left[ \begin{array}{cc}
a& b\\
c&d
\end{array}
\right],
\end{align}
where each of the entries $a,b,c,d$ is a random complex number.  The eigenvalues of $\bM_0$ are
\begin{align}
\lambda_{\pm, 0}= \frac{a+d \pm \sqrt{(a-d)^2 +4bc}}{2}.
\end{align}
Now we add the perturbing matrix
\begin{align}
\bS= \left[ \begin{array}{cc}
s& 0\\
0&-s
\end{array}
\right]
\end{align}
and calculate the eigenvalues of $\bM= \bM_0 +t \bS$, to give:
\begin{align}
\lambda_{\pm, 1} &= \pm \frac{d-a} {\sqrt{(a-d)^2 +4bc}} s\\
&= \pm \frac{d-a} { 2\lambda_{ \pm, 0} -a-d} s.
\end{align}

\subsubsection{Vortex velocity for N=3}

Define the fixed matrix
\begin{align}
\bM_0= \left[ \begin{array}{ccc}
a& b &c\\
d&e & f\\
g& h &k
\end{array}
\right],
\end{align}
where, again, each of the entries is a random complex number. For the odd-sized matrices, we need to modify the definition in Eq.~(\ref{e:Hmat}) of the deformation matrix, taking the following as the definition:
\begin{align}
\bS= \left[ \begin{array}{ccc}
s& 0&0\\
0&0&0\\
0&0& -s
\end{array}
\right].
\end{align}
In principle, we can write down the exact solution for cubic equations and so there are exact expressions for the eigenvalues of $\bM_0$ \cite[\S 1.11]{NIST_DLMF} (these expressions date back to at least Cardano in 1545), however for our purposes, the leading order in $t$ will suffice. Using Eq.~(\ref{e:linev}), we obtain
{\small\begin{align}
\lambda_{j,1} = \frac{[ae +fh -ek -bd + (k-a) \lambda_{j, 0}]s} {bd-ae+cg+fh -ak -ek +2\lambda_{j, 0} (a+e+k) -3 \lambda_{j, 0}^2}
\end{align}}
for $j=1,2,3$.

\subsubsection{Vortex velocity for arbitrary N}

Now consider vortex (eigenvalue) velocity for arbitrary $N$, focussing on the velocity of a particular eigenvalue, $\lambda_j(t)$. The time-dependent characteristic polynomial can always be factored as (cf. Eq.~(2) in \citeauthor{Groszek2018} \cite{Groszek2018}): 
\begin{align}
\chi(\lambda;t)=[\lambda_j(t)-\lambda]\tilde{\chi}_j(\lambda;t).
\label{eq:EigenvalueVelocity1}
\end{align}
The above expression defines the modulating function (envelope)  $\tilde{\chi}_j(\lambda;t)$, which (for a polynomial) we can write down explicitly
\begin{align}
\tilde{\chi}_j(\lambda;t)= \prod_{k\neq j} [ \lambda_k(t)- \lambda ].
\end{align}	
Differentiating with respect to $t$ gives $\dot{\chi}(\lambda;t)$:  
\begin{align}
\bM(t)\longrightarrow \chi(t)\longrightarrow\dot{\chi}(t).
\label{eq:EigenvalueVelocity2}
\end{align}
Note that Jacobi's formula for the derivative of a determinant may be used to evaluate $\dot{\chi}(\lambda;t)$ as:
\begin{align}
\dot{\chi}(\lambda;t)=\textrm{tr}\left\{\dot{\bf{M}}(t) \textrm{adj} [{\bf M} (t) - \lambda \bf{I}] \right\},
\label{eq:EigenvalueVelocity3}
\end{align}
where tr denotes matrix trace, and adj denotes the adjugate matrix (i.e. the transpose of the co-factor matrix).  Regardless of how one chooses to calculate $\dot{\chi}(\lambda;t)$, differentiating Eq.~(\ref{eq:EigenvalueVelocity1}) with respect to time gives: 
\begin{align}
\dot{\chi}(\lambda;t)=\dot{\lambda}_j(t)\tilde{\chi}_j(\lambda;t)+[\lambda_j(t)-\lambda]\dot{\tilde{\chi}}_j(\lambda;t)
\label{eq:EigenvalueVelocity4}
\end{align}
(If $\chi$ is polynomial then the quantities in Eq.~(\ref{eq:EigenvalueVelocity4}) are manifestly differentiable.)

Now consider a (time-dependent) small open disk $\Omega_j(t)\subset \mathbb{R}^2$ around the eigenvalue $\lambda_j(t)$, such that there is no other eigenvalue $\lambda_k$ in $\Omega_j(t)$. This implies that $\tilde{\chi}_j(z;t)$ is non-vanishing for all $z \in \Omega_j(t)$. Note that for Ginibre matrices such an open disk almost surely exists, since one can assume the zeros of the characteristic polynomial to be isolated, up to an irrelevant set of measure zero.  Evaluate Eq.~(\ref{eq:EigenvalueVelocity4}) at the vortex core $\lambda=x+iy=\lambda_j(t)\in \Omega_j(t)$, thereby eliminating the second term of this equation. Since $\tilde{\chi}_j(\lambda;t) \neq 0$ on $\Omega_j(t)$ we can divide through by this quantity to give the eigenvalue velocity (vortex velocity):       
\begin{align}
\nonumber \lambda_{j,1} &= \left[\frac{\dot{\chi}(\lambda;t)}{\tilde{\chi}_j(\lambda;t)}\right]_{\lambda=\lambda_j(t)} \\ &= \left. \left[\frac{\textrm{tr}\left\{\dot{\bf{M}}(t) \textrm{adj} [{\bf M} (t) - \lambda \bf{I}] \right\}}{\textrm{det} [{\bf M} (t) - \lambda {\bf{I}}]/[\lambda_j(t)-\lambda]}\right] \right|_{\lambda=\lambda_j(t)}.
\label{e:EigenValueVelocity}
\end{align}

The above calculation harmonises with the idea that vortices and anti-vortices may be considered as quasi-particles---e.g. it is a direct analog of the result in Eq.~(9) of \citeauthor{Groszek2018} \cite{Groszek2018}, for a point-vortex-model velocity associated with screw-type phase defects in the solutions to the Gross--Pitaevskii equation. Our point-like objects in 2D have space-time trajectories such as that in Fig.~\ref{f:NodalEv1}.  The associated force that a given vortex experiences at a given instant of time will then be proportional to the derivative of the eigenvalue velocity with respect to time; this eigenvalue acceleration may in turn be associated with a ``field'' with which the ``particle'' locally interacts.  It is natural that such a particle-like quality to the vortex trajectories should emerge as a simple consequence of the formalism outlined in the present paper, since such a connection between vortices and associated quasi-particles is well known in the literature on vortical wave functions: see e.g. \citeauthor{Groszek2018} \cite{Groszek2018}, and references therein.  

\section{Evolving-matrix model for vortex--anti-vortex gas}\label{s:QEvs}

Notwithstanding the previously mentioned list of physical systems whose wave functions admit vortices of only one sign, there is a much  wider class of vortical systems where windings of both signs are present. Examples include (2+1)-dimensional coherent optical speckle beams \cite{Nye1999}, chaotic wave-packet evolution in a Buminovich stadium \cite{Berggren2001}, paraxial propagation of coherent x-rays scattered by spatially random media \cite{Kitchen2004} and turbulent Bose--Einstein condensates \cite{Ruben2008}. This prompts us to generalise our formalism to this broader class of matrix-induced polynomial wave functions.  The present section therefore considers polynomial wave functions with an equal number of vortices and anti-vortices, with the subsequent section further generalising to the case where the number of vortices and anti-vortices is not equal.

\subsection{Generalized characteristic-polynomial wave functions}\label{s:genwave}

We incorporate anti-vortices by forming polynomials in a complex variable and its complex conjugate \cite{Freund1999a,Freund1999b,SmitGbur2016}, 
\begin{align}
\label{e:quatfact} \prod_{j=1}^{N/2} (\lambda- \beta_j) (\lambda^* - \alpha_j).
\end{align}
The associated polynomial wave function
\begin{align}
\nonumber &\prod_{j=1}^{N/2}\{[x-\Re(\beta_j)]+i [y- \Im(\beta_j)]\} \\
\label{e:quatfact1} &\times \{[x-\Re(\alpha_j)]-i [y+ \Im(\alpha_j)]\}
\end{align}
will have $N/2$ vortices at positions $(\Re(\beta_j), \Im(\beta_j))$ and $N/2$ anti-vortices at positions $(\Re(\alpha_k), - \Im(\alpha_k))$.

In analogy with Eq.~(\ref{e:charpoly0}) we would like a determinantal representation of this wave function and so we define
\begin{align}
\label{e:quatdet} \chi(\lambda, \lambda^*; t)= \det(\bM(t) - \bLam_{N/2}),
\end{align}
where $\bM$ is an even dimensional complex Ginibre matrix as in Eq.~(\ref{d:Ginmat}) and
\begin{align}
\bLam_{N/2}= \bI \otimes \left[ \begin{array}{cc}
\lambda & 0\\
0& \lambda^*
\end{array}\right].
\end{align}
We then define the wave function by
\begin{align}
\label{e:Psi0} \Psi_{N, 0} (x,y;t)= \chi(\lambda, \lambda^*; t),
\end{align}
noting that there is now a winding number of zero when tracing any closed contour that contains all zeros of Eq.~(\ref{e:Psi0}). This determinant representation is reminiscent of quaternionic structures that exist in random matrix theory---see details regarding quaternionic matrix ensembles in Appendix \ref{a:quats}. Quaternionic ensembles, also termed ``symplectic ensembles'', are one of the three classic universality classes identified by Dyson in his seminal sequence of papers from 1962 \cite{Dyso1962I, Dyso1962II, Dyso1962III, Dyso1962ThreeFold}.

From Appendix \ref{a:quats} we see that $\bLam_{N/2}$ is a matrix whose diagonal $2\times 2$ blocks are of the form in Eq.~(\ref{e:22q}), with $\alpha= \lambda\in \mathbb{C}, \beta=0$. Hence we can view the function $\chi$ in Eq.~(\ref{e:quatdet}) as mapping between planes that are embedded in four-dimensional quaternionic space, and each of these is isomorphic to the complex plane.

Interestingly, there is not always $N$ zeros $(x,y) \in \mathbb{R}^2$ of Eq.~(\ref{e:Psi0}). In Sec.~\ref{s:fundproc} we will explicitly make use of the quaternion embedding to access the solutions in the case that these zeros do not exist. As shall be seen, this potential lack of a full set of solutions leads naturally to vortex--anti-vortex pair creation and pair annihilation processes, which further leads us to consideration of an additional topological network associated with maxima, minima and  saddles in the phase of the generalised characteristic-polynomial wave function.  This additional network has its own topological conservation laws, and interacts with the previously considered nodal-line network in a well defined manner.

A point we would like to stress is that the wave-function zeros in this section (and the following section) are no longer eigenvalues, as Eq.~\eqref{e:quatdet} is no longer the characteristic polynomial of the matrix.

\subsection{Simulations}\label{s:Qsims}

\begin{figure}[h]
\includegraphics[scale=0.44]{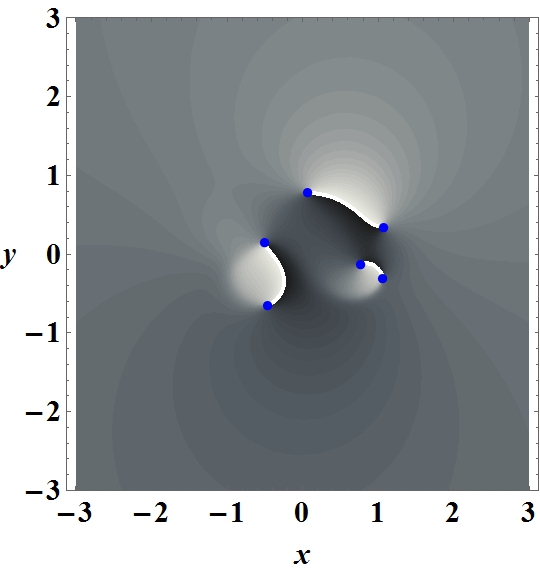}
\caption{\label{f:PhaseQ1}(Color online) Plot of $\Phi_{10, 0} (x,y;0.7485)= \arg[\Psi_{10, 0} (x, y; 0.7485)]$, where $\bM (t)$ is from Eq.~(\ref{e:Hmat}) and the matrix $\bM_0$ is the same as that used in Figs.~\ref{f:PhaseEv1} and \ref{f:NodalEv1}. The blue dots are the zeros of $\Psi_{10, 0} (x, y; 0.7485)$.}
\end{figure}

Here we use the same matrix as that used to generate Figs.~\ref{f:PhaseEv1} and \ref{f:NodalEv1}, together with the same matrix evolution law as in Eq.~(\ref{e:Hmat}).  However, we now use the wave function $\Psi_{10, 0} (x,y;t)$ from Eq.~(\ref{e:Psi0}). We see in Fig.~\ref{f:PhaseQ1} that at $t=0.7485$ there are six solutions, three of which are vortices and three of which are anti-vortices. Each solution is either a vortical or anti-vortical branch point for the multi-valued phase of the determinantal polynomial wave function, with half having winding number $m=1$ and half having winding number $m=-1$.  The absence of a net phase winding is evident, since no branch lines cross the edge of the sampled domain.  Once again, the specific locations of the branch lines themselves have no physical meaning.

\begin{figure}[h]
\includegraphics[scale=0.5]{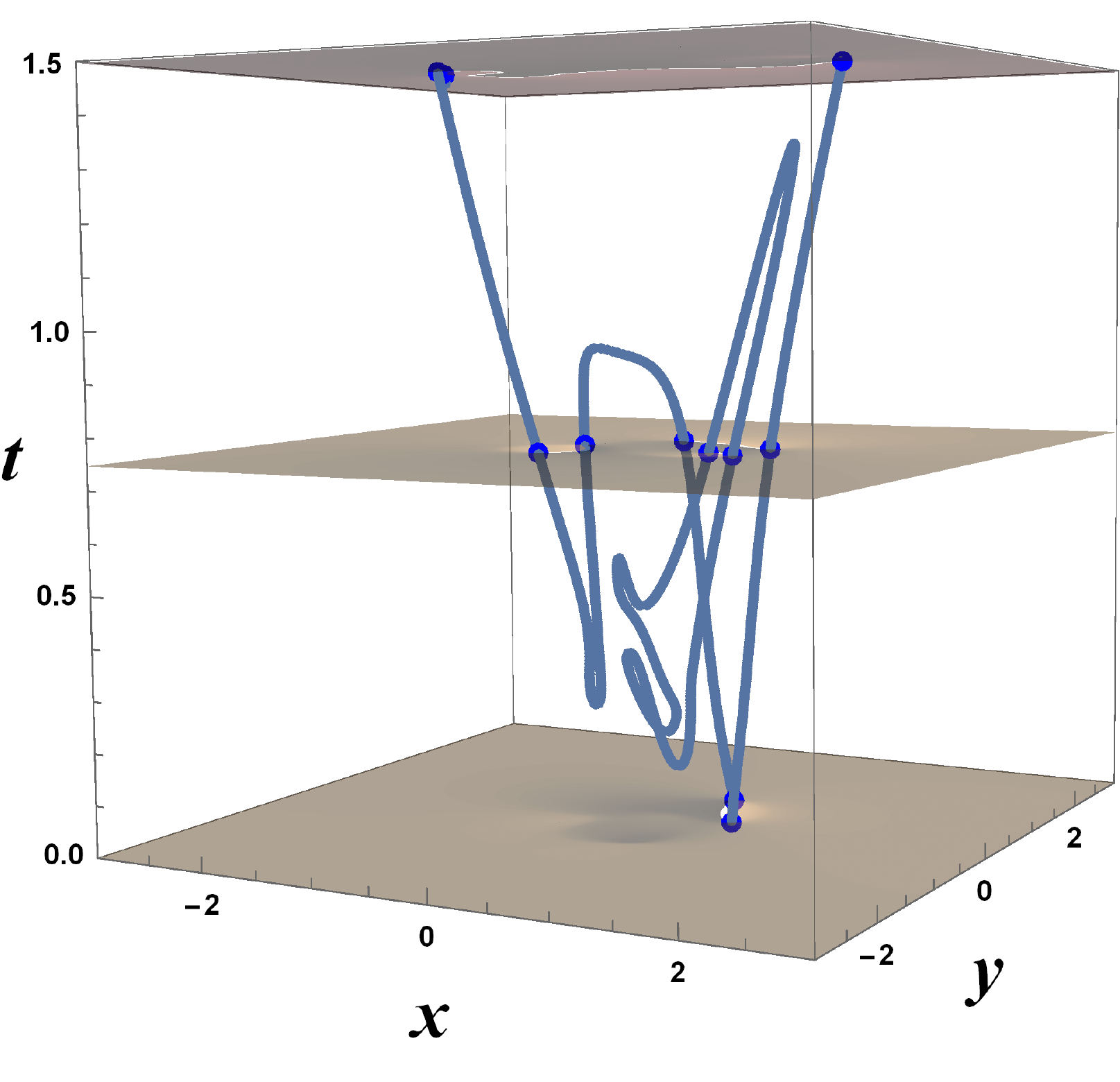}
\caption{\label{f:NodalQ1}(Color online) Trajectories of the zeros of $\Psi_{10, 0} (x, y; t)$ from Fig.~\ref{f:PhaseQ1}. See Supplemental Material at [psi10-0\_500frame50co100dpi.avi] for a video of this system.}
\end{figure}

When the nodal lines threading the vortex cores of $\Psi_{10, 0} (x,y;t)$ are plotted in the three-dimensional space-time volume, Fig.~\ref{f:NodalQ1} results.  Note that Fig.~\ref{f:PhaseQ1} is the middle layer of this diagram. At $t=0$ there is one vortex--anti-vortex pair (dipole), and the vortex merely moves transversely as $t$ increases from $0$ to $1.5$.  The anti-vortex at $t=0$ traces out a hairpin structure in space-time, corresponding to a vortex--anti-vortex pair being created, with the created vortex annihilating the anti-vortex that was initially present, leaving the created anti-vortex to evolve until $t=1.5$.  There is also a closed nodal-line loop, corresponding to several events of spontaneous creation and subsequent annihilation of vortex--anti-vortex pairs.  A maximum of three vortices and three anti-vortices is seen at any one time in this simulation, of which the diagram in Fig.~\ref{f:PhaseQ1} is an example. Note that the winding number $w=0$ is invariant over time.

\section{Determinantal model for an arbitrary number of vortices and anti-vortices}\label{s:GenEvs}

Having introduced the possibility of anti-vortices we generalize the wave functions in Eqs.~(\ref{e:wavfn}) and (\ref{e:Psi0}) further, to admit an arbitrary number of vortices and anti-vortices. To this end, we define
\begin{align}
\label{e:gendet} \Psi_{N,2\xi -N} (x,y;t)= \chi (\lambda, \lambda^*; t)= \det (\bM - \bLam_\xi)
\end{align}
where 
\begin{align}
\bLam_\xi = \mathrm{diag} (\underbrace{\lambda, \dots, \lambda}_{\footnotesize \mbox{$\xi$ copies}}, \underbrace{ \lambda^*, \dots, \lambda^*}_{ \footnotesize \mbox{$N- \xi$ copies}}),
\end{align}
and so $\xi$ is the number of vortices, and the winding number is $2\xi- N$, which is the difference between the number of vortices and anti-vortices.

In the case that there are $N$ zeros $\{(x_j,y_j)\in \mathbb{R}^2\}_{j=1,\dots N}$ then
\begin{align}
\label{e:genfact} \chi (\lambda, \lambda^*; t) = \prod_{j=1}^{\xi} (\lambda- \beta_j) \prod_{k=1}^{N- \xi} (\lambda^* - \alpha_k)
\end{align}
with associated polynomial wave function 
\begin{align}
\nonumber \Psi_{N,2\xi -N} (x,y;t)&=\prod_{j=1}^{\xi} \{[x-\Re(\beta_j)]+i [y- \Im(\beta_j)]\} \\
\label{e:quatfact2} &\times \prod_{k=1}^{N- \xi}\{[x-\Re(\alpha_k)]-i [y+ \Im(\alpha_k)]\},
\end{align}
and so we interpret $\Psi_{N,2\xi -N}$ as a wave function with (up to) $\xi$ vortices and $N- \xi$ anti-vortices \cite{Freund1999a,Freund1999b,SmitGbur2016}. The cases in Eqs.~(\ref{e:wavfact}) and (\ref{e:quatdet}) are then given by the specialisations $\xi=N$ and $\xi=N/2$ respectively. (We note that when $\xi= N/2$ an equal number of elementary row and column swaps is required to convert the matrix in Eq.~(\ref{e:gendet}) to that in Eq.~(\ref{e:quatdet}), and so the determinant is preserved.)

\subsection{Simulations}

\begin{figure}[h]
\includegraphics[scale=0.4]{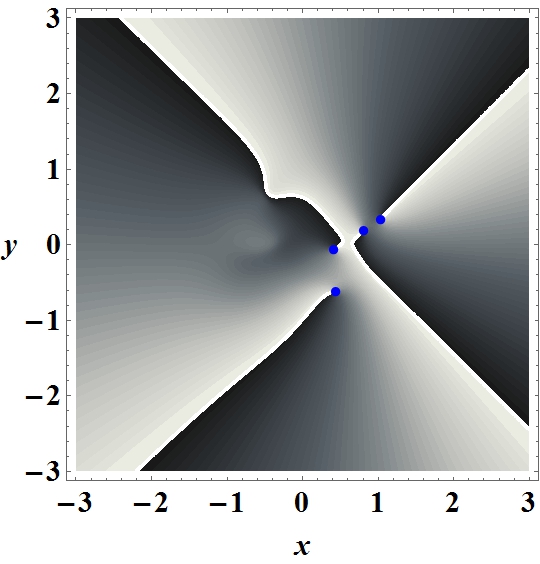}
\caption{\label{f:PhaseGen1}(Color online) Plot of $\Phi_{10, 4} (x,y;0.499)= \arg[\Psi_{10, 4} (x, y; 0.499)]$ from Eq.~(\ref{e:gendet}) with $\xi =7$, where $\bM (t)$ is from Eq.~(\ref{e:Hmat}) and the matrix $\bM_0$ is the same as that used in Figs.~\ref{f:PhaseEv1} and \ref{f:NodalEv1}. At this time, there are four zeros (the blue dots), all of which are vortices. The kink in the line of phase discontinuity in the top left is indicative of a vortex--anti-vortex creation event happening in the near future. A local phase maximum can be seen nearby which will become another pair.}
\end{figure}

\begin{figure}[h]
\includegraphics[scale=0.44]{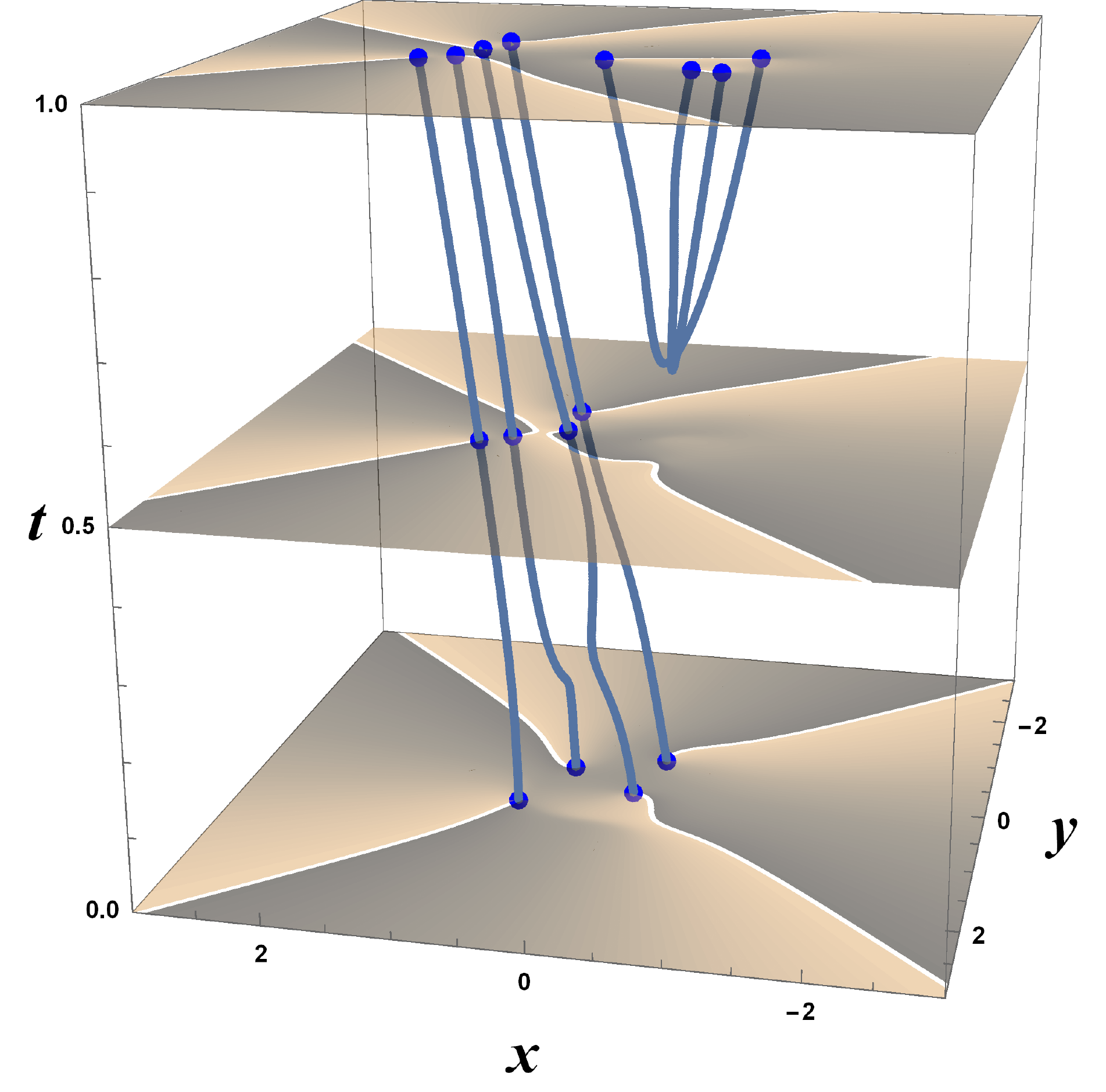}
\caption{\label{f:NodalGen1}(Color online) Trajectories of the zeros of $\Psi_{10, 4} (x, y; t)$ from Fig.~\ref{f:PhaseGen1}. Note that Fig.~\ref{f:PhaseGen1} is the middle layer of this diagram. (The figure has been rotated with respect to the orientation of Fig.~\ref{f:PhaseGen1} to make the hairpin structures clear.) Notice that there are four vortices and no anti-vortices at $t=0$, then two vortex--anti-vortex pairs are created at $t\approx 0.6$, which preserves the total winding (or topological charge) of $+4$. See Supplemental Material at [psi10-4\_500frame50co100dpi.avi] for a video of this system.}
\end{figure}

We again use the same matrix that was used to generate the simulations in Figs.~\ref{f:PhaseEv1} and \ref{f:NodalEv1}.  We calculate the wave function via the determinant in Eq.~(\ref{e:gendet}), with $\xi= 7$, and $\bM (t)$ given by Eq.~(\ref{e:Hmat}). This gives Figs.~\ref{f:PhaseGen1} and \ref{f:NodalGen1}.

\section{Topological features of defect-line interactions}\label{s:Top}

If vortices of both signs can be supported, there are topological conservation laws associated with the vortices, anti-vortices, maxima, minima and saddle points of the phase of the determinantal polynomial and its associated polynomial wave function \cite{Nye1988,Freund1995}. In our (2+1)-dimensional framework, all of the previously-mentioned topological defects are zero-dimensional, tracing out one-dimensional defect lines in $2+1$ dimensions.  We henceforth use the term {\em defect lines} to refer collectively to the 1D trajectories in $2+1$ dimensions, of the phase vortices, anti-vortices, maxima, minima and saddles.  The term {\em nodal lines} refers to the subset of the defect lines, associated with vortices and anti-vortices (which are the points at which the wave function vanishes).  We do not consider domain walls, namely jumps of $\pi$ radians in the phase of the wave function in a given $xy$ plane of constant $t$, since these can be viewed as an unstable special case of the 1D nodal line associated with vortices and anti-vortices, embedded in $2+1$ dimensions, having part of the said nodal line lying within the particular $xy$ plane. 

\subsection{Topological conservation laws}

We follow Dirac \cite{Dirac1931}---who considered the special case of defect lines, namely nodal lines, associated with vortices and anti-vortices---in assuming only the continuity in spatial and temporal variables of the polynomial wave function, together with its single-valuedness.  We also follow Maxwell \cite{Maxwell1870}, who in a seminal paper contributing to the development of what is now known as Morse theory \cite{Nash1992}, considered the complementary case of maxima, minima and saddle points.  Based on the assumption of a single-valued continuous complex polynomial wave function, one can obtain the following topological conservation laws, irrespective of the particular laws governing the spatio-temporal evolution of the wave function:

\begin{enumerate}
\item Vortices and anti-vortices may only be created in $m= \pm 1$ pairs, with total winding number always conserved as $t$ evolves, and only $m=\pm 1$ dislocations (i.e.~a unit winding number, as defined in Eq.~(\ref{e:PhaseWinding})) being topologically stable.  In the (2+1)-dimensional space with coordinates $(x,y,t)$ the associated vortex-cores and anti-vortex cores trace out continuous 1D nodal lines that may neither begin nor end at any finite $(x,y,t)$ coordinate, hence they may either form closed (and possibly knotted) loops or extend to spatial and/or temporal infinity \cite{Dirac1931,Freund1999a}. 
\item The system of 1D nodal lines (where the wave function vanishes), as described above, has a complement in the system of continuous 1D lines associated with saddle points, maxima and minima in the phase of the complex wave function \cite{Nye1988,Freund1995}.  These defect lines, which are not nodal lines (i.e. the wave function does not vanish at these points), may also only form closed continuous loops or extend to spatial and/or temporal infinity.  The topological properties of this network of non-nodal defect lines may be determined by applying Morse theory \cite{Nash1992} to the manifold of $xyt$ points corresponding to all permissible $(x,y,t_s)$ coordinates, for some fixed time $t_s$, with the points corresponding to (vortical) nodal lines being {\em removed}.   These defect lines have the property that: if a saddle-line reverses direction in $t$, it will be transformed to either a local phase maximum or a local phase minimum  line.  Similarly, if a local phase maximum or a local phase minimum line reverses direction in $t$, it will be transformed to a saddle line \cite{Maxwell1870,Arnold1985}. Note that these saddle--extrema creation and annihilation events can also be seen analytically---we discuss a canonical example in Appendix \ref{a:saddlemax}.
\item The previously-mentioned two classes of defect line---namely the nodal lines associated with phase vortices and phase anti-vortices, and the defect lines associated with phase maxima, phase minima, and phase saddles---are coupled to one another \cite{Nye1988,Freund1995,Freund1999a}.  This coupling occurs due to the fact that at $(x,y,t)$ points where a nodal line reverses direction, a maximum--minimum--saddle defect line must pass through the same point.  If the nodal line and the non-nodal defect line respectively occupy the future and the past of the vertex, the defect line will be a maximum-minimum pair.  Any deformation of the temporal sense of the non-nodal defect line, e.g. by reversing the temporal sense of either or both defect lines emanating from the point, transform maxima and minima (i.e. phase extrema) into saddles.
\end{enumerate}

In the above topological conservation laws, reference has been made to vortices, anti-vortices, local maxima and minima, and saddle points.  As previously discussed, both vortices and anti-vortices correspond to topological defects in the phase of the complex wave function, associated with a non-vanishing value for $m$ in Eq.~(\ref{e:PhaseWinding}). However, the value of $m$ is zero for the phase maximum, the phase minimum and the saddle point; this is a direct consequence of the fact that the phase is smooth, continuous and single-valued at phase maxima, phase minima and phase saddles.  The topological character of these three non-vortical defects is associated with a second measure (i.e., in addition to that in Eq.~(\ref{e:PhaseWinding})), associated with a non-vanishing value for the integer $n$ defined by \cite{Nye1988,Freund1995}   
\begin{align}
\frac{1}{2\pi}\oint_{\Gamma}d\theta = n.
\label{eq:PhaseGradientWinding}
\end{align}
Here, for any fixed $t$, $\theta(x,y;t)$ is the angle that $\nabla\Phi(x,y;t)$ makes with respect to the positive-$x$ axis, with the gradient operator $\nabla$ being with respect to the $x$ and $y$ coordinates; $\Gamma$ is a small simple smooth closed curve in the $xy$ plane which encloses a (critical) $xy$ point at which $\nabla\Phi(x,y;t)$ vanishes.  We then have $(m,n)=(1,1)$ for a phase vortex, $(m,n)=(-1,1)$ for a phase anti-vortex, $(m,n)=(0,1)$ for both phase maxima and phase minima, and $(m,n)=(0,-1)$ for phase saddles \cite{Freund1995}.  Points where $(m,n) \ne (0,0)$ can be used to classify all phase defects considered in this paper; the ambiguity between phase maxima and phase minima is resolvable via the sign of the phase Laplacian at $xy$ points where $(m,n)=(0,1)$. Exceptions to these rules correspond to a set with measure zero, and will not be considered here.

The nomenclature for the quantities $m$ and $n$ is not universally agreed upon.  We will use the term \textit{topological charge} for $m$ and the term \textit{topological index} for $n$, which is the convention used e.g. in \citeauthor{Scho2005} \cite{Scho2005}.

The consequence of the discussion so far is that the two topological quantities \cite{Nye1988}
\begin{align}
\label{e:topcharge} w&= \sum m, &&\mbox{\text{(topological charge)}}\\
\label{e:topindex} \chi&= \sum n, &&\mbox{\text{(topological index)}}
\end{align}
are both conserved by continuous deformation of the phase surface. The topological conservation laws 1, 2 and 3 listed above are consequences of the invariance of $w$ and $\chi$, both globally and locally at any interaction. As an important aside, we denote the sum over the topological indices by $\chi$ since, by the Poincar\'{e}--Hopf theorem \cite{Haze1991}, this sum is equal to the Euler characteristic for the manifold, which is a conserved topological quantity.  

\subsection{Primitive vertices in defect-line graphs}

\begin{figure}[h]
\begin{center}
\includegraphics[scale=0.62, clip=true, trim= 0 0 0 0]{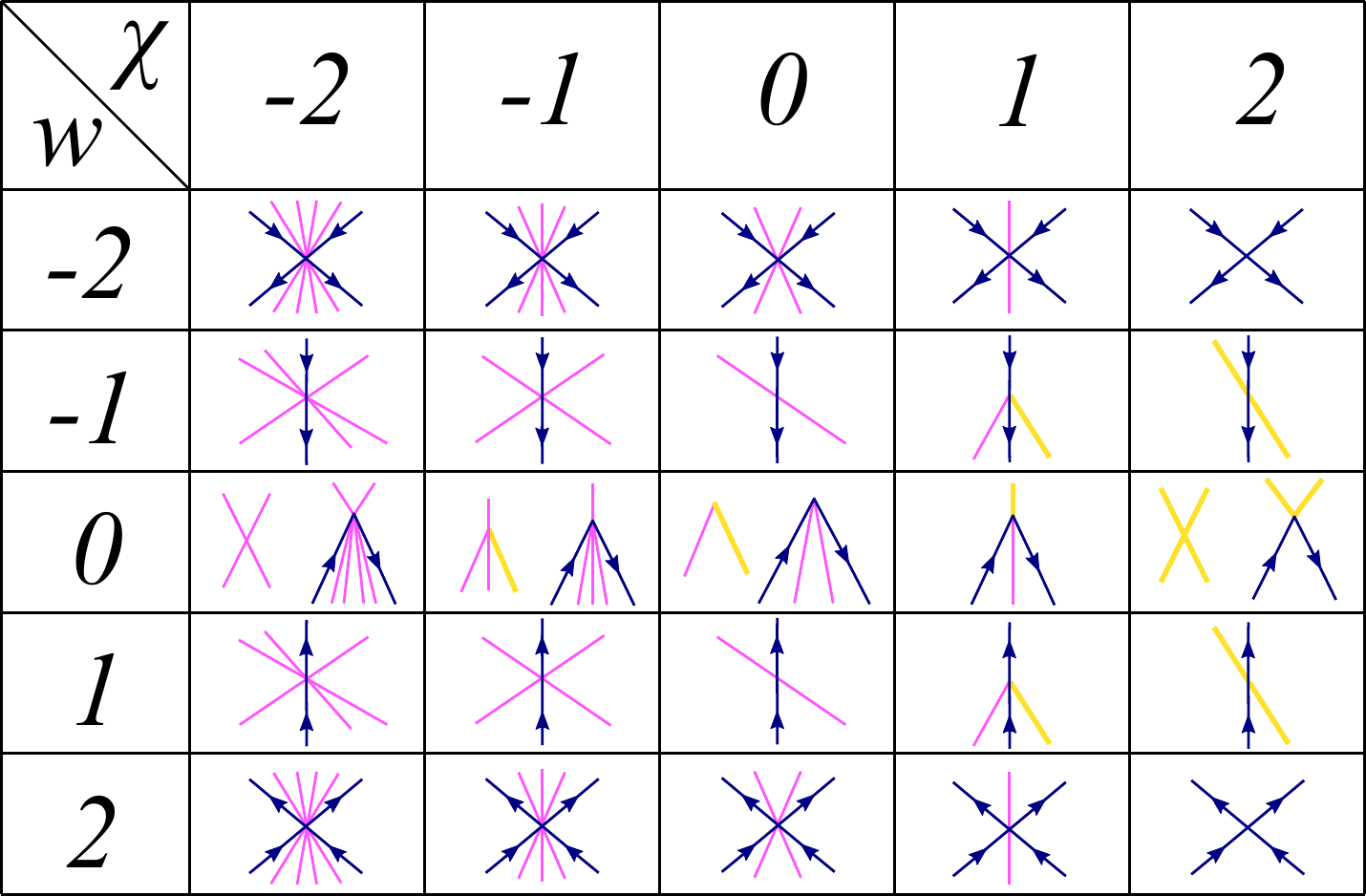}
\caption{\label{f:FundamentalProcesses}(Color online) All minimal interactions with $w,\chi \in \{ -2,-1,0 ,1,2 \}$ between the topological features of $\arg(\Psi)= \Phi$. The time arrow runs from bottom to top. The colors match those of the nodal-line figures: vortices and anti-vortices are marked with arrows in blue (black), saddles are magenta (gray) lines and maxima and minima (extrema) are yellow (light gray) lines.}
\end{center}
\end{figure}

These observations lead immediately to the question of {\em primitive vertices}, namely the idea that all topological reactions of defect lines are ultimately reducible to reactions involving a small number of  such lines.  Conservation of both total topological charge $w$ and total topological index $\chi$, at a given space-time vertex where multiple defect lines converge, implies that one can classify the set of all possible vertices into a set of equivalence classes, here termed ``$(w,\chi)$  events''.  There is an infinite hierarchy of such topological reactions.  Figure~\ref{f:FundamentalProcesses} shows the processes corresponding to $w,\chi\in \{-2,-1,0,1,2\}$.  This figure adopts the defect-line coloring used throughout the paper, but adds an arrow to the nodal-line trajectories to distinguish a vortex (blue or black arrow pointing from past to future) from an anti-vortex (blue or black arrow pointing from future to past).  Note the {\em crossing symmetries} evident in this figure, which can be used to deform certain entries into one another: e.g. reversing the time-sense in which a nodal line evolves converts vortices into anti-vortices and vice versa, and reversing the time-sense in which a saddle moves converts it into a maximum or a minimum, and vice versa.    

If one is considering a given space-time volume and its associated defect-line network, for an ensemble of random matrices at $t=0$ and a specified evolution law, then ensemble averaging will in general induce a statistical weight (probability of occurrence) for each of the minimal interactions in Fig.~\ref{f:FundamentalProcesses}.  We conjecture that this probability of occurrence will typically decrease with increasing magnitude of the topological charge and topological index.  This point will not be further explored in the present paper, but we do note here that particular processes will be more likely to occur than others, consistent with the fact that some but not all were observed in the numerical experiments presented here.  We also note that this table of point interactions could be augmented by a table of possible defect line topologies, although with the exception of the trefoil nodal-line knot (which we discuss below) the question of defect-line knots will not be further considered here.

\subsection{Some defect-line topological reactions}

\begin{figure}
\includegraphics[width=8cm]{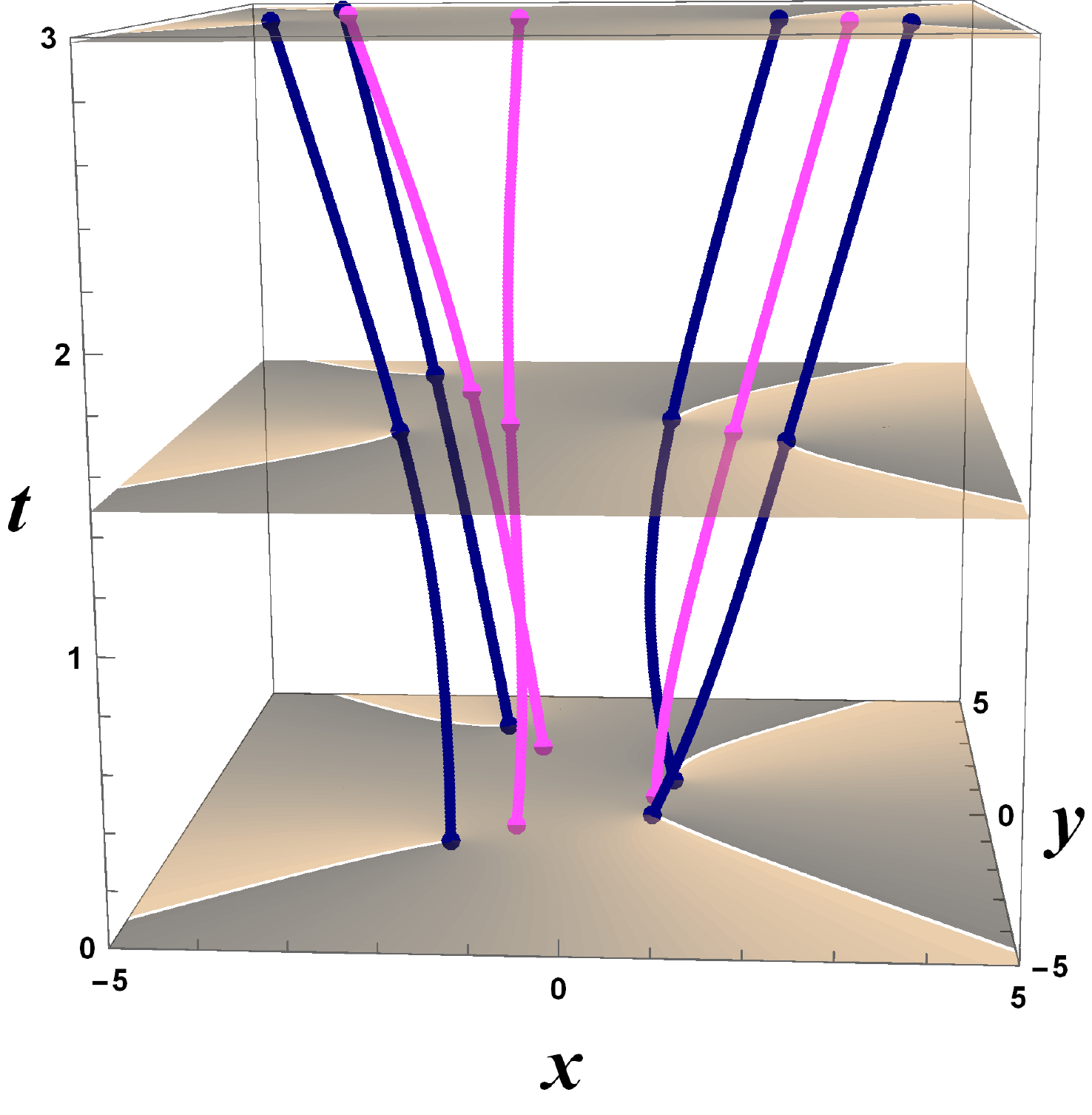}
\caption{\label{f:NodalEv1Phi}(Color online) The blue (black) lines are the eigenvalues of a $4\times 4$ matrix (i.e. the zeros of  $\Psi_{4,4}(x,y;t)$), which are seen to be all vortices. The magenta (gray) lines are the zeros of $\nabla \Phi$, all of which are saddles.}
\end{figure}

\begin{figure}
\begin{center}
\includegraphics[width=8cm]{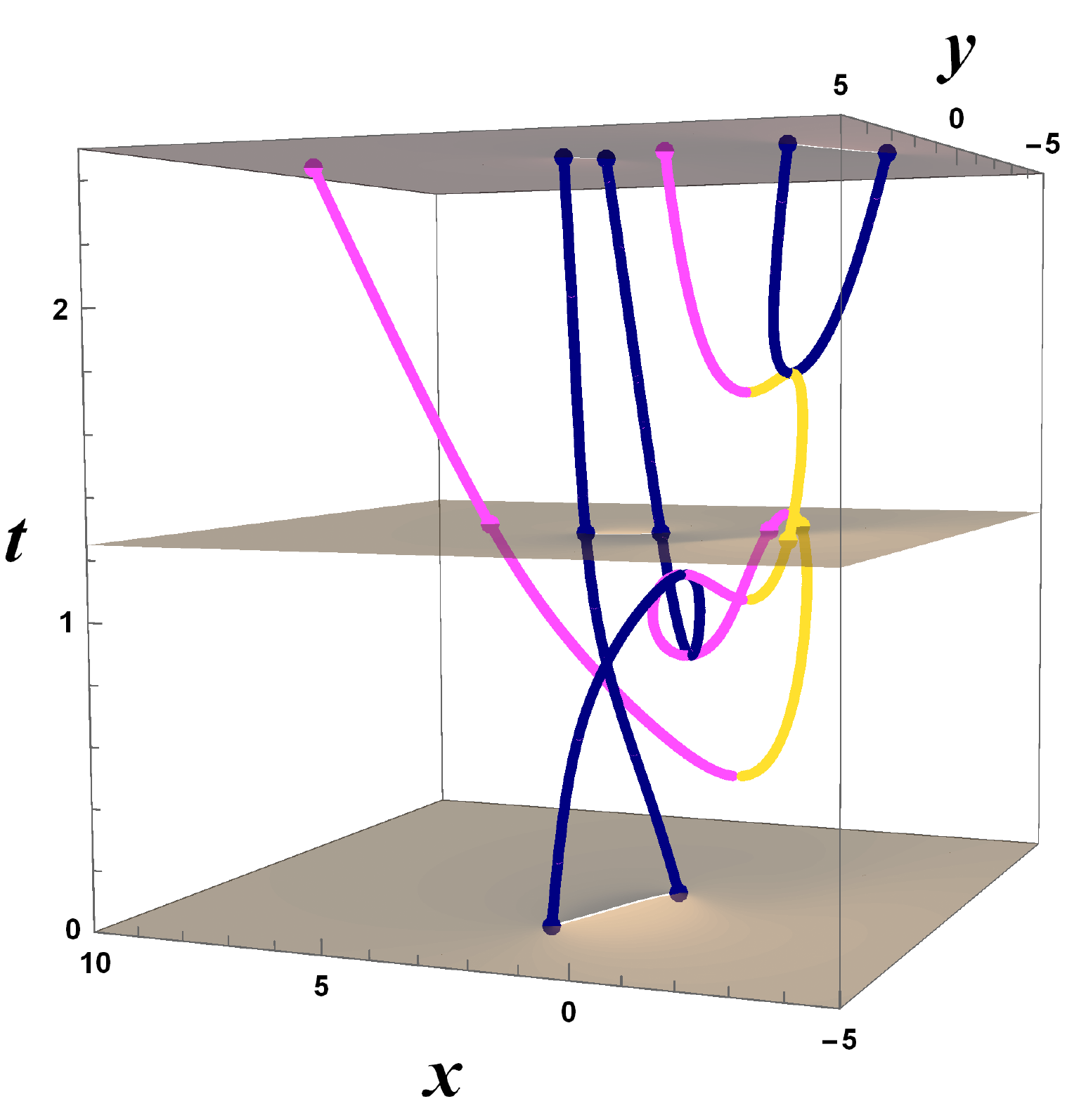}\\
(a)\\
\includegraphics[width=8cm]{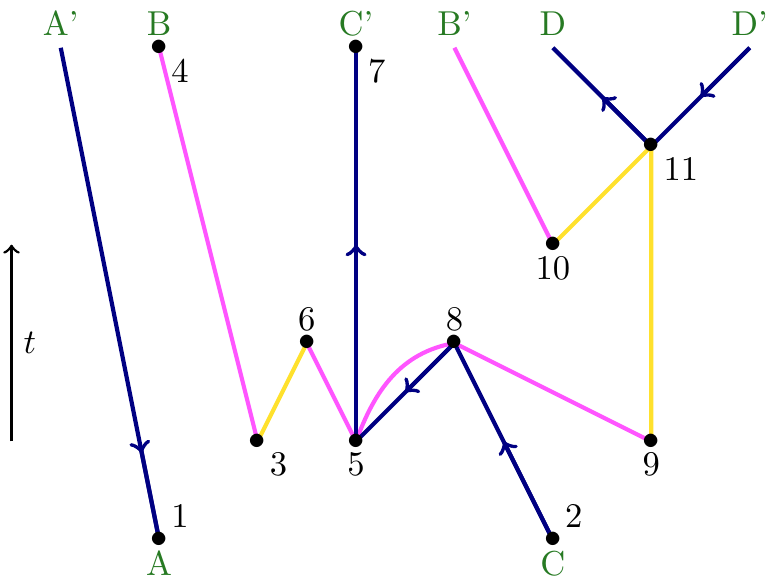}\\
(b)\\
\caption{\label{f:NodalQ1Phi}(Color online) (a) Same matrix as Fig.~\ref{f:NodalEv1Phi}, but with $\Psi_{4,0} (x,y;t)$, giving vortex--anti-vortex pairs. Blue (black) lines are the vortices--anti-vortices (zeros of $\Psi$), magenta (gray) lines are the saddles of $\Phi$ and yellow (light gray) lines are the maxima and minima of $\Phi$.  See Supplemental Material at [psi4-0\_500frame50co100dpi.avi] for a video of this system. (b) Representation of the same topological reaction using a planar graph.}
\end{center}
\end{figure}

\begin{figure}
\includegraphics[width=8cm]{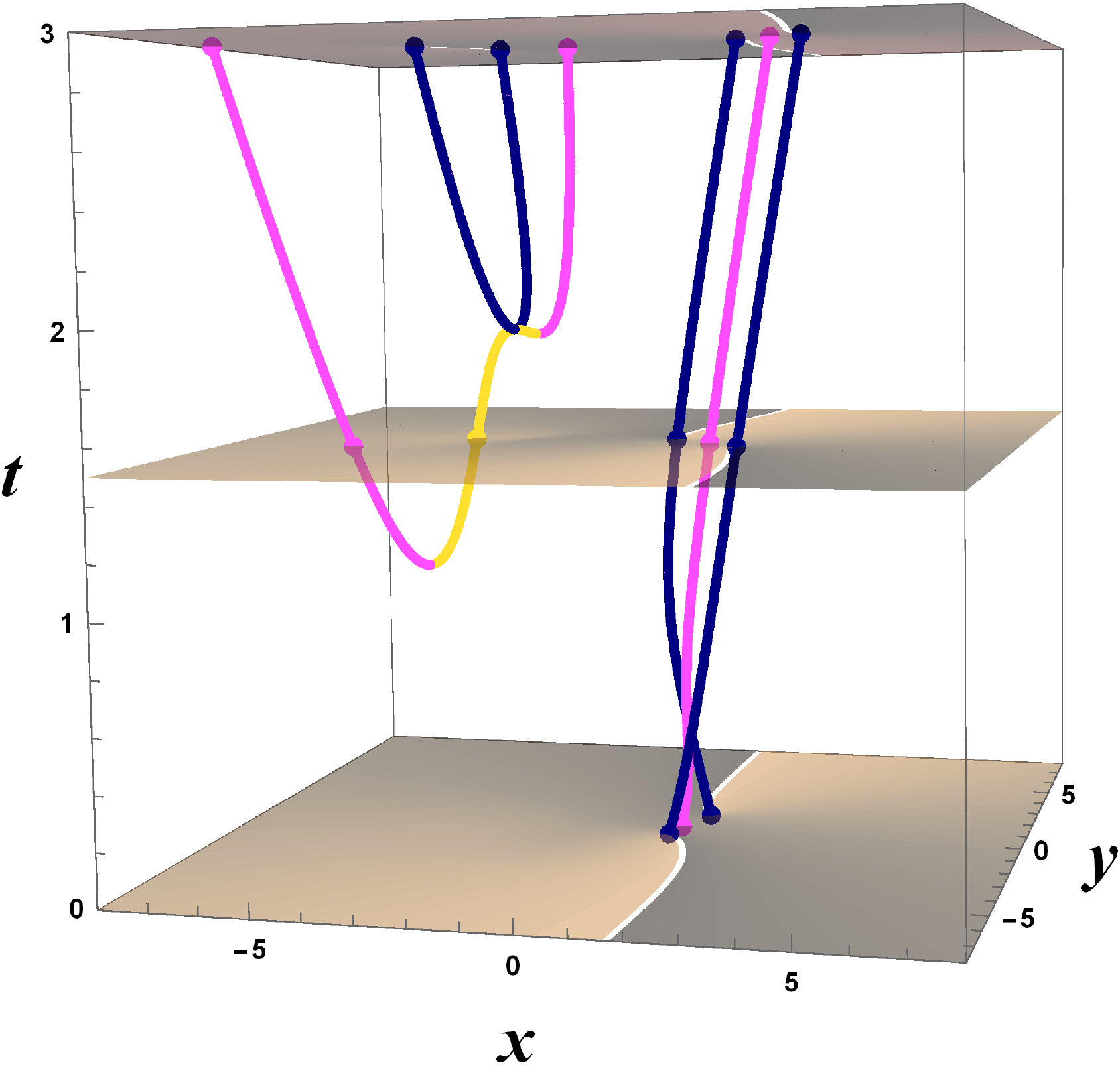}
\caption{\label{f:NodalGen1Phi}(Color online) Same matrix as Fig.~\ref{f:NodalEv1Phi}, but with $\xi =3$. Blue (black) lines are the vortices and anti-vortices (zeros of $\Psi$), magenta (gray) lines are the saddles of $\Phi$ and yellow (light gray) lines are the maxima and minima of $\Phi$.}
\end{figure}

We can study the interactions in Fig.~\ref{f:FundamentalProcesses} using the matrix model Eq.~(\ref{e:Hmat}). For example, using a single $4\times 4$ matrix for $\bM_0$ we plot the zeros of $\Psi_{4,4} (x,y;t)$ in Fig.~\ref{f:NodalEv1Phi}, $\Psi_{4,0}$ in Fig.~\ref{f:NodalQ1Phi} and $\Psi_{4,3}$ in Fig.~\ref{f:NodalGen1Phi}. In the same figures, we have also plotted the zeros of $\nabla \Phi$, the gradient of the phase of the respective wave functions \footnote{To avoid branch cuts in the phase, we use the identity $|\Psi|^2 \nabla \Phi= \Im(\Psi^* \nabla \Psi)$ to access the phase gradient, however finding the zeros of this expression is more computationally intensive than finding the zeros of $\Psi$, and so we do not have the full topological description for the $10\times 10$ matrix system displayed above in Fig.~\ref{f:NodalEv1Phi}, hence the use of the smaller $4\times 4$ systems in this section.}.
The magenta (gray) and yellow (light gray) lines in these figures represent all possible zeros of $\nabla \Phi$: maxima (yellow), minima (yellow) and saddle points (magenta). We clearly see that vortex--anti-vortex creation and annihilation events are mediated by interactions with zeros of the phase gradient.

In Fig.~\ref{f:NodalQ1Phi}(a) we find significant complexity and identify three different interactions that were tabulated in Fig.~\ref{f:FundamentalProcesses}: 
\begin{itemize}
\item{$w=0, \chi=0$: one instance of vortex--anti-vortex--saddle creation and then one instance of the reverse interaction (annihilation)};
\item{$w=0, \chi=0$: three instances of extremum--saddle creation and one annihilation;}
\item{$w=0, \chi=2$: one instance of extremum annihilation leading to vortex--anti-vortex creation.}
\end{itemize}
Figure~\ref{f:NodalQ1Phi}(a) is represented as a planar graph in Fig.~\ref{f:NodalQ1Phi}(b).  There are only four defect lines $AA'$, $BB'$, $CC'$, $DD'$. The (vortical) nodal line $AA'$ passing through point 1 experiences no topological reaction. The second initial vortex, passing through point 2, has a much more complex evolution.  An extremum--saddle pair is created at point 3 \cite{Arnold1985}, with the saddle persisting through to point 4.  Two saddles together with a vortex--anti-vortex pair are created at point 5; one of these saddles annihilates (at 6) with the extremum created at 3; the vortex created at 5 persists until point 7; the other saddle plus the anti-vortex (from point 5) then annihilate (at 8) both the initial vortex (that passed through 2) together with the saddle arising from the saddle--extremum creation at point 9. The extremum created at 9 meets with another extremum arising from the extremum--saddle creation event at 10, to generate a vortex--anti-vortex pair at 11. So the net topological reaction for the vortex at point 2 is
\begin{align}
v\rightarrow v+v+ v^*+s+s,
\end{align}
which is identical to the reaction $0\rightarrow  v+v^*+s+s$ in \citeauthor{Nye1988} \cite{Nye1988} if a $v$ is cancelled from both sides. Here, $v$ denotes a vortex and $v^*$ denotes an anti-vortex.      

Figure~\ref{f:NodalEv1Phi} appears to have no interactions, and so we may suspect that systems with only one species of particle (eigenvalues or vortices in this case) have no interactions. However, in light of our table of interactions, we reconsider Fig.~\ref{f:NodalEv1} above, where the highlighted eigenvalue interactions are perhaps examples of $(w, \chi)= (2,2)$ in Fig.~\ref{f:FundamentalProcesses}. The plot in Fig.~\ref{f:NodalGen1Phi} does not exhibit any new interactions, but is clearly seen to contain features of both the one-species and two-species systems.

\subsection{Defect-line knots}

\begin{figure}
\includegraphics[scale=0.4]{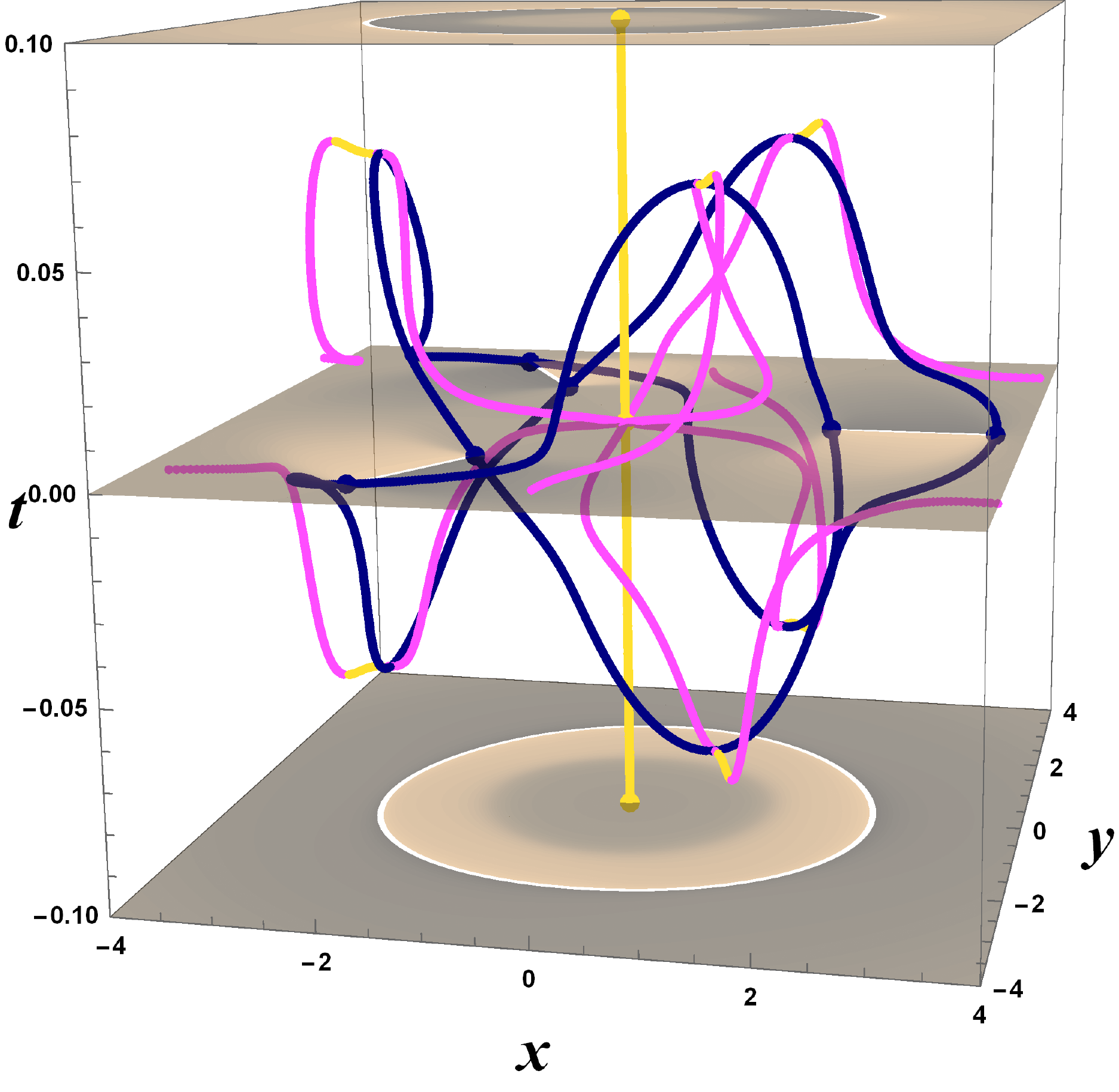}
\caption{\label{f:Trefoil}(Color online) Defect-line knot, plotting the defect lines of the wave function given in \cite[Eq.~17]{PagaBeltPete2018}. The closed blue (black) nodal line, which forms a trefoil knot, traces the zeros of the wave function (which are the same points as plotted in Fig.~5 of \cite{PagaBeltPete2018}), the magenta (gray) points are the saddles of the phase and the yellow (light gray) points are maxima and minima of the phase.}
\end{figure}

Although we have not yet found a realization using the determinantal formalism, we can identify another interaction in Fig.~\ref{f:FundamentalProcesses} using the wave function given by \cite[Eq.~17]{PagaBeltPete2018}, which we have plotted in Fig.~\ref{f:Trefoil}. As pointed out in that paper, which plotted only the nodal lines, the nodal lines of zeros of the wave function form a trefoil knot in (2+1)-dimensional space (cf. the earlier paper by \citeauthor{Leach2005} \cite{Leach2005}, and references therein).  However we can now see additional structure, associated with a ``scaffolding'' of phase extrema and saddles. We see that at each of the vortex--anti-vortex creation and annihilation events we obtain the interaction with $(w,\chi)= (0,1)$ of Fig.~\ref{f:FundamentalProcesses}: the collision of a vortex--anti-vortex pair and a phase saddle to create a phase extremum (or the reverse process).  The possibility of such knotted nodal lines has been previously considered by Freund \cite{Freund2000}, as well as being achieved in experiment using visible light \cite{Leach2005} and water \cite{Kleckner2013}, all of this work having parallels with Lord Kelvin's defunct yet fruitful model of atoms as knotted vortex rings \cite{Kelvin1867}.  One can think of nodal-line knots, such as that shown in Fig.~\ref{f:Trefoil}, as a ``topology of topologies'' insofar as they constitute topologically-nontrivial constructs comprised of one-dimensional manifolds which are themselves topological in origin (cf. \citeauthor{Mawson2018} \cite{Mawson2018}).

\subsection{Defect-line bubbles}

Return to the $(w,\chi)=(0,0)$ cell in Fig.~\ref{f:FundamentalProcesses}.  All such defect complexes can be excited out of the topological vacuum $\varnothing$---i.e. the topologically trivial uniform phase background---and then decay back to $\varnothing$.  The set of all such complexes, in their creation and subsequent decay, comprise an infinite set of topological-vacuum fluctuations containing no external lines.  

\begin{figure}
\begin{center}
\includegraphics[width=8cm]{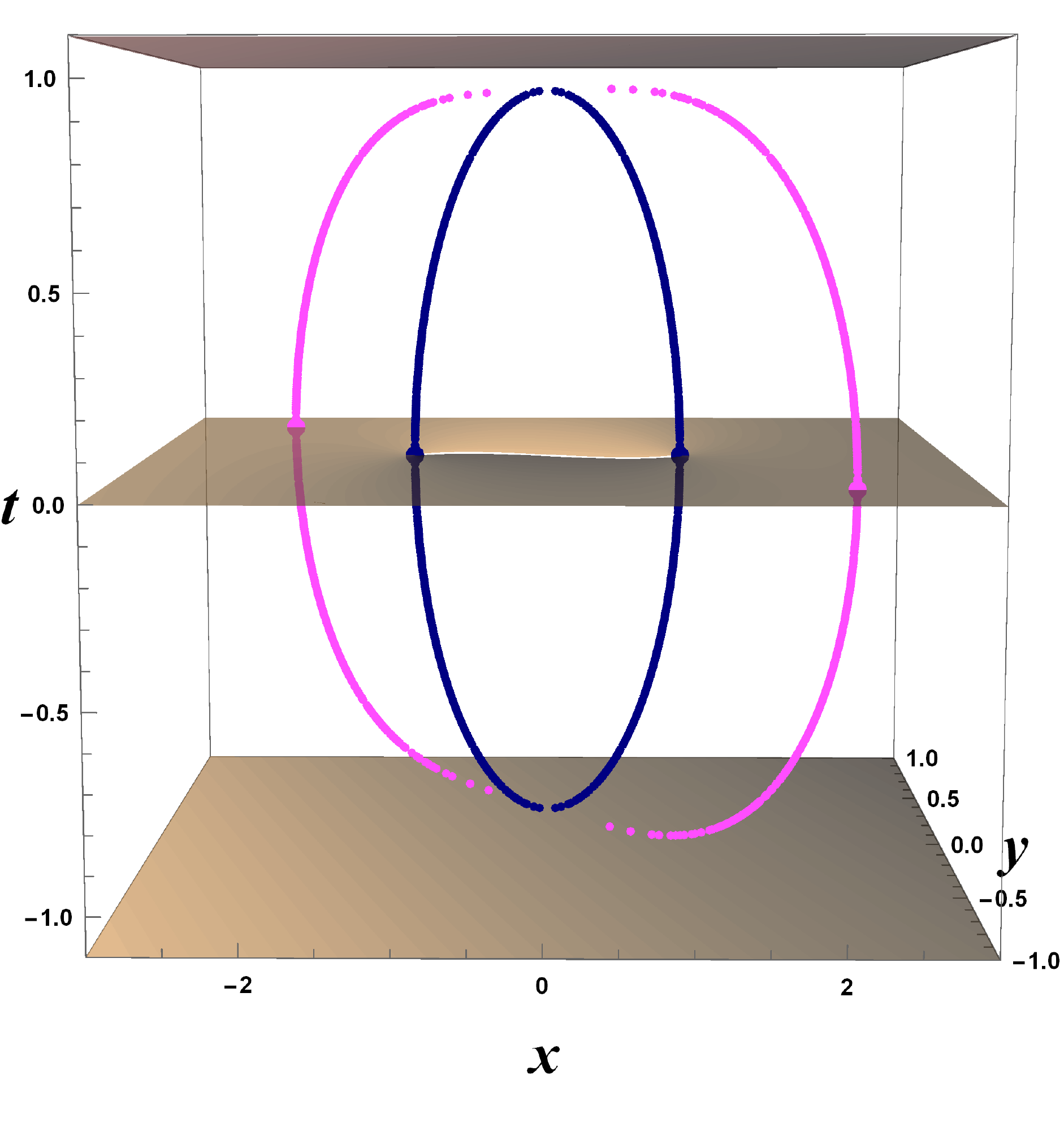}
\caption{\label{f:VortSaddle1}(Color online) Topological-vacuum fluctuation (vacuum diagram), namely a defect-line complex containing no external lines.  This is obtained as a plot of the zeros of $\Psi(x,y)$ [blue (black)] and $\nabla \Phi(x,y)$ [magenta (gray)] from Eq.~(\ref{e:f:VS1}), with $T=1$, $-1.1\leq t \leq 1.1$.}
\end{center}
\end{figure}

For an example of such a defect-line bubble, consider the polynomial wave function
\begin{align}
\label{e:f:VS1} \Psi(x,y;t)= \left(x+i y -X_0 \right) \left(x-i y -X_0 \right) (1- i (x+y)),
\end{align}
where $X_0= \sqrt{T^2 -t^2}, T>0$. In Fig.~\ref{f:VortSaddle1} we see that this wave function has locally flat (although tilted) phase surface for $t<-T$, at which point a vortex--anti-vortex pair is created and the phase surface becomes discontinuous. This  creation is simultaneous with the creation of a pair of phase saddles \cite{Nye1988,Freund1995} at position $(0,0)$ of Fig.~\ref{f:FundamentalProcesses}. The vortex--anti-vortex pair and the phase saddles all annihilate at $t=T$, with the phase surface returning to local flatness.  

Such localised defect-line structures containing no external lines, being topologically allowed, would be expected to occur with non-vanishing probability in ensembles of random polynomial fields generated e.g.~by suitable random-matrix ensembles. The possibility even exists for defect-line bubbles to be knotted or braided. All of this is in line with the familiar precept that ``...any process which is not forbidden by a conservation law actually does take place'' \cite{Gell-Mann1956}.  

In analogy to particle physics, one could consider such structures as polarizing the (topological) vacuum \cite{MartinShaw1997}, since external lines may have their trajectories (and indeed their topologies) influenced by such localised defect-line structures.  For example, return to Fig.~\ref{f:NodalQ1Phi}b and consider vertex 5.  The two magenta (gray) lines and two blue (black) lines nucleated at vertex 5 are a topological-vacuum fluctuation since they are precisely what is seen at $t=-1$ in Fig.~\ref{f:VortSaddle1}. However, unlike the case in Fig.~\ref{f:VortSaddle1}, the four defect lines nucleated at point 5 in Fig.~\ref{f:NodalQ1Phi}b do not all mutually annihilate, but rather couple to external lines such as the one between vertices 2 and 8.  The $x<0$ space-time volume of Fig.~\ref{f:NodalGen1Phi} gives another example of a topological-vacuum fluctuation, nucleated from $\varnothing$, with the resulting defect-line network persisting until the final time shown ($t=3$).   

\section{\label{s:AlgebraicStructureAndSupermultiplets}Algebraic structure: defect-line interactions and defect-complex super multiplets}

\begin{figure}
\begin{center}
\includegraphics[width=7cm]{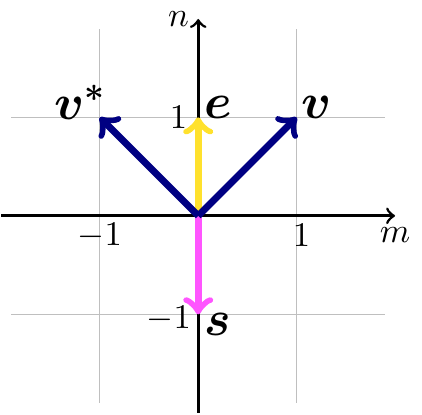}
\caption{\label{f:mnvectors}(Color online) Vector representation of the topological charge ($m$) and topological (Poincar\'{e}) index ($n$) of each zero in $\Psi$ and $\nabla\Phi$. The vortices and anti-vortices ($v$ and $v^*$) are in blue (black), the saddles ($s$) are in magenta (gray) and the extrema ($e$) are yellow (light gray).}
\end{center}
\end{figure}

We can think of the topological quantities (vortices $v$, anti-vortices $v^*$, saddles $s$ and extrema $e$) as defining vectors in the $m,n$-plane plotted in Fig.~\ref{f:mnvectors}.
This representation allows us to see that every vector has an additive inverse, e.g. $-v^* = v +2s$. In fact, the topological rules discussed above define the commutative defect group
\begin{align}
\label{e:TopGrp}D= \Big\langle v, v^*, s, e\; \Big| \; v+v^* =2e, e+s=0 \Big\rangle,
\end{align}
where $v, v^*, s, e$ are the group generators, the group operation is vector addition and the relations between the group elements are given by $v+ v^* =2e, e+ s=0$.

The existence of an algebraic structure leads to an obvious parallel with the hadron-spectroscopy Eightfold Way of Gell-Mann and Ne'eman \cite{Gell-Mann1962} and the associated quark model \cite{Gell-Mann1964}.  One may form an infinite hierarchy of all possible topologically allowed defect complexes such as $v+ v,~v+ v^*,v+ v+ s$, etc.  Each such complex is an element of the group $D$ and may be labelled by the triplet $[w, \chi, P]$, where $w$ is the net topological charge of the defect complex, $\chi$ is its net topological index and $P$ is the number of elementary defects (chosen from the set $\{v, v^*, e, s\}$) in the complex. Each topologically-allowed defect complex may decay to one or more defect complexes, in any topological reaction that conserves both net topological charge and net topological index.  Whether such a topologically-allowed process is energetically possible will depend on the particular Hamiltonian used to evolve the system.  More precisely, depending on the energetics, some defect-complexes will be stable and others will be unstable.  Some defect complexes are more ``natural'' than others, e.g.~(i) as noted by Maxwell \cite{Maxwell1870} a saddle will often naturally occur between two maxima; (ii) the first loop rule of Freund \cite{Freund1995} gives another natural association of saddles with extrema; (iii) the enlarged sign principle of Freund \cite{Freund1995} gives a natural association of alternating-sign point vortices.  The $P=1$ defect-complexes are always topologically stable, since they have no lower-$P$ complexes to decay to. 

\begin{figure}[h]
\begin{center}
\includegraphics[width=3cm]{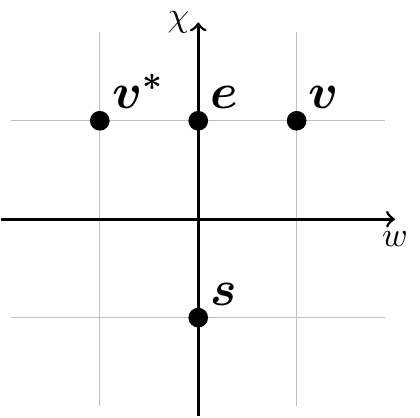}

(a) $P=1$, quartet

\includegraphics[width=5cm]{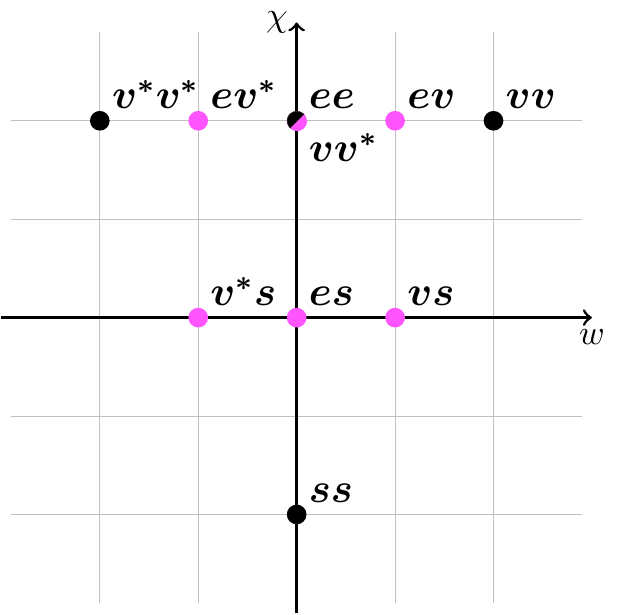}

(b) $P=2$, decuplet

\includegraphics[width=7cm]{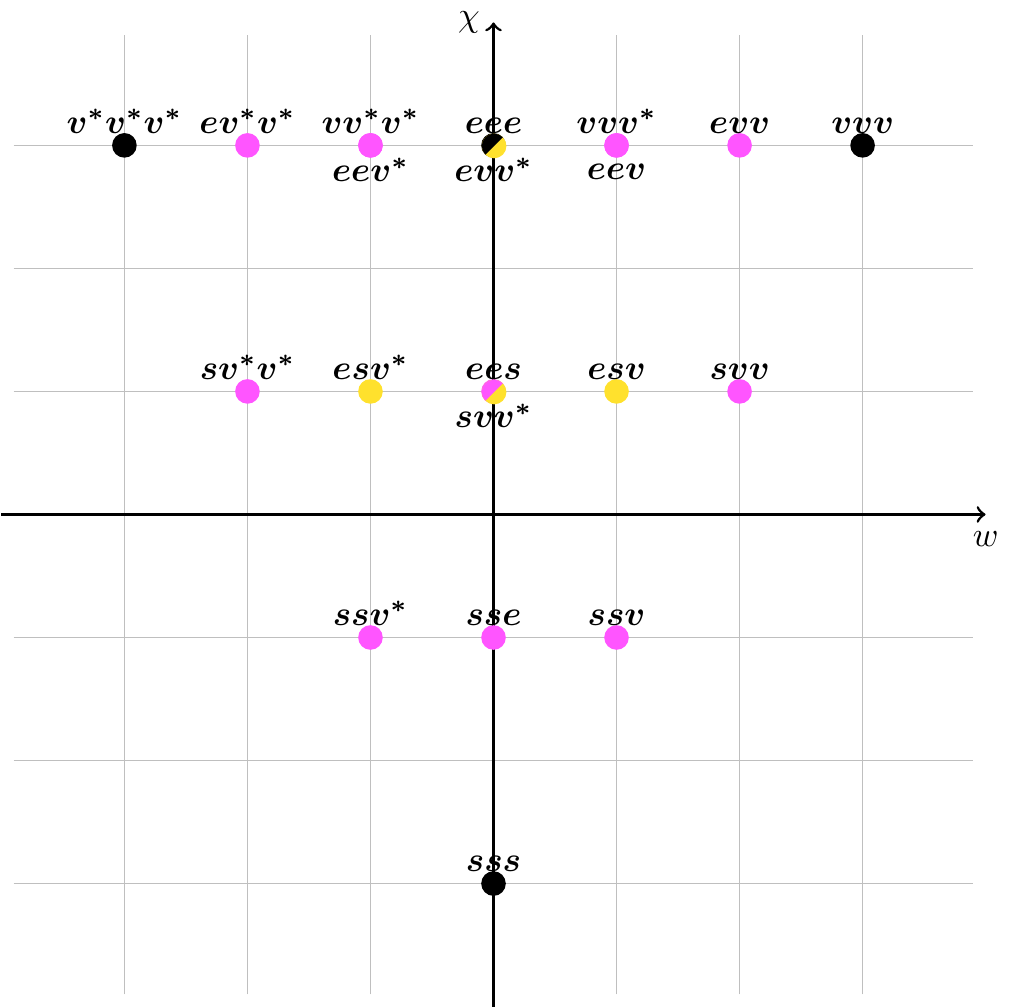}

(c) $P=3$, 20-plet

\caption{\label{f:EightfoldWay}(Color online) All possible $[w , \chi, P]$ defect complexes from the group $D$ with $P\le 3$, arranged according to ascending $P$: (a) $P=1$ quartet; (b) $P=2$ decuplet; (c) $P=3$ 20-plet.  Defect complexes with one species of defect are in black, those with two species of defect are in magenta (gray), and those with three species of defect are in yellow (light gray). (Note that we have suppressed the ``+'' signs of the complexes in the figure to save space.)}
\end{center}
\end{figure}

This parallel with hadron spectroscopy is evident when comparing Fig.~\ref{f:EightfoldWay} to baryon super multiplets in the quark model of hadrons \cite{MartinShaw1997}.  Figure~\ref{f:EightfoldWay} sketches the set of all defect complexes consisting of single defects ($P=1$ quartet, panel a), defect pairs ($P=2$ decuplet, panel b) and defect triplets ($P=3$ 20-plet, panel c). These are the sets of all elements of $D$ using $1,2$ and $3$ generators respectively.  Stacking these defect complexes in a hierarchy of ascending $P$ generates the super-multiplet analog of baryon super multiplets, with the $P$-axis denoting the number of defects in the complex.  Some examples of such defect complexes and their reactions include the $v+ v+ v+ v$ structure on the right of Figs.~\ref{f:PhaseEv1} and \ref{f:NodalEv1}, the $v+ s+ v$ complex on the right of Fig.~\ref{f:NodalEv1Phi}, the topological reaction $(v+ s+ s+ v) \rightarrow (v+ s+ v) +s$ on the left of Fig.~\ref{f:NodalEv1Phi}, and the creation and subsequent annihilation of the $s+ v+ v^* +s$ defect complex in Fig.~\ref{f:VortSaddle1}.  Further examples of defect complexes in $2+1$ dimensions include Onsager vortex clusters  \cite{Groszek2016}, $(v+v^*)$-dipoles in BECs \cite{Ruben2008}, $(2v+2v^*)$-quadrupoles in BECs \cite{Ruben2008}, paired Skyrmions in thin-film ferromagnets \cite{Pinna2018} and oscillon aggregates $(e+e+\cdots)$ in sinusoidally driven granular layers \cite{Oscillons1996}.  Note also the interesting linguistic coincidence, that the $s$ used to label saddles in Fig.~\ref{f:EightfoldWay}, directly parallels the use of $s$ to label strange quarks in the baryon super multiplets \cite{MartinShaw1997}.  This whimsical connection with baryon strangeness motivates an alternative term for the vertical axis of Fig.~\ref{f:EightfoldWay}, which may be spoken of as ``saddleness''.  

Combining the possible interaction vertices tabulated in Fig.~\ref{f:FundamentalProcesses} with the defect complexes listed in Fig.~\ref{f:EightfoldWay}, one could also classify all possible inelastic processes in which the number and/or nature of one or more defect complexes changes as a result of their mutual interaction. For instance, we see in Fig.~\ref{f:EightfoldWay}(b) that $e+e= v+ v^*$, which gives us the corresponding interaction with $(w, \chi)= (2,0)$ of Fig.~\ref{f:FundamentalProcesses}.

\section{vortex--anti-vortex fundamental processes and quaternionic solutions}\label{s:fundproc}

\begin{figure}
\includegraphics[width=8cm]{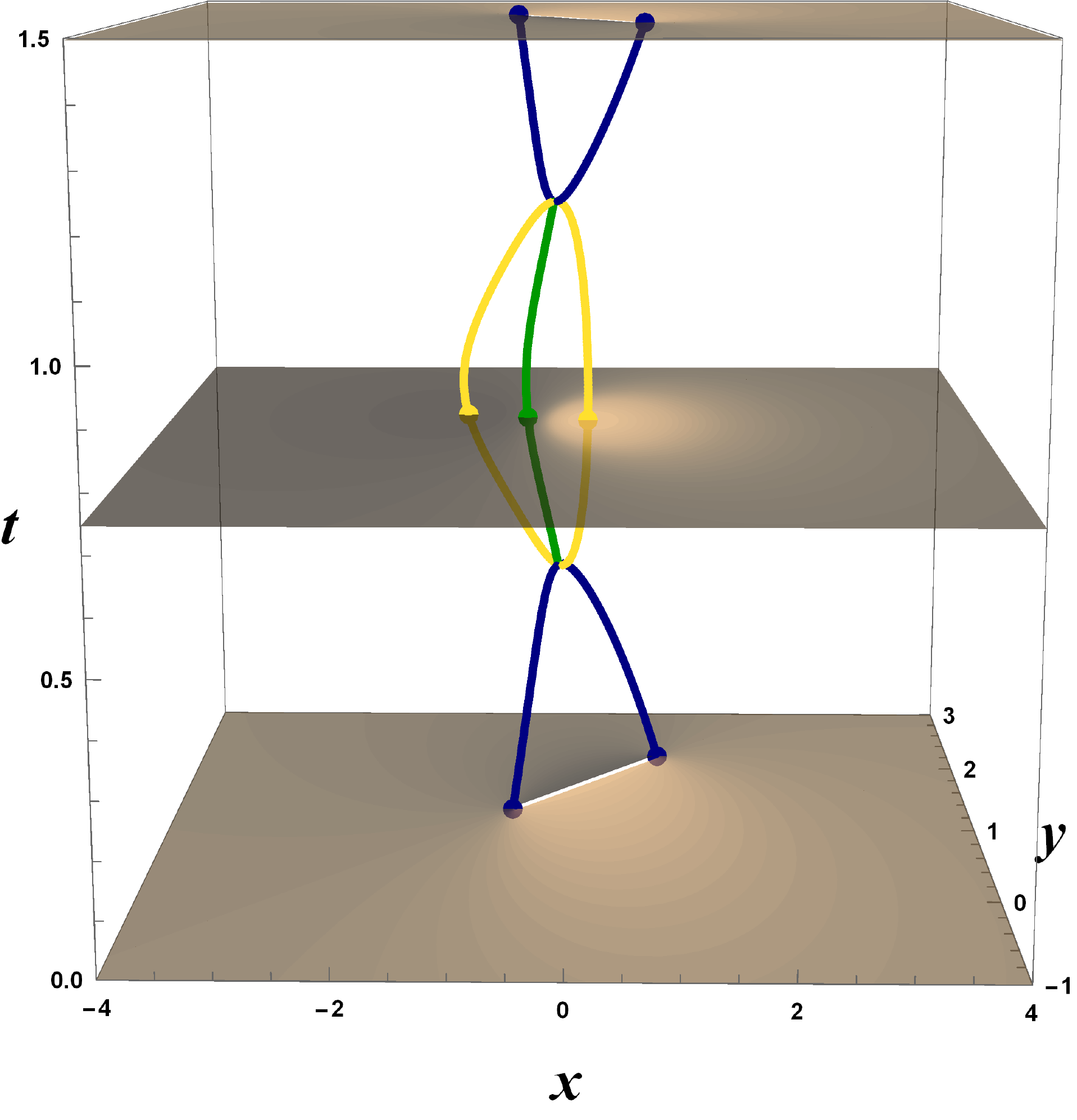}
\caption{\label{f:Qn2}(Color online) Using Eq.~(\ref{e:Psi0}). Blue (black): $\Psi_{2,0} (x,y)=0$. Yellow (light gray): $\nabla \Phi_{2,0} (x,y)=0$, where the stationary point is a maximum or a minimum. The green (gray) line is a representation of a virtual particle, and the color matches that of the quaternionic zeros in Fig.~\ref{f:nodal1} as this virtual particle is (in some sense) a shadow of those zeros.}
\end{figure}

We can see from Secs.~\ref{s:QEvs} and \ref{s:GenEvs} that a key feature of systems given by Eqs.~(\ref{e:Psi0}) and (\ref{e:gendet}) is the vortex--anti-vortex creation and annihilation events. Locally, these can be described by a $2\times 2$ system
\begin{align}
\nonumber &\Psi_{2,0}(x,y;t)= \det \left( \left[ \begin{array}{cc}
a& b\\
c&d
\end{array}
\right] - \left[ \begin{array}{cc}
\lambda & 0\\
0& \lambda^*
\end{array}
\right] \right)\\
\nonumber &= |\lambda|^2 - a\lambda^* - d\lambda +ad -bc\\
\label{e:quatcharp} &= x^2 +y^2 -(a+d)x +(a-d)i y +ad -bc.
\end{align}
We plot an example of such a system in Fig.~\ref{f:Qn2}.

Setting this determinant equal to 0 and taking the real and imaginary parts (assuming $x,y\in \mathbb{R}$) gives us a pair of coupled equations for $x$ and $y$. As we assumed, we are seeking solutions $x,y\in \mathbb{R}$ such that we obtain the $N=2$ case of Eq.~(\ref{e:quatfact1}), and we then have a vortex--anti-vortex pair. However, this need not always yield a pair of equations that is solvable over the real numbers.  As a trivial example, take $a=c=1, b=-2, d=2$, to give the solutions $x= (3\pm i \sqrt{7})/2, y=0$. From a purely algebraic perspective, the real and imaginary parts of Eq.~(\ref{e:quatcharp}) result in two real functions of $x$ and $y$, one linear and one quadratic, so we should not be surprised that in general we do not have real solutions.

We know from Fig.~\ref{f:FundamentalProcesses} that the annihilation of a vortex--anti-vortex pair should be accompanied by the simultaneous annihilation of two saddles or the creation of two maxima. (In Appendix~\ref{a:vortmaxmin} we show that in the $2\times 2$ case it must be the latter.) One way to understand this process is to analyze the pair of equations that result by taking the real and imaginary parts of $\Psi_{2,0} (x,y;t)$ (recalling that we assume $x,y\in \mathbb{R}$).  We obtain a quadratic and a linear equation
\begin{widetext}
{\small \begin{align}
\label{e:phaseR} 0&= x^2+y^2 - \big[ \Re(a) +\Re(d) \big]x + \big[ \Im(d) -\Im(a) \big] y +\Re(a)\Re(d) -\Im(a)\Im(d) +\Im(b) \Im(c) -\Re(b)\Re(c),\\
\label{e:phaseI} 0&= \big[ \Re(a)- \Re(d) \big] y - \big[ \Im(a) +\Im(d) \big] x +\Re(a) \Im(d) +\Im(a) \Re(d) - \Re(b) \Im(c) - \Im(b) \Re(c).
\end{align}}
\end{widetext}

This is equivalent to looking for the intersection between an ellipse and a line, which we depict schematically in Fig.~\ref{f:lineelipse} of Appendix~\ref{a:vortmaxmin}. However, when there is no intersection between the curves this does not mean that there is no zero but it indicates that taking real and imaginary parts is not a well-defined operation in this context, and suggests that the solutions are not in the complex plane. Indeed, we can view our solutions as existing in a four dimensional space $\mathbb{C}\times \mathbb{C}$. This is analogous to the situation for complex polynomials in a single complex variable---not all roots are real, however they are all complex (the complex numbers are the \textit{splitting field} for complex polynomials). In our case, the complex plane only contains some of our solutions, with the rest being in a four-dimensional space, within which the complex plane is embedded. In light of the discussions above about the determinant in Sec.~\ref{s:genwave} having a quaternionic structure, it is not surprising that we are only looking at a 2D subspace of a larger 4D space of solutions.

We can access these solutions by explicitly embedding our matrix in quaternion space using the representation in Eq.~(\ref{e:22q}), however for our purposes it is more straightforward to work in the traditional quaternion representation and calculate the determinant
\begin{align}
\label{e:4Dquatdet} &\det \left( \left[ \begin{array}{cc}
a_q& b_q\\
c_q&d_q
\end{array}
\right] - \left[ \begin{array}{cc}
\lambda_q & 0\\
0& \hat{\lambda}_q
\end{array}
\right] \right),
\end{align}
where $a_q= \Re(a)+ i\Im(a) +0j +0k$ and similarly for $b_q,c_q,d_q$ and 
$\lambda_q= x+ iy +jz +kw, \hat{\lambda}_q= x- iy +jz -kw$. (Note that $\hat{\lambda}_q$ is not the usual quaternion conjugate---we can think of it as a form of Eq.~(\ref{e:22q}), where the two complex numbers $\alpha, \beta$ are in different copies of $\mathbb{C}$. In terms of the Cayley--Dickson \cite{Baez2002} construction of the quaternions we write $\lambda_q = (x+iy)1 + (z+iw)j$ and $\hat{\lambda}_q = (x-iy)1 + (z-iw)j$; so we take a plane isomorphic to $\mathbb{C}$ and to every real and imaginary part we attach another independent copy of $\mathbb{C}$.) In Appendix \ref{app:qdet} we write out this determinant explicitly and we find the four expressions in Eqs.~(\ref{e:qr})--(\ref{e:qk}), which respectively correspond to the real, $i$, $j$ and $k$ components of Eq.~(\ref{e:4Dquatdet}), which we set all equal to zero. With this structure we find that our example above ($a=c=1, b=-2, d=2$) yields the two zeros $\left(\frac{3}{2}, 0, \pm \frac{\sqrt{7}}{2}, 0 \right)\in \mathbb{R}^4$.

\begin{figure}[h]
\includegraphics[scale=0.4]{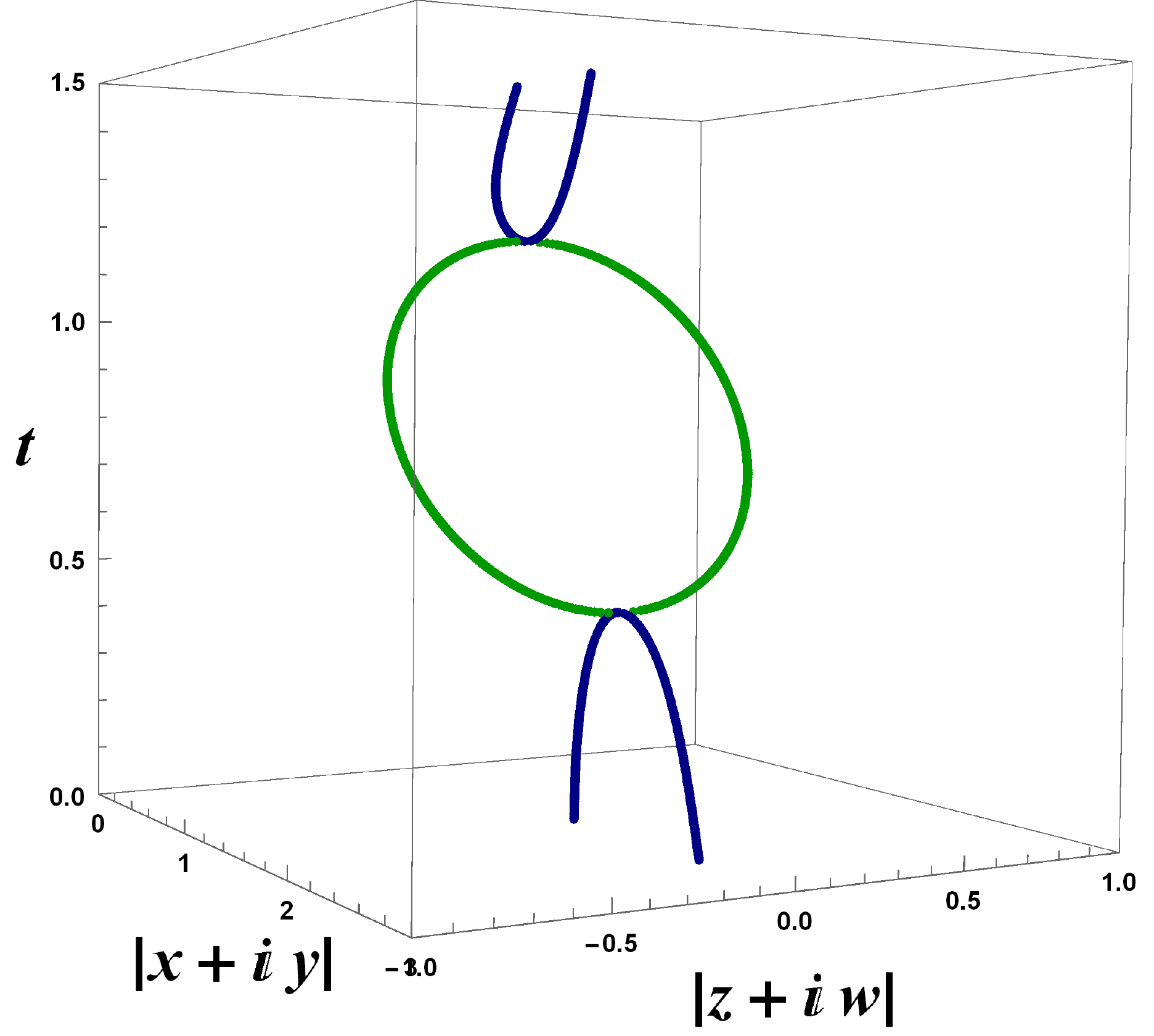}
\caption{\label{f:nodal1}(Color online) Solving for quaternionic zeros $\lambda= x+i y+ jz + k w$ of Eq.~(\ref{e:4Dquatdet}) with the matrix that was used to generate Fig.~\ref{f:Qn2}. The blue (black) lines in this one-loop diagram correspond to solutions where $z=w=0$, which are the blue (black) solutions plotted in Fig.~\ref{f:Qn2}. The green (gray) lines correspond to the solutions with $z\neq 0$ and $w\neq0$, which correspond to the yellow (light gray) points in Fig.~\ref{f:Qn2}. The $z,w$ components come in $\pm$ pairs so the quaternionic solutions give the points $(|x+i y|,|z+i w|, t)$ and $(|x+i y|,-|z+i w|, t)$.}
\end{figure}

We find that two roots $(x_1,y_1), (x_2,y_2)\in \mathbb{R}^2$ can collide and produce two new solutions $(x_3,y_3,z_3,w_3), (x_4,y_4,z_4,w_4) \in \mathbb{R}^4$. This results in the exchange of a topologically stable complex (a vortex--anti-vortex pair) with an intermediate state (a maximum and a minimum) that rapidly coalesce to yield a vortex--anti-vortex pair once again.  The two-component intermediate state, besides being a consequence of the previously discussed topological conservation laws, is seen to be directly connected with the fact that the quaternionic roots are observed to always have $j, k$ components in $\pm$ pairs, that is $z_1=-z_2$ and $w_1=-w_2$. We depict these four dimensional solutions in Fig.~\ref{f:nodal1}, which is the counterpart to Fig.~\ref{f:Qn2}.

Figures~\ref{f:Qn2} and \ref{f:nodal1} represent the topological reaction
\begin{align}
v+ v^*\rightarrow I \rightarrow v + v^*,
\end{align}
where $I$ may be interpreted as a transient intermediate state \footnote{This situation is rather analogous to the annihilation channel for Bhabha scattering, namely $e^+ +e^-\rightarrow\gamma\rightarrow e^+ +e^-$, where $\gamma$ is an intermediate virtual photon.} associated with the annihilation of zeros in the space of complex solutions.  In light of the preceding paragraphs, this may be viewed as scattering of the vortex--anti-vortex complex zeros into a transiently-excited zero associated with the quaternionic degrees of freedom (see esp. Fig.~\ref{f:nodal1}). 

\begin{figure}
\begin{center}
\includegraphics[scale=0.6]{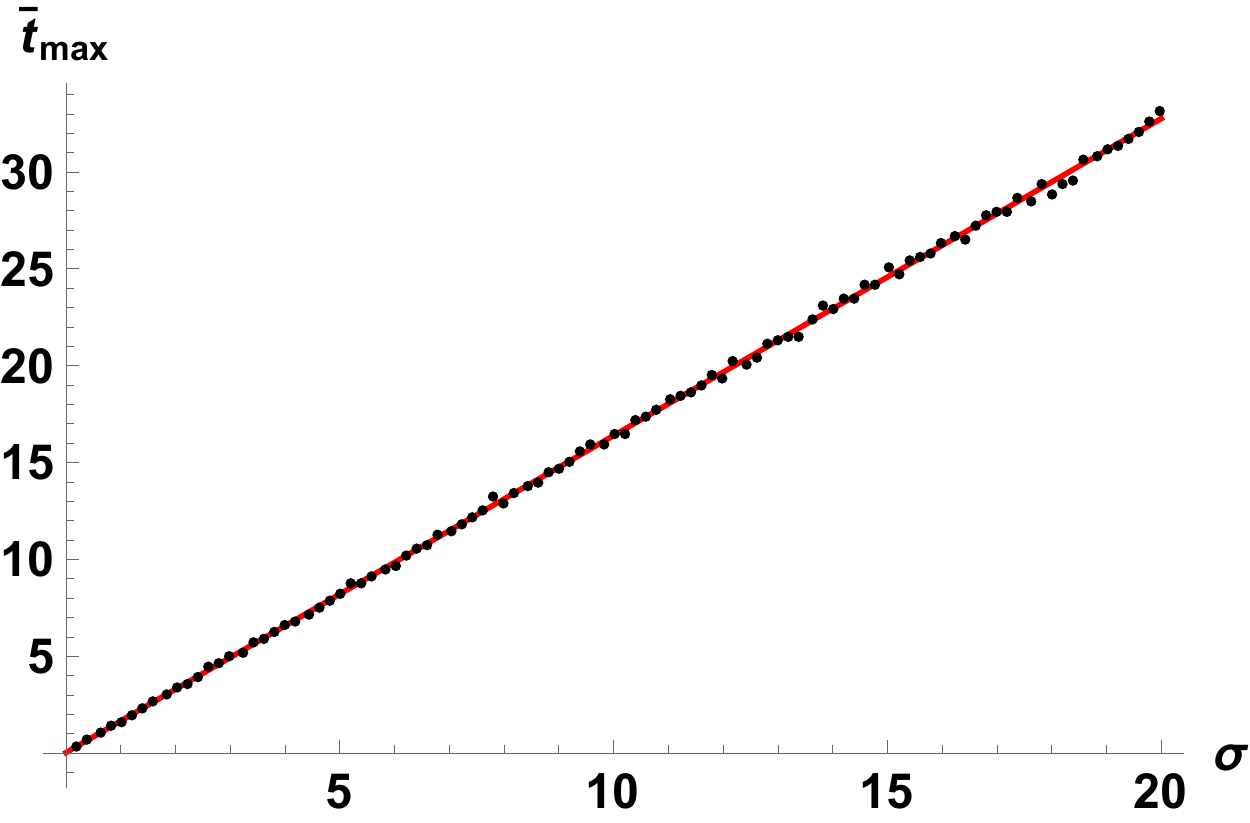}
\caption{\label{f:Tmax_versus_sigma_simulation}(Color online) $\overline{t}_{max}(\sigma)$ is the mean of the lifetimes $t_{max}$ of the quaternionic-zero transients sketched in Figs.~\ref{f:Qn2} and \ref{f:nodal1}.  For each $\sigma$, an ensemble of 5000 instances was generated, of $2\times 2$ matrices with entries having real and imaginary parts independently and identically distributed as a normal distribution of mean zero and standard deviation $\sigma$. The line of best fit is given by $\overline{t}_{max}= 0.027266 +1.63642 \sigma$.}
\end{center}
\end{figure}

These observations motivate investigation of the lifetime $t_{\rm{max}}$ of the intermediate state evident in Figs.~\ref{f:Qn2} and \ref{f:nodal1}.  Accordingly, an ensemble of $2\times 2$ Ginibre matrices of the form given by Eq.~(\ref{d:Ginmat}) was generated, each element of each matrix having a real and an imaginary component chosen from a Gaussian distribution with standard deviation $\sigma$.  For each $\sigma$, 5000 random matrices were generated, each of which was used as the matrix ${\bf M}_0$ in Eq.~(\ref{e:Hmat}), with the resulting time-dependent matrix ${\bf M}(t)$ being used to generate a determinantal wave function using Eq.~(\ref{e:Psi0}).   When the intermediate quaternionic-transient state in Figs.~\ref{f:Qn2} and \ref{f:nodal1} occurred, its lifetime $t_{\rm{max}}$ was recorded.  The mean of this lifetime, denoted by $\overline{t}_{\rm{max}}(\sigma)$, was then calculated for a range of $\sigma$ values in the range $0\le\sigma\le 20$ via an ensemble average for the 5000 random matrices generated for each $\sigma$. The simulated results are shown in Fig.~\ref{f:Tmax_versus_sigma_simulation}.  

From the straight-line fit one obtains
\begin{align}
\label{e:EmergentUncertaintyPrinciple}
\overline{t}_{\rm{max}}(\sigma) \, (1.6\,\sigma)^{-1}\approx 1.0.
\end{align}
This strengthens the analogy drawn between the intermediate quaternionic states and virtual particles, since Eq.~(\ref{e:EmergentUncertaintyPrinciple}) is identical in form to the usual time--energy uncertainty-principle estimate for the mean lifetime of a virtual particle, if $\sigma^{-1}$ is interpreted as an energy scale, and natural units are employed.  We note, in this context, that $\sigma$ is the only natural scale present in the model, and that on dimensional grounds it must have the units of a reciprocal energy.  Note also that, in the limit as the energy scale $\sigma^{-1}$ becomes sufficiently large, the green loop in Fig.~\ref{f:nodal1} may be considered to contract to a point, hence the topological reaction $v+ v^*\rightarrow v+ v^*$ may be considered as approaching a contact interaction in this limit.         

\section{\label{s:Discussion}Discussion}

An immediate avenue for future work is to  more thoroughly investigate the utilisation of ensemble averaging, over the set of random matrices belonging to a particular class, in the formalism utilised in the present paper.  An obvious example is to take the modelling at the end of the previous section, and work with higher-dimensional matrices (e.g.~Ginibre matrices with $N\ge2$). One could choose an evolution law consistent with a given (2+1)-dimensional classical field theory, such as the (2+1)-dimensional Goldstone model \cite{Maggiore2005}, the (2+1)-dimensional Gross--Pitaevskii equation \cite{Pitaevskii2003}, (2+1)-dimensional linear theories with arbitrary propagators \cite{PagaBeltPete2018} etc.  From a numerical perspective, the ensemble of starting matrices (which may have equal statistical weights, but not necessarily) would then generate a series of defect-line topological reactions in the interaction region of Fig.~\ref{f:GenericScatteringScenario}, which, upon time evolution, could be sorted into equivalence classes.  A measure could then be associated with each equivalence class of topological reaction, which would be rather analogous to a set of Feynman diagrams exploring all possible paths \cite{FeynmanHibbs1965} for all interactions permitted for a specified system.  The associated measure would be equal to a probability of occurrence for each particular topological reaction, which could be numerically tabulated via an appropriate histogram, as the topologically distinct reactions are sorted.  Related quantities such as scattering cross sections and lifetimes of particular transient defect complexes, could then be readily computed.  The scattering cross sections could be computed both for collisions between defect complexes {\em in vacuo}, and also for scattering of one defect complex from an introduced scattering potential.  In addition to the indicated numerical study, the questions outlined in the present paragraph could be addressed from an analytical perspective.  

The quaternionic transients, and their connection to an associated energy--lifetime uncertainty principle, are a fascinating outcome of the present study.  The $2\times 2$ cases studied in Sec.~\ref{s:fundproc} (with one vortex and one anti-vortex) provide examples of ``fundamental processes'' in these vortex--anti-vortex systems. The fact that the $x$ and $y$ coordinates of the zeros of the wave function do not always have solutions over $\mathbb{R}$ motivates the study of quaternionic solutions to the determinantal equation, as in Eq.~(\ref{e:4Dquatdet}). Further, we are guaranteed to obtain solutions in the 4-dimensional quaternion space $\mathbb{H}$ by the structure of the real and imaginary parts of the determinantal equation (a quadratic and a linear function, respectively). The appearance of quaternionic solutions ($\mathbb{H} \backslash \mathbb{C}$) corresponds to the annihilation of the vortex--anti-vortex pair, and the creation of a maximum--minimum pair in the phase of the wave function. 

One interpretation of this event may be that the pair of zeros exist naturally in the 2-dimensional (complex) plane, and then a collision event scatters them off the plane into 4-dimensional space. However, we conjecture that motion in these other dimensions is typically transient so the particle trajectories bring them back together and they then re-scatter onto the 2-dimensional plane---in real space this appears as annihilation and creation. The associated lifetime--energy uncertainty principle, given in Eq.~(\ref{e:EmergentUncertaintyPrinciple}), is particularly interesting.  How general is this result?  Can analogous results be obtained for all unstable defect complexes, thereby generating a hierarchy of unstable-defect-complex half-lives?   The above four-dimensional interpretation seems to naturally lend itself to quaternionic calculations (as discussed in Sec.~\ref{s:fundproc}); so we may, in some sense, call the virtual particles ``quaternionic quasi-particles''. This provides an intriguing connection to the quaternionic random matrix ensembles (beyond just the mathematical formalism of Sec.~\ref{s:genwave} and Appendix \ref{a:quats}), which are part of Dyson's Threefold Way \cite{Dyso1962ThreeFold}, and as such are of fundamental concern in random matrix theory. Although we can shed no more light on this at the moment, we feel that this connection is worthy of further study.  We also remark regarding the quaternionic transients, that the initial field generated by the random matrix can be viewed as a perturbation  with energy scale $\sigma^{-1}$, whose subsequent evolution may generate the quaternionic transient.  Some parallels may exist here with the creation of closed nodal-line loops in the vicinity of caustics that spontaneously form in coherent optical \cite{Berry1979} and matter-wave \cite{Petersen2013} fields.

Next, we return to a point made early in the paper.  We saw in Sec.~\ref{s:Evs} that an interacting set of particles can be implemented by calculating eigenvalues of a time dependent matrix. Since the eigenvalues are given by the characteristic polynomial, this effectively gives another representation of a system of particles defined as the zeros of a polynomial wave function, which undergoes (non-linear) evolution. The relationship is represented in Fig.~\ref{f:commdiag} and is a manifestation of the duality between the Heisenberg and Schr\"{o}dinger pictures. The lack of bijectivity between the two representations also naturally leads to a type of gauge freedom, with the number of degrees of freedom given by the difference between the number of roots of the polynomial and the number of independent entries of the corresponding matrix. This can be encoded explicitly via matrix conjugation, as in Eq.~(\ref{e:matconj}).

The systems formed in this way are, however, somewhat limited as each eigenvalue represents a vortex, which all have the same charge, and so they form a gas of mutually repelling particles.  This can be clearly seen in the eigenvalue joint probability density function, Eq.~(\ref{e:evjpdf}). There is, of course, the possibility of degenerate eigenvalues (having multiplicity greater than one), but if the matrix is randomly drawn from a continuous distribution and the matrix update rule is generic, then such a system exhibiting eigenvalue creation or annihilation is highly specialized and artificial. This is confirmed by the simulation shown in Fig.~\ref{f:NodalEv1}, which is typical of such determinantal systems using Ginibre matrices, Eq.~(\ref{d:Ginmat}).

A natural way to introduce more complex interactions is to note that polynomials in a complex variable and its complex conjugate (such as Eq.~(\ref{e:genfact})) produce zeros which behave like oppositely-charged particles. So if some number of complex-conjugate variables is introduced into the characteristic polynomial of a random matrix (as we did in Secs.~\ref{s:QEvs} and \ref{s:GenEvs}), we obtain the same behavior. These zeros are no longer eigenvalues, but they are still calculated via a determinantal polynomial.  Mathematically, since we no longer have a polynomial in a single variable, we are no longer guaranteed to find a full set of solutions in the complex plane (even degenerate ones). This allowed for the possibility that these particles and anti-particles will coalesce and annihilate, or that the wave function may nucleate particles.

These particle--anti-particle interactions change the topological landscape of the wave-function phase, yet they arise from continuous transformations, and so any topological invariants must be preserved. These invariants are given by Eqs.~(\ref{e:topcharge}) and (\ref{e:topindex}), the latter of which is the Euler characteristic. Only interactions that preserve the invariants are allowed and we have tabulated some of the lowest order ones in Fig.~\ref{f:FundamentalProcesses}. Several of these events were realized in our simulations, although most are not. We conjecture that the probability of locally observing a particular interaction decreases as the quantity $|w|+|\chi|$ increases. Indeed, given that we are working with systems generated by polynomials (having a finite degree), there is a finite upper bound on $|w|$, given by the difference between the number of variables and their complex conjugates in the determinantal expressions. The topological values of each of the topological points on the phase surface can be represented via the vector diagram in Fig.~\ref{f:mnvectors}, which naturally leads to the group structure given in Eq.~(\ref{e:TopGrp}). We note that this group, along with the Poincar\'{e}--Hopf theorem provides a connection between the topological, analytic and algebraic structures of these systems, described by the Euler characteristic, the Poincar\'{e} index and the group in Eq.~(\ref{e:TopGrp}) respectively.

Conservation of the Euler characteristic and Poincar\'{e} index at any primitive vertex, implies the crossing symmetry previously observed in this paper.  This refers to the topological deformation of one allowed vertex into another allowed vertex.  For example, if a vertex has a magenta (yellow) line connected to it, the said line may be moved from a past-directed line to a future-directed line upon changing magenta (yellow) to yellow (magenta). A similar crossing symmetry holds if past and future are interchanged in the preceding sentence.  We also saw that blue lines entering or leaving a vertex can be changed from past-pointing to future-pointing, or conversely, provided that the arrow associated with the blue line is maintained.  These crossing symmetries refer to the time-reversal operation, but one may also consider the associated deformations of primitive vertices under other transformations such as spatial reflection and parity transformations.  For example, under the parity transformation $(x,y)\rightarrow(x,-y)$, vortices and anti-vortices are interchanged, phase maxima and phase minima are unchanged, and saddles are unchanged modulo a topologically irrelevant rotation.

As we have already argued, the set of all possible topological defect-line reactions is ultimately reducible to reactions involving the hierarchy of primitive vertices given in Fig.~\ref{f:FundamentalProcesses}. In the context of random matrices and their associated determinantal wave functions, all permissible processes should be generated via an ensemble of random matrices of large enough dimension.  This would form another interesting avenue for future work.

It is also worth commenting on the ``topological defects as particles'' theme running throughout the paper.  This idea is well explored in the context of particle physics---see e.g.~\citeauthor{VilenkinShellard1994} \cite{VilenkinShellard1994} and \citeauthor{Volovik2003} \cite{Volovik2003}, together with references therein. Moreover, the theme has already been justified in the more limited classical-optics context of the present paper, on the topological grounds we have already given.  However two further motivations are worth mentioning.  (i) For (2+1)-dimensional incompressible frictionless fluids, Onsager showed that the resulting complex of vortices possesses a particle-like Hamiltonian depending only on the location of the vortex cores \cite{Onsager1949}.  This has subsequently been developed into the point-vortex model for vortices in (2+1)-dimensional non-linear complex order-parameter fields (see \citeauthor{Groszek2018}  \cite{Groszek2018} and references therein), in which  vortex cores are again treated as evolving point particles. (ii) In situations with symmetry-breaking potentials $V$ such as \cite{VilenkinShellard1994,Maggiore2005} 
\begin{equation}
V(|\Psi|)=\mu(|\Psi|^2-\eta^2)^2, \quad \mu\in\mathbb{R}^+,\quad\eta\in\mathbb{R},
\end{equation}
the wave-function zeros that are trapped in vortex cores comprise a false vacuum (even in the purely classical-field setting of the present paper) in the sense that the above potential is a local maximum when $|\Psi|=0$, achieving its global minimum when $|\Psi|=|\eta|$.  For such fields, therefore, a vortex or an anti-vortex would be particle-like in the sense of trapping a certain positive energy associated with the false vacuum, with such a wave-function zero typically being embedded in a true-vacuum background for which $|\Psi|=\eta$.  Such a phenomenon, which is well known \cite{VilenkinShellard1994,Volovik2003}, gives further impetus for considering the vortex and anti-vortex cores as quasi-particles, as has been done in the present paper.  From the same perspective, the phase maxima, phase minima and saddles also assume a particle-like identity, not because they trap false vacuum, but because they are vacuum excitations (cf. e.g. \citeauthor{Sinha1976} \cite{Sinha1976,Sinha1978}).  Throughout this paragraph, ``vacuum'' is taken to be synonymous with ``zero-potential ground-state background field $\Psi(x,y,t)=\eta\exp[i\alpha(x,y,t)]$ in which $\alpha$ is any smooth real function of $(x,y,t)$.''    

Finally, we emphasize that all of the results in the present paper, that depend purely on topological arguments regarding the continuity and single-valued nature of a complex wave function in $2+1$ dimensions, are applicable beyond the particular model we have developed.  Such model-independent aspects include the set of all allowed defect-line reactions, the super multiplet of possible defect complexes, remarks regarding knotted and braided structures in defect-line networks, and the set of allowed isolated defect-line networks.  All of these concepts are applicable to the previously mentioned (2+1)-dimensional continuous complex classical fields, whether they obey linear or non-linear differential equations (including non-linear equations admitting spontaneous symmetry breaking), integral equations, or integro-differential equations.  Moreover, generalisation to defect networks in spinor and tensor fields is evident if one first computes the admissible topological defects via calculation of their associated homotopy groups \cite{VilenkinShellard1994, SethnaBook} and notes the associated topological conservation laws.

\section{Conclusion}\label{Sec:Conclusion}

We have set up a determinantal correspondence between (i) random matrices and their associated temporal evolution rule, and (ii) polynomial wave functions and their associated Hamiltonian; in contrast to earlier studies, the model that we developed is entirely deterministic. Particular attention was paid to the phase defects of the induced wave functions, namely the phase maxima and phase minima, together with phase saddles, phase vortices and phase anti-vortices.  The defect line dynamics, induced by temporal evolution of the random matrix, were considered.  Such dynamics involve an interpenetrating network of defects, in which the nodal lines interact with lines associated with phase maxima, phase minima and saddles.  Local closed-form analyses were given, for several possible allowed defect-line processes.  All possible defect composites were classified, and their associated commutative group structure written down.  Low-order multiplets of defect aggregates were specified, although multiplets of all orders are immediately implied by our formalism. Allowed topological reactions of defect-line complexes were considered, as well as knotted defect-line structures and totally-closed defect line structures.  The role of quaternionic degrees of freedom, together with their associated transients and an induced uncertainty principle, was also treated.

\section*{Acknowledgements}
We acknowledge useful discussions with Nicholas Beaton, Mario Beltran, Tim Davis, Peter Forrester, Kieran Larkin, Tim Petersen, Tapio Simula and Paul Zinn-Justin. A.M. and A.P. are supported by the Australian Research Council (ARC) Centre of Excellence for the Mathematical and Statistical Frontiers (ACEMS), ARC Grant No. CE140100049.

\appendix

\section{Proof of vortex nature of eigenvalues}\label{a:vorPf}

We show that the winding number of every zero, for a wave function that is a polynomial in $\lambda=x+iy$, must be positive.  Hence all such zeros are vortices.  

A vortex is defined by a positive topological charge. In our context, where the characteristic equation is treated as a wave function as in Eq.~(\ref{e:wavfn}), the sign of the topological charge of any zero is given by the sign of the Jacobian \cite{Freund1994,Freund1995,Rothschild2012,Werdiger2016}
\begin{align}
\label{d:Jac0} J(x,y)= \det \left[ \begin{array}{cc}
\frac{\partial \Re(\Psi_{N,N})} {\partial x}& \frac{\partial \Im(\Psi_{N,N})} {\partial x}\\
\frac{\partial \Re(\Psi_{N,N})} {\partial y}& \frac{\partial \Im(\Psi_{N,N})} {\partial y}
\end{array}\right]
\end{align}
when it is evaluated at the zero.

\subsection{$N=2$}

For the case of a $2\times 2$ matrix, with characteristic polynomial $\chi= (x+iy- \lambda_1) (x+iy - \lambda_2)$, we can calculate the Jacobian in Eq.~(\ref{d:Jac0}) explicitly
{\small \begin{align}
\nonumber\label{d:Jac} J(x,y)&= \Big( (x- \lambda_1^r ) +(x- \lambda_2^r ) \Big)^2 +\Big( (y- \lambda_1^i ) +(y- \lambda_2^i ) \Big)^2\\
&>0, \qquad \forall (x,y) \in \mathbb{R}^2,
\end{align}}where $\lambda_j^r$ and $\lambda_j^i$ are the real and imaginary parts of $\lambda$. Hence both zeros (eigenvalues) are vortices in this system, and in fact, they have identical Jacobians
\begin{align}
J(\lambda_{1,2}^r, \lambda_{1,2}^i)= (\lambda_1^r- \lambda_2^r )^2 + (\lambda_1^i- \lambda_2^i )^2.
\end{align}

\subsection{General $N$}

Note that we can write $\Psi_N$ in the form
\begin{align}
\label{e:PsiFact} \Psi_{N,N} (\lambda)= \Psi_{N-1,N-1}(\lambda) (\lambda - \lambda_N)
\end{align}
then
\begin{align}
\nonumber\Re(\Psi_{N, N}) &= \Re(\Psi_{N-1, N-1}) (x - \lambda_N^r) \\
&\quad- \Im(\Psi_{N-1, N-1}) (y- \lambda_N^i)\\
\nonumber\Im(\Psi_{N, N}) &= \Re(\Psi_{N-1, N-1}) (y - \lambda_N^i) \\
&\quad+ \Im(\Psi_{N-1, N-1}) (x- \lambda_N^r)
\end{align}
so
\begin{widetext}
\begin{align}
\frac{\partial } {\partial x} \Re(\Psi_{N,N}) &= \Re(\Psi_{N-1, N-1}) + (x- \lambda_N^r) \frac{\partial } {\partial x} \Re(\Psi_{N-1, N-1})- (y- \lambda_N^i) \frac{\partial } {\partial x} \Im(\Psi_{N-1, N-1})\\
\frac{\partial } {\partial y} \Re(\Psi_{N,N}) &= -\Im(\Psi_{N-1, N-1}) - (y- \lambda_N^i) \frac{\partial } {\partial y} \Im(\Psi_{N-1, N-1})- (x- \lambda_N^r) \frac{\partial } {\partial y} \Re(\Psi_{N-1, N-1})\\
\frac{\partial } {\partial x} \Im(\Psi_{N,N}) &= \Im(\Psi_{N-1, N-1})+ (y- \lambda_N^i) \frac{\partial } {\partial x} \Re(\Psi_{N-1, N-1})+(x- \lambda_N^r) \frac{\partial } {\partial x} \Im(\Psi_{N-1, N-1})\\
\frac{\partial } {\partial y} \Im(\Psi_{N,N}) &= \Re(\Psi_{N-1, N-1}) + (x- \lambda_N^r) \frac{\partial } {\partial x} \Im(\Psi_{N-1, N-1})- (y- \lambda_N^i) \frac{\partial } {\partial x} \Re(\Psi_{N-1, N-1}).
\end{align}
\end{widetext}
Then we see that substituting in $x= \lambda_N^r$, $y= \lambda_N^i$ the only terms that survive are $\Re(\Psi_{N-1, N-1}) (\lambda_N)$ and $\Im(\Psi_{N-1, N-1}) (\lambda_N)$. Calculating the determinant in Eq.~(\ref{d:Jac}) gives us
\begin{align}
\nonumber J(\lambda_{N}^r, \lambda_{N}^i)&= \Big[ \Re(\Psi_{N-1, N-1}) (\lambda_N) \Big]^2\\
\nonumber &+\Big[ \Im(\Psi_{N-1, N-1}) (\lambda_N) \Big]^2\\
&= \big| \Psi_{N-1, N-1} (\lambda_N) \big|^2.
\end{align}
Since we could have factored out any of the $N$ factors in Eq.~(\ref{e:PsiFact}), we obtain the Jacobian evaluated at any of the zeros
\begin{align}
\label{e:JN} J(\lambda_{j}^r, \lambda_{j}^i)&= \prod_{k=1 \atop k\neq j}^N (\lambda_j^r- \lambda_k^r )^2 + (\lambda_j^i- \lambda_k^i )^2\\
&>0, \qquad \forall j
\end{align}
and so all zeros have positive winding number.  Hence they are all vortices. Thus the topological charge of $\Psi_{N, N} (\lambda)$ is $+N$.

From this proof we see that any (single-variable) polynomial wave function $\Psi_{N,N} (\lambda)$ has positive topological charge, not just characteristic polynomials.

\section{Quaternionic ensembles}\label{a:quats}

We here review some of the formalism of quaternions as used in random matrix theory. We define a quaternion as a number with four independent real components,
\begin{align}
q= q_0 +q_1 i +q_2 j + q_3 k, \qquad q_0, q_1, q_2, q_3\in \mathbb{R}
\end{align}
with the properties $i^2= j^2= k^2= ijk=-1$. If we write $q$ as a pair of standard complex numbers $\alpha= q_0+i q_1, \beta= q_2+iq_3$ (where $i$ is the usual imaginary unit) then we have the equivalent representation
\begin{align}
\label{e:22q} q= \left[ \begin{array}{cc}
\alpha & \beta\\
-\beta^*& \alpha^*
\end{array}\right]
\end{align}
in which case the basis elements $1,i,j,k$ are represented by $\bI_{2}, i \sigma_z, i\sigma_y, i\sigma_x$ respectively, where $\bI_{2}$ is the $2\times 2$ identity matrix and $\sigma_x, \sigma_y, \sigma_z$ are the Pauli matrices. We denote by $\mathbb{H}$ the (four dimensional) span of $1, i, j, k$.

In \citeauthor{Dyso1962ThreeFold} \cite{Dyso1962ThreeFold} the three primary universality classes of random matrix theory are identified as part of his Threefold Way: matrices with real, complex or quaternion entries. Traditionally, when dealing with quaternionic ensembles, one looks for solutions to Eq.~(\ref{e:charpoly0}) where $\bM$ is a $2N \times 2N$ matrix, where each $2\times 2$ block is of the form Eq.~(\ref{e:22q}).  This gives complex-conjugate paired eigenvalues for a complex representation of a quaternionic matrix.

\section{Analysis of the collision of a phase extremum and a phase saddle} \label{a:saddlemax}

\begin{figure}
\begin{center}
\includegraphics[width=8cm,clip=true, trim= 160 0 160 0]{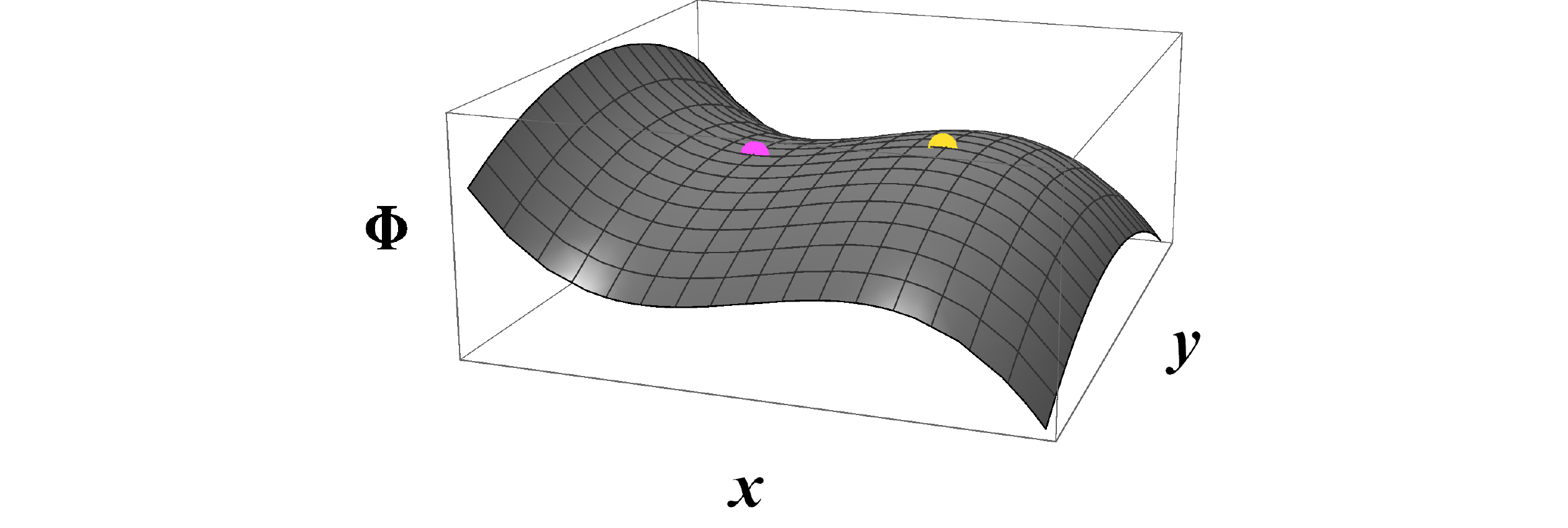}
\caption{\label{f:extsad}(Color online) A schematic of the phase surface Eq.~(\ref{e:phasesurf}), showing the saddle point in magenta (gray) and a local maximum in yellow (light gray) before collision.}
\end{center}
\end{figure}

We can describe a generic phase surface containing an extremum and a saddle point with the equation
\begin{align}
\label{e:phasesurf}\Phi= \arg(\Psi) = \epsilon x -y^2 -x^3,
\end{align}
where a saddle point is located at $(-\sqrt{\epsilon/3}, 0)$ and a local maximum is at $(\sqrt{\epsilon/3}, 0)$. A plot of this function is given in Fig.~\ref{f:extsad}. Letting $\epsilon \to 0$ has the effect that the stationary points coalesce, and then with $\epsilon$ becoming negative the stationary points disappear, which corresponds to what we see in (for example) Fig.~\ref{f:NodalGen1Phi} when a magenta (gray) and a yellow (light gray) thread meet \cite{Arnold1985}.

\section{Analysis of creation and annihilation events} \label{a:vortmaxmin}

As seen in Eqs.~(\ref{e:phaseR}) and (\ref{e:phaseI}) we obtain a quadratic function and a linear function for the real and imaginary parts (respectively) of the phase of the wave function $\Psi_{2,2} (x,y;t)$. By suitable topological deformation, we can describe any interaction of this type in the same way,
\begin{align}
\label{e:inteqn} \Psi(x,y;\epsilon)= y- \epsilon+ i \left[ x^2+(y-1)^2 -1 \right]
\end{align}
and so the linear function corresponding to the real part shifts vertically with changing $\epsilon$. (Note that, for simplicity, we have made the real part linear and the imaginary part quadratic, which is opposite to the situation in Eq.~(\ref{e:quatcharp}).)

\begin{figure}
\begin{center}
\includegraphics[width=8cm]{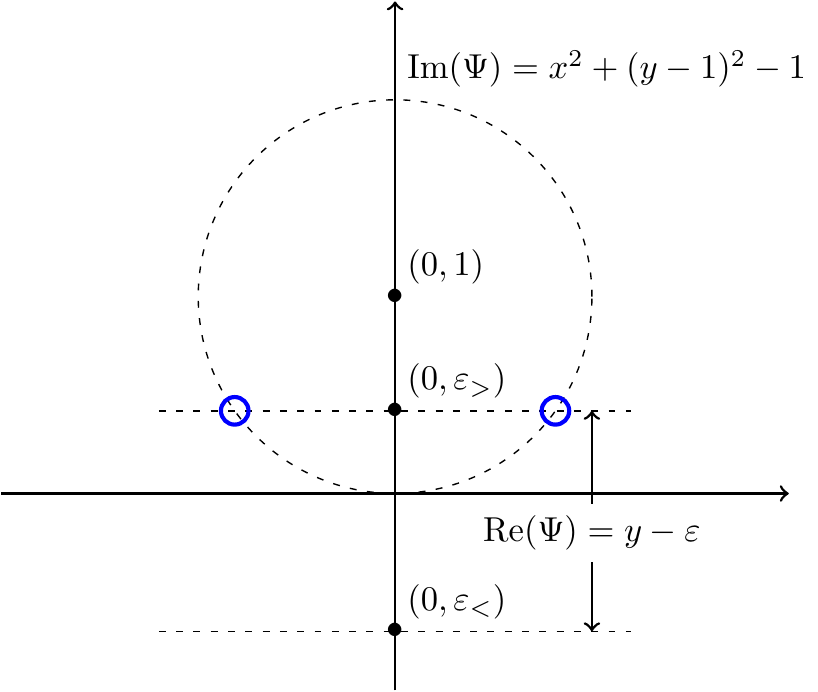}
\caption{\label{f:lineelipse}(Color online) Plots of the real and imaginary parts of Eq.~(\ref{e:inteqn}). When $\epsilon= \epsilon_{>}$ there are two intersections between the curves, which correspond to a vortex--anti-vortex pair, marked by small circles. When $\epsilon= \epsilon_{<}$ then there is no intersection and we obtain a maximum--minimum pair in the phase.}
\end{center}
\end{figure}

The wave function vanishes when both real and imaginary parts are equal to zero, which we plot in Fig.~\ref{f:lineelipse}. For $\epsilon_{>}= \epsilon> 0$ we have $y= \epsilon_{>}$ and $x= \pm \sqrt{2\epsilon_{>} -\epsilon_{>}^2}$: the two lines intersect twice and we obtain the locations of our vortex and anti-vortex cores $(-\sqrt{2\epsilon_{>} -\epsilon_{>}^2}, \epsilon_{>})$ and $(\sqrt{2\epsilon_{>} -\epsilon_{>}^2}, \epsilon_{>})$, which are marked by blue circles in Fig.~\ref{f:lineelipse}. When $\epsilon_{<}= \epsilon< 0$ then the linear curve lies underneath the quadratic one and there is no solution to $\Re(\Psi)= \Im(\Psi) =0$. The argument function and its derivative are given by
\begin{align}
\arg(\Psi) &= \arctan\left[ \frac{x^2+(y-1)^2 -1} {y- \epsilon_{<}} \right]\\
\frac{\partial}{\partial x} \arg(\Psi)&= \frac{2x (y- \epsilon_{<})} {(y- \epsilon_{<})^2+ [x^2+ (y-1)^2 -1]^2}\\
\frac{\partial}{\partial y} \arg(\Psi)&=\frac{2(y-1) (y- \epsilon_{<}) - (x^2+(y-1)^2 -1)}{(y- \epsilon_{<})^2 +[x^2+(y-1)^2 -1]^2}
\end{align}
and then
\begin{align}
&\frac{\partial}{\partial x} \arg(\Psi)=0 \quad \Rightarrow \quad x=0\\
&\frac{\partial}{\partial y} \arg(\Psi)=0 \quad \Rightarrow \quad y= \epsilon_{<} \pm \sqrt{\epsilon_{<}^2 -2\epsilon_{<}}
\end{align}
and taking $\epsilon_{<}\to 0^-$ we find that there are two stationary points of the phase, $(0, \sqrt{-2\epsilon_{<}})$ and $(0,- \sqrt{-2\epsilon_{<}})$. Taking second derivatives we can confirm that the first stationary point is a minimum and the second is a maximum.

\section{Determinant using a quaternionic variable}
\label{app:qdet}

Writing out Eq.~(\ref{e:4Dquatdet}) we have
\begin{widetext}
\begin{align}
\nonumber &\det \left( \left[ \begin{array}{cc}
a& b\\
c&d
\end{array}
\right] - \left[ \begin{array}{cc}
x+iy +jz +kw & 0\\
0& x-iy+jz-kw
\end{array}
\right] \right)\\
\nonumber &= ad -bc - a (x-iy+jz-kw) - (x+iy +jz +kw) d+(x+iy +jz +kw) (x-iy+jz-kw)\\
\nonumber &=ad -bc +x^2 +y^2 -z^2 +w^2 +2iwz+ 2jxz+2kyz -ax+ aiy- ajz+ akw- dx- diy- djz -dkw\\
&= \chi_q
\end{align}
\end{widetext}
then the real, $i,j$, and $k$ components are given by
\begin{widetext}
\begin{align}
\label{e:qr} (\chi_q)_r&= \Re(ad- bc) + x^2 +y^2 -z^2 +w^2 - \Re(a) x- \Im(a) y -\Re(d)x +\Im(d)y\\
\label{e:qi} (\chi_q)_i&= \Im(ad- bc) -2wz- \Im(a) x+ \Re(a) y -\Im(d)x -\Re(d)y\\
\label{e:qj} (\chi_q)_j&= 2xz - \Re(a)z - \Im(a)w -\Re(d)z -\Im(d)w\\
\label{e:qk} (\chi_q)_k&= 2yz- \Im(a)z + \Re(a)w + \Im(d)z- \Re(d)w
\end{align}
\end{widetext}
respectively. Setting these all equal to zero gives a set of four coupled equations in the variables $x,y,z,w$.

\newpage

\bibliography{BEC}

\begin{thebibliography}{84}%
\makeatletter
\providecommand \@ifxundefined [1]{%
 \@ifx{#1\undefined}
}%
\providecommand \@ifnum [1]{%
 \ifnum #1\expandafter \@firstoftwo
 \else \expandafter \@secondoftwo
 \fi
}%
\providecommand \@ifx [1]{%
 \ifx #1\expandafter \@firstoftwo
 \else \expandafter \@secondoftwo
 \fi
}%
\providecommand \natexlab [1]{#1}%
\providecommand \enquote  [1]{``#1''}%
\providecommand \bibnamefont  [1]{#1}%
\providecommand \bibfnamefont [1]{#1}%
\providecommand \citenamefont [1]{#1}%
\providecommand \href@noop [0]{\@secondoftwo}%
\providecommand \href [0]{\begingroup \@sanitize@url \@href}%
\providecommand \@href[1]{\@@startlink{#1}\@@href}%
\providecommand \@@href[1]{\endgroup#1\@@endlink}%
\providecommand \@sanitize@url [0]{\catcode `\\12\catcode `\$12\catcode
  `\&12\catcode `\#12\catcode `\^12\catcode `\_12\catcode `\%12\relax}%
\providecommand \@@startlink[1]{}%
\providecommand \@@endlink[0]{}%
\providecommand \url  [0]{\begingroup\@sanitize@url \@url }%
\providecommand \@url [1]{\endgroup\@href {#1}{\urlprefix }}%
\providecommand \urlprefix  [0]{URL }%
\providecommand \Eprint [0]{\href }%
\providecommand \doibase [0]{http://dx.doi.org/}%
\providecommand \selectlanguage [0]{\@gobble}%
\providecommand \bibinfo  [0]{\@secondoftwo}%
\providecommand \bibfield  [0]{\@secondoftwo}%
\providecommand \translation [1]{[#1]}%
\providecommand \BibitemOpen [0]{}%
\providecommand \bibitemStop [0]{}%
\providecommand \bibitemNoStop [0]{.\EOS\space}%
\providecommand \EOS [0]{\spacefactor3000\relax}%
\providecommand \BibitemShut  [1]{\csname bibitem#1\endcsname}%
\let\auto@bib@innerbib\@empty
%</preamble>
\bibitem [{\citenamefont {Annett}(2004)}]{Annett2004}%
  \BibitemOpen
  \bibfield  {author} {\bibinfo {author} {\bibfnamefont {J.~F.}\ \bibnamefont
  {Annett}},\ }\href@noop {} {\emph {\bibinfo {title} {Superconductivity,
  Superfluids, and Condensates}}}\ (\bibinfo  {publisher} {Oxford University
  Press},\ \bibinfo {address} {Oxford},\ \bibinfo {year} {2004})\BibitemShut
  {NoStop}%
\bibitem [{\citenamefont {Volovik}(2003)}]{Volovik2003}%
  \BibitemOpen
  \bibfield  {author} {\bibinfo {author} {\bibfnamefont {G.~E.}\ \bibnamefont
  {Volovik}},\ }\href@noop {} {\emph {\bibinfo {title} {The Universe in a
  Helium Droplet}}}\ (\bibinfo  {publisher} {Oxford University Press},\
  \bibinfo {address} {Oxford},\ \bibinfo {year} {2003})\BibitemShut {NoStop}%
\bibitem [{\citenamefont {Ashcroft}\ and\ \citenamefont
  {Mermin}(1976)}]{AshcroftMermin}%
  \BibitemOpen
  \bibfield  {author} {\bibinfo {author} {\bibfnamefont {N.~W.}\ \bibnamefont
  {Ashcroft}}\ and\ \bibinfo {author} {\bibfnamefont {N.~D.}\ \bibnamefont
  {Mermin}},\ }\href@noop {} {\emph {\bibinfo {title} {Solid State Physics}}}\
  (\bibinfo  {publisher} {Thomson Learning},\ \bibinfo {address} {Singapore},\
  \bibinfo {year} {1976})\BibitemShut {NoStop}%
\bibitem [{\citenamefont {Umbanhowar}\ \emph {et~al.}(1996)\citenamefont
  {Umbanhowar}, \citenamefont {Melo},\ and\ \citenamefont
  {Swinney}}]{Oscillons1996}%
  \BibitemOpen
  \bibfield  {author} {\bibinfo {author} {\bibfnamefont {P.~B.}\ \bibnamefont
  {Umbanhowar}}, \bibinfo {author} {\bibfnamefont {F.}~\bibnamefont {Melo}}, \
  and\ \bibinfo {author} {\bibfnamefont {H.~L.}\ \bibnamefont {Swinney}},\
  }\href@noop {} {\bibfield  {journal} {\bibinfo  {journal} {Nature}\ }\textbf
  {\bibinfo {volume} {382}},\ \bibinfo {pages} {793} (\bibinfo {year}
  {1996})}\BibitemShut {NoStop}%
\bibitem [{\citenamefont {Drazin}\ and\ \citenamefont
  {Johnson}(1989)}]{Drazin1989}%
  \BibitemOpen
  \bibfield  {author} {\bibinfo {author} {\bibfnamefont {P.~G.}\ \bibnamefont
  {Drazin}}\ and\ \bibinfo {author} {\bibfnamefont {R.~S.}\ \bibnamefont
  {Johnson}},\ }\href@noop {} {\emph {\bibinfo {title} {Solitons: An
  Introduction}}}\ (\bibinfo  {publisher} {Cambridge University Press},\
  \bibinfo {address} {Cambridge},\ \bibinfo {year} {1989})\BibitemShut
  {NoStop}%
\bibitem [{\citenamefont {Vilenkin}\ and\ \citenamefont
  {Shellard}(1994)}]{VilenkinShellard1994}%
  \BibitemOpen
  \bibfield  {author} {\bibinfo {author} {\bibfnamefont {A.}~\bibnamefont
  {Vilenkin}}\ and\ \bibinfo {author} {\bibfnamefont {E.~P.~S.}\ \bibnamefont
  {Shellard}},\ }\href@noop {} {\emph {\bibinfo {title} {Cosmic Strings and
  Other Topological Defects}}}\ (\bibinfo  {publisher} {Cambridge University
  Press},\ \bibinfo {address} {Cambridge},\ \bibinfo {year} {1994})\BibitemShut
  {NoStop}%
\bibitem [{\citenamefont {Dirac}(1931)}]{Dirac1931}%
  \BibitemOpen
  \bibfield  {author} {\bibinfo {author} {\bibfnamefont {P.~A.~M.}\
  \bibnamefont {Dirac}},\ }\href@noop {} {\bibfield  {journal} {\bibinfo
  {journal} {Proc. Roy. Soc. A}\ }\textbf {\bibinfo {volume} {133}},\ \bibinfo
  {pages} {60} (\bibinfo {year} {1931})}\BibitemShut {NoStop}%
\bibitem [{\citenamefont {Freund}(1999{\natexlab{a}})}]{Freund1999b}%
  \BibitemOpen
  \bibfield  {author} {\bibinfo {author} {\bibfnamefont {I.}~\bibnamefont
  {Freund}},\ }\href@noop {} {\bibfield  {journal} {\bibinfo  {journal} {Opt.
  Commun.}\ }\textbf {\bibinfo {volume} {163}},\ \bibinfo {pages} {230}
  (\bibinfo {year} {1999}{\natexlab{a}})}\BibitemShut {NoStop}%
\bibitem [{\citenamefont {Nye}(1999)}]{Nye1999}%
  \BibitemOpen
  \bibfield  {author} {\bibinfo {author} {\bibfnamefont {J.~F.}\ \bibnamefont
  {Nye}},\ }\href@noop {} {\emph {\bibinfo {title} {Natural Focusing and Fine
  Structure of Light: Caustics and Wave Dislocations}}}\ (\bibinfo  {publisher}
  {Institute of Physics Publishing},\ \bibinfo {address} {Bristol},\ \bibinfo
  {year} {1999})\BibitemShut {NoStop}%
\bibitem [{\citenamefont {Berry}(1981)}]{BerryLesHouches}%
  \BibitemOpen
  \bibfield  {author} {\bibinfo {author} {\bibfnamefont {M.}~\bibnamefont
  {Berry}},\ }\href@noop {} {\emph {\bibinfo {title} {Les Houches Lecture
  Series Session XXXV}}},\ edited by\ \bibinfo {editor} {\bibfnamefont
  {R.}~\bibnamefont {Balian}}, \bibinfo {editor} {\bibfnamefont
  {M.}~\bibnamefont {Kl\'{e}man}}, \ and\ \bibinfo {editor} {\bibfnamefont
  {J.-P.}\ \bibnamefont {Poirier}}\ (\bibinfo  {publisher} {North--Holland},\
  \bibinfo {address} {Amsterdam},\ \bibinfo {year} {1981})\ pp.\ \bibinfo
  {pages} {453--543}\BibitemShut {NoStop}%
\bibitem [{\citenamefont {Nye}\ and\ \citenamefont
  {Berry}(1974)}]{NyeBerry1974}%
  \BibitemOpen
  \bibfield  {author} {\bibinfo {author} {\bibfnamefont {J.~F.}\ \bibnamefont
  {Nye}}\ and\ \bibinfo {author} {\bibfnamefont {M.~V.}\ \bibnamefont
  {Berry}},\ }\href@noop {} {\bibfield  {journal} {\bibinfo  {journal} {Proc.
  Roy. Soc. Lond. A}\ }\textbf {\bibinfo {volume} {336}},\ \bibinfo {pages}
  {165} (\bibinfo {year} {1974})}\BibitemShut {NoStop}%
\bibitem [{\citenamefont {Freund}(1999{\natexlab{b}})}]{Freund1999a}%
  \BibitemOpen
  \bibfield  {author} {\bibinfo {author} {\bibfnamefont {I.}~\bibnamefont
  {Freund}},\ }\href@noop {} {\bibfield  {journal} {\bibinfo  {journal} {Opt.
  Commun.}\ }\textbf {\bibinfo {volume} {159}},\ \bibinfo {pages} {99}
  (\bibinfo {year} {1999}{\natexlab{b}})}\BibitemShut {NoStop}%
\bibitem [{\citenamefont {Dennis}\ \emph {et~al.}(2011)\citenamefont {Dennis},
  \citenamefont {G\"{o}tte}, \citenamefont {King}, \citenamefont {Morgan},\
  and\ \citenamefont {Alonso}}]{Dennis2011}%
  \BibitemOpen
  \bibfield  {author} {\bibinfo {author} {\bibfnamefont {M.~R.}\ \bibnamefont
  {Dennis}}, \bibinfo {author} {\bibfnamefont {J.~R.}\ \bibnamefont
  {G\"{o}tte}}, \bibinfo {author} {\bibfnamefont {R.~P.}\ \bibnamefont {King}},
  \bibinfo {author} {\bibfnamefont {M.~A.}\ \bibnamefont {Morgan}}, \ and\
  \bibinfo {author} {\bibfnamefont {M.~A.}\ \bibnamefont {Alonso}},\
  }\href@noop {} {\bibfield  {journal} {\bibinfo  {journal} {Opt. Lett.}\
  }\textbf {\bibinfo {volume} {36}},\ \bibinfo {pages} {4452} (\bibinfo {year}
  {2011})}\BibitemShut {NoStop}%
\bibitem [{\citenamefont {Paganin}\ \emph
  {et~al.}(2018{\natexlab{a}})\citenamefont {Paganin}, \citenamefont
  {Beltran},\ and\ \citenamefont {Petersen}}]{PagaBeltPete2018}%
  \BibitemOpen
  \bibfield  {author} {\bibinfo {author} {\bibfnamefont {D.~M.}\ \bibnamefont
  {Paganin}}, \bibinfo {author} {\bibfnamefont {M.~A.}\ \bibnamefont
  {Beltran}}, \ and\ \bibinfo {author} {\bibfnamefont {T.~C.}\ \bibnamefont
  {Petersen}},\ }\href@noop {} {\bibfield  {journal} {\bibinfo  {journal} {Opt.
  Lett.}\ }\textbf {\bibinfo {volume} {43}},\ \bibinfo {pages} {975} (\bibinfo
  {year} {2018}{\natexlab{a}})}\BibitemShut {NoStop}%
\bibitem [{\citenamefont {Bohigas}\ \emph {et~al.}(2012)\citenamefont
  {Bohigas}, \citenamefont {De~Carvalho},\ and\ \citenamefont
  {Pato}}]{BohiDeCaPato2012}%
  \BibitemOpen
  \bibfield  {author} {\bibinfo {author} {\bibfnamefont {O.}~\bibnamefont
  {Bohigas}}, \bibinfo {author} {\bibfnamefont {J.~X.}\ \bibnamefont
  {De~Carvalho}}, \ and\ \bibinfo {author} {\bibfnamefont {M.~P.}\ \bibnamefont
  {Pato}},\ }\href {\doibase 10.1103/PhysRevE.86.031118} {\bibfield  {journal}
  {\bibinfo  {journal} {Phys. Rev. E}\ }\textbf {\bibinfo {volume} {86}},\
  \bibinfo {pages} {031118} (\bibinfo {year} {2012})}\BibitemShut {NoStop}%
\bibitem [{\citenamefont {Wigner}(1955)}]{Wigner1955}%
  \BibitemOpen
  \bibfield  {author} {\bibinfo {author} {\bibfnamefont {E.}~\bibnamefont
  {Wigner}},\ }\href@noop {} {\bibfield  {journal} {\bibinfo  {journal} {Ann.
  Math.}\ }\textbf {\bibinfo {volume} {62}},\ \bibinfo {pages} {548} (\bibinfo
  {year} {1955})}\BibitemShut {NoStop}%
\bibitem [{\citenamefont {Dyson}(1962{\natexlab{a}})}]{Dyso1962I}%
  \BibitemOpen
  \bibfield  {author} {\bibinfo {author} {\bibfnamefont {F.~J.}\ \bibnamefont
  {Dyson}},\ }\href@noop {} {\bibfield  {journal} {\bibinfo  {journal} {J.
  Math. Phys.}\ }\textbf {\bibinfo {volume} {3}},\ \bibinfo {pages} {140}
  (\bibinfo {year} {1962}{\natexlab{a}})}\BibitemShut {NoStop}%
\bibitem [{\citenamefont {Dyson}(1962{\natexlab{b}})}]{Dyso1962II}%
  \BibitemOpen
  \bibfield  {author} {\bibinfo {author} {\bibfnamefont {F.~J.}\ \bibnamefont
  {Dyson}},\ }\href@noop {} {\bibfield  {journal} {\bibinfo  {journal} {J.
  Math. Phys.}\ }\textbf {\bibinfo {volume} {3}},\ \bibinfo {pages} {157}
  (\bibinfo {year} {1962}{\natexlab{b}})}\BibitemShut {NoStop}%
\bibitem [{\citenamefont {Dyson}(1962{\natexlab{c}})}]{Dyso1962III}%
  \BibitemOpen
  \bibfield  {author} {\bibinfo {author} {\bibfnamefont {F.~J.}\ \bibnamefont
  {Dyson}},\ }\href@noop {} {\bibfield  {journal} {\bibinfo  {journal} {J.
  Math. Phys.}\ }\textbf {\bibinfo {volume} {3}},\ \bibinfo {pages} {166}
  (\bibinfo {year} {1962}{\natexlab{c}})}\BibitemShut {NoStop}%
\bibitem [{\citenamefont {Dyson}(1962{\natexlab{d}})}]{Dyso1962ThreeFold}%
  \BibitemOpen
  \bibfield  {author} {\bibinfo {author} {\bibfnamefont {F.~J.}\ \bibnamefont
  {Dyson}},\ }\href@noop {} {\bibfield  {journal} {\bibinfo  {journal} {J.
  Math. Phys.}\ }\textbf {\bibinfo {volume} {3}},\ \bibinfo {pages} {1199}
  (\bibinfo {year} {1962}{\natexlab{d}})}\BibitemShut {NoStop}%
\bibitem [{\citenamefont {Berry}(1987)}]{Berr1987}%
  \BibitemOpen
  \bibfield  {author} {\bibinfo {author} {\bibfnamefont {M.~V.}\ \bibnamefont
  {Berry}},\ }\href@noop {} {\bibfield  {journal} {\bibinfo  {journal} {Proc.
  Roy. Soc. Lond. A}\ }\textbf {\bibinfo {volume} {413}},\ \bibinfo {pages}
  {183} (\bibinfo {year} {1987})}\BibitemShut {NoStop}%
\bibitem [{\citenamefont {Bohigas}\ \emph {et~al.}(1984)\citenamefont
  {Bohigas}, \citenamefont {Giannoni},\ and\ \citenamefont
  {Schmit}}]{BohiGianSchm1984}%
  \BibitemOpen
  \bibfield  {author} {\bibinfo {author} {\bibfnamefont {O.}~\bibnamefont
  {Bohigas}}, \bibinfo {author} {\bibfnamefont {M.~J.}\ \bibnamefont
  {Giannoni}}, \ and\ \bibinfo {author} {\bibfnamefont {C.}~\bibnamefont
  {Schmit}},\ }\href {\doibase 10.1103/PhysRevLett.52.1} {\bibfield  {journal}
  {\bibinfo  {journal} {Phys. Rev. Lett.}\ }\textbf {\bibinfo {volume} {52}},\
  \bibinfo {pages} {1} (\bibinfo {year} {1984})}\BibitemShut {NoStop}%
\bibitem [{\citenamefont {Kos}\ \emph {et~al.}(2018)\citenamefont {Kos},
  \citenamefont {Ljubotina},\ and\ \citenamefont {Prosen}}]{KosLjubPros2017}%
  \BibitemOpen
  \bibfield  {author} {\bibinfo {author} {\bibfnamefont {P.}~\bibnamefont
  {Kos}}, \bibinfo {author} {\bibfnamefont {M.}~\bibnamefont {Ljubotina}}, \
  and\ \bibinfo {author} {\bibfnamefont {T.}~\bibnamefont {Prosen}},\
  }\href@noop {} {\bibfield  {journal} {\bibinfo  {journal} {Phys. Rev. X}\
  }\textbf {\bibinfo {volume} {8}},\ \bibinfo {pages} {021062} (\bibinfo {year}
  {2018})}\BibitemShut {NoStop}%
\bibitem [{\citenamefont {Ginibre}(1965)}]{Gini1965}%
  \BibitemOpen
  \bibfield  {author} {\bibinfo {author} {\bibfnamefont {J.}~\bibnamefont
  {Ginibre}},\ }\href@noop {} {\bibfield  {journal} {\bibinfo  {journal} {J.
  Math. Phys.}\ }\textbf {\bibinfo {volume} {6}},\ \bibinfo {pages} {440}
  (\bibinfo {year} {1965})}\BibitemShut {NoStop}%
\bibitem [{\citenamefont {Kreyszig}(1978)}]{Krey1978}%
  \BibitemOpen
  \bibfield  {author} {\bibinfo {author} {\bibfnamefont {E.}~\bibnamefont
  {Kreyszig}},\ }\href@noop {} {\emph {\bibinfo {title} {Introductory
  Functional Analysis with Applications}}}\ (\bibinfo  {publisher} {John Wiley
  \& Sons},\ \bibinfo {address} {New York},\ \bibinfo {year}
  {1978})\BibitemShut {NoStop}%
\bibitem [{\citenamefont {Girko}(1985)}]{Girk1984}%
  \BibitemOpen
  \bibfield  {author} {\bibinfo {author} {\bibfnamefont {V.~L.}\ \bibnamefont
  {Girko}},\ }\href@noop {} {\bibfield  {journal} {\bibinfo  {journal} {Theory
  Probab. Its Appl.}\ }\textbf {\bibinfo {volume} {29}},\ \bibinfo {pages}
  {694} (\bibinfo {year} {1985})}\BibitemShut {NoStop}%
\bibitem [{\citenamefont {Tao}\ \emph {et~al.}(2010)\citenamefont {Tao},
  \citenamefont {Vu},\ and\ \citenamefont {Krishnapur}}]{TaoVuKris2010}%
  \BibitemOpen
  \bibfield  {author} {\bibinfo {author} {\bibfnamefont {T.}~\bibnamefont
  {Tao}}, \bibinfo {author} {\bibfnamefont {V.}~\bibnamefont {Vu}}, \ and\
  \bibinfo {author} {\bibfnamefont {M.}~\bibnamefont {Krishnapur}},\
  }\href@noop {} {\bibfield  {journal} {\bibinfo  {journal} {Ann. Probab.}\
  }\textbf {\bibinfo {volume} {38}},\ \bibinfo {pages} {2023} (\bibinfo {year}
  {2010})}\BibitemShut {NoStop}%
\bibitem [{\citenamefont {Janik}\ \emph {et~al.}(1997)\citenamefont {Janik},
  \citenamefont {Nowak}, \citenamefont {Papp},\ and\ \citenamefont
  {Zahed}}]{JaniNowaPappZahe1997a}%
  \BibitemOpen
  \bibfield  {author} {\bibinfo {author} {\bibfnamefont {R.~A.}\ \bibnamefont
  {Janik}}, \bibinfo {author} {\bibfnamefont {M.~A.}\ \bibnamefont {Nowak}},
  \bibinfo {author} {\bibfnamefont {G.}~\bibnamefont {Papp}}, \ and\ \bibinfo
  {author} {\bibfnamefont {I.}~\bibnamefont {Zahed}},\ }\href@noop {}
  {\bibfield  {journal} {\bibinfo  {journal} {Nucl. Phys. B}\ }\textbf
  {\bibinfo {volume} {501}},\ \bibinfo {pages} {603} (\bibinfo {year}
  {1997})}\BibitemShut {NoStop}%
\bibitem [{\citenamefont {Janik}\ \emph {et~al.}(2002)\citenamefont {Janik},
  \citenamefont {Nowak}, \citenamefont {Papp},\ and\ \citenamefont
  {Zahed}}]{JaniNowaPappZahe2002}%
  \BibitemOpen
  \bibfield  {author} {\bibinfo {author} {\bibfnamefont {R.~A.}\ \bibnamefont
  {Janik}}, \bibinfo {author} {\bibfnamefont {M.~A.}\ \bibnamefont {Nowak}},
  \bibinfo {author} {\bibfnamefont {G.}~\bibnamefont {Papp}}, \ and\ \bibinfo
  {author} {\bibfnamefont {I.}~\bibnamefont {Zahed}},\ }\href@noop {}
  {\bibfield  {journal} {\bibinfo  {journal} {New Developments in Quantum Field
  Theory: NATO Science Series B}\ }\textbf {\bibinfo {volume} {366}},\ \bibinfo
  {pages} {297} (\bibinfo {year} {2002})}\BibitemShut {NoStop}%
\bibitem [{\citenamefont {Feinberg}\ and\ \citenamefont
  {Zee}(1997)}]{FeinZee1997}%
  \BibitemOpen
  \bibfield  {author} {\bibinfo {author} {\bibfnamefont {J.}~\bibnamefont
  {Feinberg}}\ and\ \bibinfo {author} {\bibfnamefont {A.}~\bibnamefont {Zee}},\
  }\href@noop {} {\bibfield  {journal} {\bibinfo  {journal} {Nucl. Phys. B}\
  }\textbf {\bibinfo {volume} {504}},\ \bibinfo {pages} {579} (\bibinfo {year}
  {1997})}\BibitemShut {NoStop}%
\bibitem [{\citenamefont {Hough}\ \emph {et~al.}(2009)\citenamefont {Hough},
  \citenamefont {Krishnapur}, \citenamefont {Peres},\ and\ \citenamefont
  {Vir{\'{a}}g}}]{HougKrisPereVira2009}%
  \BibitemOpen
  \bibfield  {author} {\bibinfo {author} {\bibfnamefont {J.~B.}\ \bibnamefont
  {Hough}}, \bibinfo {author} {\bibfnamefont {M.}~\bibnamefont {Krishnapur}},
  \bibinfo {author} {\bibfnamefont {Y.}~\bibnamefont {Peres}}, \ and\ \bibinfo
  {author} {\bibfnamefont {B.}~\bibnamefont {Vir{\'{a}}g}},\ }\href@noop {}
  {\emph {\bibinfo {title} {Zeros of Gaussian Analytic Functions and
  Determinantal Point Processes}}},\ \bibinfo {series} {University Lecture
  Series}, Vol.~\bibinfo {volume} {51}\ (\bibinfo  {publisher} {American
  Mathematical Society},\ \bibinfo {address} {Providence},\ \bibinfo {year}
  {2009})\BibitemShut {NoStop}%
\bibitem [{\citenamefont {Forrester}(2010)}]{Forr2010}%
  \BibitemOpen
  \bibfield  {author} {\bibinfo {author} {\bibfnamefont {P.~J.}\ \bibnamefont
  {Forrester}},\ }\href@noop {} {\emph {\bibinfo {title} {Log-Gases and Random
  Matrices}}},\ \bibinfo {series} {London Mathematical Society Monographs},
  Vol.~\bibinfo {volume} {34}\ (\bibinfo  {publisher} {Princeton University
  Press},\ \bibinfo {address} {Princeton},\ \bibinfo {year} {2010})\BibitemShut
  {NoStop}%
\bibitem [{\citenamefont {Dyson}(1962{\natexlab{e}})}]{Dyso1962BrownianMotion}%
  \BibitemOpen
  \bibfield  {author} {\bibinfo {author} {\bibfnamefont {F.~J.}\ \bibnamefont
  {Dyson}},\ }\href@noop {} {\bibfield  {journal} {\bibinfo  {journal} {J.
  Math. Phys.}\ }\textbf {\bibinfo {volume} {3}},\ \bibinfo {pages} {1191}
  (\bibinfo {year} {1962}{\natexlab{e}})}\BibitemShut {NoStop}%
\bibitem [{\citenamefont {Dyson}(1972)}]{Dyso1972}%
  \BibitemOpen
  \bibfield  {author} {\bibinfo {author} {\bibfnamefont {F.~J.}\ \bibnamefont
  {Dyson}},\ }\href@noop {} {\bibfield  {journal} {\bibinfo  {journal} {J.
  Math. Phys.}\ }\textbf {\bibinfo {volume} {13}},\ \bibinfo {pages} {90}
  (\bibinfo {year} {1972})}\BibitemShut {NoStop}%
\bibitem [{\citenamefont {Smolyarenko}\ and\ \citenamefont
  {Simons}(2003{\natexlab{a}})}]{SmolSimo2003a}%
  \BibitemOpen
  \bibfield  {author} {\bibinfo {author} {\bibfnamefont {I.~E.}\ \bibnamefont
  {Smolyarenko}}\ and\ \bibinfo {author} {\bibfnamefont {B.~D.}\ \bibnamefont
  {Simons}},\ }\href@noop {} {\bibfield  {journal} {\bibinfo  {journal} {J.
  Phys. A}\ }\textbf {\bibinfo {volume} {36}},\ \bibinfo {pages} {3551}
  (\bibinfo {year} {2003}{\natexlab{a}})}\BibitemShut {NoStop}%
\bibitem [{\citenamefont {Smolyarenko}\ and\ \citenamefont
  {Simons}(2003{\natexlab{b}})}]{SmolSimo2003b}%
  \BibitemOpen
  \bibfield  {author} {\bibinfo {author} {\bibfnamefont {I.~E.}\ \bibnamefont
  {Smolyarenko}}\ and\ \bibinfo {author} {\bibfnamefont {B.~D.}\ \bibnamefont
  {Simons}},\ }\href@noop {} {\bibfield  {journal} {\bibinfo  {journal} {Phys.
  Rev. E}\ }\textbf {\bibinfo {volume} {67}},\ \bibinfo {pages} {025202}
  (\bibinfo {year} {2003}{\natexlab{b}})}\BibitemShut {NoStop}%
\bibitem [{\citenamefont {Blaizot}\ \emph {et~al.}(2016)\citenamefont
  {Blaizot}, \citenamefont {Grela}, \citenamefont {Nowak}, \citenamefont
  {Tarnowski},\ and\ \citenamefont {Warcho{\l{}}}}]{BlaiGrelNowaTarnWarc2016}%
  \BibitemOpen
  \bibfield  {author} {\bibinfo {author} {\bibfnamefont {J.-P.}\ \bibnamefont
  {Blaizot}}, \bibinfo {author} {\bibfnamefont {J.}~\bibnamefont {Grela}},
  \bibinfo {author} {\bibfnamefont {M.~A.}\ \bibnamefont {Nowak}}, \bibinfo
  {author} {\bibfnamefont {W.}~\bibnamefont {Tarnowski}}, \ and\ \bibinfo
  {author} {\bibfnamefont {P.}~\bibnamefont {Warcho{\l{}}}},\ }\href@noop {}
  {\bibfield  {journal} {\bibinfo  {journal} {J. Stat. Mech. Theory E.}\
  }\textbf {\bibinfo {volume} {2016}},\ \bibinfo {pages} {054037} (\bibinfo
  {year} {2016})}\BibitemShut {NoStop}%
\bibitem [{\citenamefont {Fyodorov}\ \emph {et~al.}(1997)\citenamefont
  {Fyodorov}, \citenamefont {Khoruzhenko},\ and\ \citenamefont
  {Sommers}}]{FyodKhorSomm1996}%
  \BibitemOpen
  \bibfield  {author} {\bibinfo {author} {\bibfnamefont {Y.~V.}\ \bibnamefont
  {Fyodorov}}, \bibinfo {author} {\bibfnamefont {B.~A.}\ \bibnamefont
  {Khoruzhenko}}, \ and\ \bibinfo {author} {\bibfnamefont {H.-J.}\ \bibnamefont
  {Sommers}},\ }\href@noop {} {\bibfield  {journal} {\bibinfo  {journal} {Phys.
  Rev. Lett.}\ }\textbf {\bibinfo {volume} {79}},\ \bibinfo {pages} {557}
  (\bibinfo {year} {1997})}\BibitemShut {NoStop}%
\bibitem [{\citenamefont {Fyodorov}\ \emph {et~al.}(1998)\citenamefont
  {Fyodorov}, \citenamefont {Sommers},\ and\ \citenamefont
  {Khoruzhenko}}]{FyodSommKhor1998}%
  \BibitemOpen
  \bibfield  {author} {\bibinfo {author} {\bibfnamefont {Y.~V.}\ \bibnamefont
  {Fyodorov}}, \bibinfo {author} {\bibfnamefont {H.-J.}\ \bibnamefont
  {Sommers}}, \ and\ \bibinfo {author} {\bibfnamefont {B.~A.}\ \bibnamefont
  {Khoruzhenko}},\ }\href@noop {} {\bibfield  {journal} {\bibinfo  {journal}
  {Annales de l'Inst. H. P.-Phys. Theor.}\ }\textbf {\bibinfo {volume} {68}},\
  \bibinfo {pages} {449} (\bibinfo {year} {1998})}\BibitemShut {NoStop}%
\bibitem [{\citenamefont {Slater}(1929)}]{Slater1929}%
  \BibitemOpen
  \bibfield  {author} {\bibinfo {author} {\bibfnamefont {J.~C.}\ \bibnamefont
  {Slater}},\ }\href@noop {} {\bibfield  {journal} {\bibinfo  {journal} {Phys.
  Rev.}\ }\textbf {\bibinfo {volume} {34}},\ \bibinfo {pages} {1293} (\bibinfo
  {year} {1929})}\BibitemShut {NoStop}%
\bibitem [{\citenamefont {Nye}\ \emph {et~al.}(1988)\citenamefont {Nye},
  \citenamefont {Hajnal},\ and\ \citenamefont {Hannay}}]{Nye1988}%
  \BibitemOpen
  \bibfield  {author} {\bibinfo {author} {\bibfnamefont {J.~F.}\ \bibnamefont
  {Nye}}, \bibinfo {author} {\bibfnamefont {J.~V.}\ \bibnamefont {Hajnal}}, \
  and\ \bibinfo {author} {\bibfnamefont {J.~H.}\ \bibnamefont {Hannay}},\
  }\href@noop {} {\bibfield  {journal} {\bibinfo  {journal} {Proc. Roy. Soc.
  Lond. A}\ }\textbf {\bibinfo {volume} {417}},\ \bibinfo {pages} {7} (\bibinfo
  {year} {1988})}\BibitemShut {NoStop}%
\bibitem [{\citenamefont {Paganin}\ \emph
  {et~al.}(2018{\natexlab{b}})\citenamefont {Paganin}, \citenamefont
  {Petersen},\ and\ \citenamefont {Beltran}}]{Paganin2018b}%
  \BibitemOpen
  \bibfield  {author} {\bibinfo {author} {\bibfnamefont {D.~M.}\ \bibnamefont
  {Paganin}}, \bibinfo {author} {\bibfnamefont {T.~C.}\ \bibnamefont
  {Petersen}}, \ and\ \bibinfo {author} {\bibfnamefont {M.~A.}\ \bibnamefont
  {Beltran}},\ }\href@noop {} {\bibfield  {journal} {\bibinfo  {journal} {Phys.
  Rev. A}\ }\textbf {\bibinfo {volume} {97}},\ \bibinfo {pages} {023835}
  (\bibinfo {year} {2018}{\natexlab{b}})}\BibitemShut {NoStop}%
\bibitem [{\citenamefont {Pitaevskii}\ and\ \citenamefont
  {Stringari}(2003)}]{Pitaevskii2003}%
  \BibitemOpen
  \bibfield  {author} {\bibinfo {author} {\bibfnamefont {L.~P.}\ \bibnamefont
  {Pitaevskii}}\ and\ \bibinfo {author} {\bibfnamefont {S.}~\bibnamefont
  {Stringari}},\ }\href@noop {} {\emph {\bibinfo {title} {Bose--Einstein
  Condensation}}}\ (\bibinfo  {publisher} {Oxford University Press},\ \bibinfo
  {address} {Oxford},\ \bibinfo {year} {2003})\BibitemShut {NoStop}%
\bibitem [{\citenamefont {Wells}\ \emph {et~al.}(2015)\citenamefont {Wells},
  \citenamefont {Pan}, \citenamefont {Wang}, \citenamefont {Fedoseev},\ and\
  \citenamefont {Hilgenkamp}}]{Wells2015}%
  \BibitemOpen
  \bibfield  {author} {\bibinfo {author} {\bibfnamefont {F.~S.}\ \bibnamefont
  {Wells}}, \bibinfo {author} {\bibfnamefont {A.~V.}\ \bibnamefont {Pan}},
  \bibinfo {author} {\bibfnamefont {X.~R.}\ \bibnamefont {Wang}}, \bibinfo
  {author} {\bibfnamefont {S.~A.}\ \bibnamefont {Fedoseev}}, \ and\ \bibinfo
  {author} {\bibfnamefont {H.}~\bibnamefont {Hilgenkamp}},\ }\href@noop {}
  {\bibfield  {journal} {\bibinfo  {journal} {Sci. Rep.}\ }\textbf {\bibinfo
  {volume} {5}},\ \bibinfo {pages} {8677} (\bibinfo {year} {2015})}\BibitemShut
  {NoStop}%
\bibitem [{\citenamefont {Groszek}\ \emph {et~al.}(2016)\citenamefont
  {Groszek}, \citenamefont {Simula}, \citenamefont {Paganin},\ and\
  \citenamefont {Helmerson}}]{Groszek2016}%
  \BibitemOpen
  \bibfield  {author} {\bibinfo {author} {\bibfnamefont {A.~J.}\ \bibnamefont
  {Groszek}}, \bibinfo {author} {\bibfnamefont {T.~P.}\ \bibnamefont {Simula}},
  \bibinfo {author} {\bibfnamefont {D.~M.}\ \bibnamefont {Paganin}}, \ and\
  \bibinfo {author} {\bibfnamefont {K.}~\bibnamefont {Helmerson}},\ }\href@noop
  {} {\bibfield  {journal} {\bibinfo  {journal} {Phys. Rev. A}\ }\textbf
  {\bibinfo {volume} {93}},\ \bibinfo {pages} {043614} (\bibinfo {year}
  {2016})}\BibitemShut {NoStop}%
\bibitem [{\citenamefont {Goldberg}\ \emph {et~al.}(1967)\citenamefont
  {Goldberg}, \citenamefont {Schey},\ and\ \citenamefont
  {Schwartz}}]{Goldberg1967}%
  \BibitemOpen
  \bibfield  {author} {\bibinfo {author} {\bibfnamefont {A.}~\bibnamefont
  {Goldberg}}, \bibinfo {author} {\bibfnamefont {H.~M.}\ \bibnamefont {Schey}},
  \ and\ \bibinfo {author} {\bibfnamefont {J.~L.}\ \bibnamefont {Schwartz}},\
  }\href@noop {} {\bibfield  {journal} {\bibinfo  {journal} {Am. J. Phys.}\
  }\textbf {\bibinfo {volume} {35}},\ \bibinfo {pages} {177} (\bibinfo {year}
  {1967})}\BibitemShut {NoStop}%
\bibitem [{\citenamefont {O'Holleran}\ \emph {et~al.}(2008)\citenamefont
  {O'Holleran}, \citenamefont {Dennis}, \citenamefont {Flossmann},\ and\
  \citenamefont {Padgett}}]{OHolleran2008}%
  \BibitemOpen
  \bibfield  {author} {\bibinfo {author} {\bibfnamefont {K.}~\bibnamefont
  {O'Holleran}}, \bibinfo {author} {\bibfnamefont {M.~R.}\ \bibnamefont
  {Dennis}}, \bibinfo {author} {\bibfnamefont {F.}~\bibnamefont {Flossmann}}, \
  and\ \bibinfo {author} {\bibfnamefont {M.~J.}\ \bibnamefont {Padgett}},\
  }\href@noop {} {\bibfield  {journal} {\bibinfo  {journal} {Phys. Rev. Lett.}\
  }\textbf {\bibinfo {volume} {100}},\ \bibinfo {pages} {053902} (\bibinfo
  {year} {2008})}\BibitemShut {NoStop}%
\bibitem [{\citenamefont {Ruben}\ \emph {et~al.}(2008)\citenamefont {Ruben},
  \citenamefont {Paganin},\ and\ \citenamefont {Morgan}}]{Ruben2008}%
  \BibitemOpen
  \bibfield  {author} {\bibinfo {author} {\bibfnamefont {G.}~\bibnamefont
  {Ruben}}, \bibinfo {author} {\bibfnamefont {D.~M.}\ \bibnamefont {Paganin}},
  \ and\ \bibinfo {author} {\bibfnamefont {M.~J.}\ \bibnamefont {Morgan}},\
  }\href@noop {} {\bibfield  {journal} {\bibinfo  {journal} {Phys. Rev. A}\
  }\textbf {\bibinfo {volume} {78}},\ \bibinfo {pages} {013631} (\bibinfo
  {year} {2008})}\BibitemShut {NoStop}%
\bibitem [{\citenamefont {Paganin}(2006)}]{Paganin2006}%
  \BibitemOpen
  \bibfield  {author} {\bibinfo {author} {\bibfnamefont {D.~M.}\ \bibnamefont
  {Paganin}},\ }\href@noop {} {\emph {\bibinfo {title} {Coherent X-Ray
  Optics}}}\ (\bibinfo  {publisher} {Oxford University Press},\ \bibinfo
  {address} {Oxford},\ \bibinfo {year} {2006})\BibitemShut {NoStop}%
\bibitem [{\citenamefont {Maggiore}(2005)}]{Maggiore2005}%
  \BibitemOpen
  \bibfield  {author} {\bibinfo {author} {\bibfnamefont {M.}~\bibnamefont
  {Maggiore}},\ }\href@noop {} {\emph {\bibinfo {title} {A Modern Introduction
  to Quantum Field Theory}}}\ (\bibinfo  {publisher} {Oxford University
  Press},\ \bibinfo {address} {Oxford},\ \bibinfo {year} {2005})\BibitemShut
  {NoStop}%
\bibitem [{\citenamefont {{National Institute of Standards and Technology
  (NIST)}}()}]{NIST_DLMF}%
  \BibitemOpen
  \bibfield  {author} {\bibinfo {author} {\bibnamefont {{National Institute of
  Standards and Technology (NIST)}}},\ }\href {https://dlmf.nist.gov/}
  {\enquote {\bibinfo {title} {Digital library of mathematical functions},}\
  }\bibinfo {note} {{h}ttps://dlmf.nist.gov/}\BibitemShut {NoStop}%
\bibitem [{\citenamefont {Groszek}\ \emph {et~al.}(2018)\citenamefont
  {Groszek}, \citenamefont {Paganin}, \citenamefont {Helmerson},\ and\
  \citenamefont {Simula}}]{Groszek2018}%
  \BibitemOpen
  \bibfield  {author} {\bibinfo {author} {\bibfnamefont {A.~J.}\ \bibnamefont
  {Groszek}}, \bibinfo {author} {\bibfnamefont {D.~M.}\ \bibnamefont
  {Paganin}}, \bibinfo {author} {\bibfnamefont {K.}~\bibnamefont {Helmerson}},
  \ and\ \bibinfo {author} {\bibfnamefont {T.~P.}\ \bibnamefont {Simula}},\
  }\href@noop {} {\bibfield  {journal} {\bibinfo  {journal} {Phys. Rev. A}\
  }\textbf {\bibinfo {volume} {97}},\ \bibinfo {pages} {023617} (\bibinfo
  {year} {2018})}\BibitemShut {NoStop}%
\bibitem [{\citenamefont {Berggren}\ \emph {et~al.}(2001)\citenamefont
  {Berggren}, \citenamefont {Sadreev},\ and\ \citenamefont
  {Starikov}}]{Berggren2001}%
  \BibitemOpen
  \bibfield  {author} {\bibinfo {author} {\bibfnamefont {K.-F.}\ \bibnamefont
  {Berggren}}, \bibinfo {author} {\bibfnamefont {A.~F.}\ \bibnamefont
  {Sadreev}}, \ and\ \bibinfo {author} {\bibfnamefont {A.~A.}\ \bibnamefont
  {Starikov}},\ }\href@noop {} {\bibfield  {journal} {\bibinfo  {journal}
  {Nanotechnology}\ }\textbf {\bibinfo {volume} {12}},\ \bibinfo {pages} {562}
  (\bibinfo {year} {2001})}\BibitemShut {NoStop}%
\bibitem [{\citenamefont {Kitchen}\ \emph {et~al.}(2004)\citenamefont
  {Kitchen}, \citenamefont {Paganin}, \citenamefont {Lewis}, \citenamefont
  {Yagi}, \citenamefont {Uesugi},\ and\ \citenamefont {Mudie}}]{Kitchen2004}%
  \BibitemOpen
  \bibfield  {author} {\bibinfo {author} {\bibfnamefont {M.~J.}\ \bibnamefont
  {Kitchen}}, \bibinfo {author} {\bibfnamefont {D.}~\bibnamefont {Paganin}},
  \bibinfo {author} {\bibfnamefont {R.~A.}\ \bibnamefont {Lewis}}, \bibinfo
  {author} {\bibfnamefont {N.}~\bibnamefont {Yagi}}, \bibinfo {author}
  {\bibfnamefont {K.}~\bibnamefont {Uesugi}}, \ and\ \bibinfo {author}
  {\bibfnamefont {S.~T.}\ \bibnamefont {Mudie}},\ }\href@noop {} {\bibfield
  {journal} {\bibinfo  {journal} {Phys. Med. Biol.}\ }\textbf {\bibinfo
  {volume} {49}},\ \bibinfo {pages} {4335} (\bibinfo {year}
  {2004})}\BibitemShut {NoStop}%
\bibitem [{\citenamefont {Smith}\ and\ \citenamefont
  {Gbur}(2016)}]{SmitGbur2016}%
  \BibitemOpen
  \bibfield  {author} {\bibinfo {author} {\bibfnamefont {M.~K.}\ \bibnamefont
  {Smith}}\ and\ \bibinfo {author} {\bibfnamefont {G.~J.}\ \bibnamefont
  {Gbur}},\ }\href@noop {} {\bibfield  {journal} {\bibinfo  {journal} {Opt.
  Lett.}\ }\textbf {\bibinfo {volume} {41}},\ \bibinfo {pages} {4979} (\bibinfo
  {year} {2016})}\BibitemShut {NoStop}%
\bibitem [{\citenamefont {Freund}(1995)}]{Freund1995}%
  \BibitemOpen
  \bibfield  {author} {\bibinfo {author} {\bibfnamefont {I.}~\bibnamefont
  {Freund}},\ }\href@noop {} {\bibfield  {journal} {\bibinfo  {journal} {Phys.
  Rev. E}\ }\textbf {\bibinfo {volume} {52}},\ \bibinfo {pages} {2348}
  (\bibinfo {year} {1995})}\BibitemShut {NoStop}%
\bibitem [{\citenamefont {Maxwell}(1870)}]{Maxwell1870}%
  \BibitemOpen
  \bibfield  {author} {\bibinfo {author} {\bibfnamefont {J.~C.}\ \bibnamefont
  {Maxwell}},\ }\href@noop {} {\bibfield  {journal} {\bibinfo  {journal} {Phil.
  Mag.}\ }\textbf {\bibinfo {volume} {40}},\ \bibinfo {pages} {421} (\bibinfo
  {year} {1870})}\BibitemShut {NoStop}%
\bibitem [{\citenamefont {Nash}\ and\ \citenamefont {Sen}(1983)}]{Nash1992}%
  \BibitemOpen
  \bibfield  {author} {\bibinfo {author} {\bibfnamefont {C.}~\bibnamefont
  {Nash}}\ and\ \bibinfo {author} {\bibfnamefont {S.}~\bibnamefont {Sen}},\
  }\href@noop {} {\emph {\bibinfo {title} {Topology and Geometry for
  Physicists}}}\ (\bibinfo  {publisher} {Academic Press, Inc.},\ \bibinfo
  {address} {New York},\ \bibinfo {year} {1983})\BibitemShut {NoStop}%
\bibitem [{\citenamefont {Arnold}\ \emph {et~al.}(1985)\citenamefont {Arnold},
  \citenamefont {Gusein-Zade},\ and\ \citenamefont {Varchenko}}]{Arnold1985}%
  \BibitemOpen
  \bibfield  {author} {\bibinfo {author} {\bibfnamefont {V.~I.}\ \bibnamefont
  {Arnold}}, \bibinfo {author} {\bibfnamefont {S.~M.}\ \bibnamefont
  {Gusein-Zade}}, \ and\ \bibinfo {author} {\bibfnamefont {A.~N.}\ \bibnamefont
  {Varchenko}},\ }\href@noop {} {\emph {\bibinfo {title} {Singularities of
  Differentiable Maps}}},\ Vol.~\bibinfo {volume} {1}\ (\bibinfo  {publisher}
  {Birkh\"{a}user},\ \bibinfo {address} {Boston},\ \bibinfo {year} {1985})\
  \bibinfo {note} {p. 165}\BibitemShut {NoStop}%
\bibitem [{\citenamefont {Schouten}(2005)}]{Scho2005}%
  \BibitemOpen
  \bibfield  {author} {\bibinfo {author} {\bibfnamefont {H.~F.}\ \bibnamefont
  {Schouten}},\ }\href@noop {} {\bibfield  {journal} {\bibinfo  {journal} {{\em
  Light Transmission through Sub-Wavelength Apertures}, PhD thesis, University
  of Amsterdam}\ } (\bibinfo {year} {2005})}\BibitemShut {NoStop}%
\bibitem [{\citenamefont {Hazewinkel}(1991)}]{Haze1991}%
  \BibitemOpen
  \bibinfo {editor} {\bibfnamefont {M.}~\bibnamefont {Hazewinkel}},\ ed.,\
  \href@noop {} {\emph {\bibinfo {title} {Encyclopaedia of Mathematics}}},\
  Vol.~\bibinfo {volume} {7}\ (\bibinfo  {publisher} {Kluwer Academic
  Publishers},\ \bibinfo {year} {1991})\BibitemShut {NoStop}%
\bibitem [{Note1()}]{Note1}%
  \BibitemOpen
  \bibinfo {note} {To avoid branch cuts in the phase, we use the identity
  $|\Psi |^2 \nabla \Phi = \protect \operatorname {Im}(\Psi ^* \nabla \Psi )$
  to access the phase gradient, however finding the zeros of this expression is
  more computationally intensive than finding the zeros of $\Psi $, and so we
  do not have the full topological description for the $10\times 10$ matrix
  system displayed above in Fig.~\ref {f:NodalEv1Phi}, hence the use of the
  smaller $4\times 4$ systems in this section.}\BibitemShut {Stop}%
\bibitem [{\citenamefont {Leach}\ \emph {et~al.}(2005)\citenamefont {Leach},
  \citenamefont {Dennis}, \citenamefont {Courtial},\ and\ \citenamefont
  {Padgett}}]{Leach2005}%
  \BibitemOpen
  \bibfield  {author} {\bibinfo {author} {\bibfnamefont {J.}~\bibnamefont
  {Leach}}, \bibinfo {author} {\bibfnamefont {M.~R.}\ \bibnamefont {Dennis}},
  \bibinfo {author} {\bibfnamefont {J.}~\bibnamefont {Courtial}}, \ and\
  \bibinfo {author} {\bibfnamefont {M.~J.}\ \bibnamefont {Padgett}},\
  }\href@noop {} {\bibfield  {journal} {\bibinfo  {journal} {New J. Phys.}\
  }\textbf {\bibinfo {volume} {7}},\ \bibinfo {pages} {55} (\bibinfo {year}
  {2005})}\BibitemShut {NoStop}%
\bibitem [{\citenamefont {Freund}(2000)}]{Freund2000}%
  \BibitemOpen
  \bibfield  {author} {\bibinfo {author} {\bibfnamefont {I.}~\bibnamefont
  {Freund}},\ }\href@noop {} {\bibfield  {journal} {\bibinfo  {journal} {Opt.
  Commun.}\ }\textbf {\bibinfo {volume} {181}},\ \bibinfo {pages} {19}
  (\bibinfo {year} {2000})}\BibitemShut {NoStop}%
\bibitem [{\citenamefont {Kleckner}\ and\ \citenamefont
  {Irvine}(2013)}]{Kleckner2013}%
  \BibitemOpen
  \bibfield  {author} {\bibinfo {author} {\bibfnamefont {D.}~\bibnamefont
  {Kleckner}}\ and\ \bibinfo {author} {\bibfnamefont {W.~T.~M.}\ \bibnamefont
  {Irvine}},\ }\href@noop {} {\bibfield  {journal} {\bibinfo  {journal} {Nat.
  Phys.}\ }\textbf {\bibinfo {volume} {9}},\ \bibinfo {pages} {253} (\bibinfo
  {year} {2013})}\BibitemShut {NoStop}%
\bibitem [{\citenamefont {{Lord Kelvin}}(1867)}]{Kelvin1867}%
  \BibitemOpen
  \bibfield  {author} {\bibinfo {author} {\bibnamefont {{Lord Kelvin}}},\
  }\href@noop {} {\bibfield  {journal} {\bibinfo  {journal} {Phil. Mag.}\
  }\textbf {\bibinfo {volume} {34}},\ \bibinfo {pages} {15} (\bibinfo {year}
  {1867})}\BibitemShut {NoStop}%
\bibitem [{\citenamefont {Mawson}\ \emph {et~al.}(2018)\citenamefont {Mawson},
  \citenamefont {Petersen},\ and\ \citenamefont {Simula}}]{Mawson2018}%
  \BibitemOpen
  \bibfield  {author} {\bibinfo {author} {\bibfnamefont {T.}~\bibnamefont
  {Mawson}}, \bibinfo {author} {\bibfnamefont {T.}~\bibnamefont {Petersen}}, \
  and\ \bibinfo {author} {\bibfnamefont {T.}~\bibnamefont {Simula}},\ }\href
  {arXiv: cond-mat.quant-gas/1805.10009} {\enquote {\bibinfo {title} {{Braiding
  and fusion of non-Abelian vortex anyons}},}\ } (\bibinfo {year} {2018}),\
  \bibinfo {note} {arXiv: cond-mat.quant-gas/1805.10009}\BibitemShut {NoStop}%
\bibitem [{\citenamefont {Gell-Mann}(1956)}]{Gell-Mann1956}%
  \BibitemOpen
  \bibfield  {author} {\bibinfo {author} {\bibfnamefont {M.}~\bibnamefont
  {Gell-Mann}},\ }\href@noop {} {\bibfield  {journal} {\bibinfo  {journal}
  {Nuovo Cimento}\ }\textbf {\bibinfo {volume} {4}},\ \bibinfo {pages} {848}
  (\bibinfo {year} {1956})}\BibitemShut {NoStop}%
\bibitem [{\citenamefont {Martin}\ and\ \citenamefont
  {Shaw}(1997)}]{MartinShaw1997}%
  \BibitemOpen
  \bibfield  {author} {\bibinfo {author} {\bibfnamefont {B.~R.}\ \bibnamefont
  {Martin}}\ and\ \bibinfo {author} {\bibfnamefont {G.}~\bibnamefont {Shaw}},\
  }\href@noop {} {\emph {\bibinfo {title} {Particle Physics}}},\ \bibinfo
  {edition} {2nd}\ ed.\ (\bibinfo  {publisher} {John Wiley \& Sons},\ \bibinfo
  {address} {Chichester},\ \bibinfo {year} {1997})\BibitemShut {NoStop}%
\bibitem [{\citenamefont {Gell-Mann}(1962)}]{Gell-Mann1962}%
  \BibitemOpen
  \bibfield  {author} {\bibinfo {author} {\bibfnamefont {M.}~\bibnamefont
  {Gell-Mann}},\ }\href@noop {} {\bibfield  {journal} {\bibinfo  {journal}
  {Phys. Rev.}\ }\textbf {\bibinfo {volume} {125}},\ \bibinfo {pages} {1067}
  (\bibinfo {year} {1962})}\BibitemShut {NoStop}%
\bibitem [{\citenamefont {Gell-Mann}(1964)}]{Gell-Mann1964}%
  \BibitemOpen
  \bibfield  {author} {\bibinfo {author} {\bibfnamefont {M.}~\bibnamefont
  {Gell-Mann}},\ }\href@noop {} {\bibfield  {journal} {\bibinfo  {journal}
  {Phys. Lett.}\ }\textbf {\bibinfo {volume} {8}},\ \bibinfo {pages} {214}
  (\bibinfo {year} {1964})}\BibitemShut {NoStop}%
\bibitem [{\citenamefont {Pinna}\ \emph {et~al.}(2018)\citenamefont {Pinna},
  \citenamefont {Abreu~Araujo}, \citenamefont {Kim}, \citenamefont {Cros},
  \citenamefont {Querlioz}, \citenamefont {Bessiere}, \citenamefont {Droulez},\
  and\ \citenamefont {Grollier}}]{Pinna2018}%
  \BibitemOpen
  \bibfield  {author} {\bibinfo {author} {\bibfnamefont {D.}~\bibnamefont
  {Pinna}}, \bibinfo {author} {\bibfnamefont {F.}~\bibnamefont {Abreu~Araujo}},
  \bibinfo {author} {\bibfnamefont {J.-V.}\ \bibnamefont {Kim}}, \bibinfo
  {author} {\bibfnamefont {V.}~\bibnamefont {Cros}}, \bibinfo {author}
  {\bibfnamefont {D.}~\bibnamefont {Querlioz}}, \bibinfo {author}
  {\bibfnamefont {P.}~\bibnamefont {Bessiere}}, \bibinfo {author}
  {\bibfnamefont {J.}~\bibnamefont {Droulez}}, \ and\ \bibinfo {author}
  {\bibfnamefont {J.}~\bibnamefont {Grollier}},\ }\href@noop {} {\bibfield
  {journal} {\bibinfo  {journal} {Phys. Rev. Appl.}\ }\textbf {\bibinfo
  {volume} {9}},\ \bibinfo {pages} {064018} (\bibinfo {year}
  {2018})}\BibitemShut {NoStop}%
\bibitem [{\citenamefont {Baez}(2002)}]{Baez2002}%
  \BibitemOpen
  \bibfield  {author} {\bibinfo {author} {\bibfnamefont {J.}~\bibnamefont
  {Baez}},\ }\href@noop {} {\bibfield  {journal} {\bibinfo  {journal} {Bull.
  Am. Math. Soc.}\ }\textbf {\bibinfo {volume} {39}},\ \bibinfo {pages} {145}
  (\bibinfo {year} {2002})}\BibitemShut {NoStop}%
\bibitem [{Note2()}]{Note2}%
  \BibitemOpen
  \bibinfo {note} {This situation is rather analogous to the annihilation
  channel for Bhabha scattering, namely $e^+ +e^-\rightarrow \gamma \rightarrow
  e^+ +e^-$, where $\gamma $ is an intermediate virtual photon.}\BibitemShut
  {Stop}%
\bibitem [{\citenamefont {Feynman}\ and\ \citenamefont
  {Hibbs}(1965)}]{FeynmanHibbs1965}%
  \BibitemOpen
  \bibfield  {author} {\bibinfo {author} {\bibfnamefont {R.~P.}\ \bibnamefont
  {Feynman}}\ and\ \bibinfo {author} {\bibfnamefont {A.~R.}\ \bibnamefont
  {Hibbs}},\ }\href@noop {} {\emph {\bibinfo {title} {Quantum Mechanics and
  Path Integrals}}}\ (\bibinfo  {publisher} {McGraw--Hill},\ \bibinfo {address}
  {New York},\ \bibinfo {year} {1965})\BibitemShut {NoStop}%
\bibitem [{\citenamefont {Berry}\ \emph {et~al.}(1979)\citenamefont {Berry},
  \citenamefont {Nye},\ and\ \citenamefont {Wright}}]{Berry1979}%
  \BibitemOpen
  \bibfield  {author} {\bibinfo {author} {\bibfnamefont {M.~V.}\ \bibnamefont
  {Berry}}, \bibinfo {author} {\bibfnamefont {J.~F.}\ \bibnamefont {Nye}}, \
  and\ \bibinfo {author} {\bibfnamefont {F.~J.}\ \bibnamefont {Wright}},\
  }\href@noop {} {\bibfield  {journal} {\bibinfo  {journal} {Phil. Trans. R.
  Soc. A}\ }\textbf {\bibinfo {volume} {291}},\ \bibinfo {pages} {453}
  (\bibinfo {year} {1979})}\BibitemShut {NoStop}%
\bibitem [{\citenamefont {Petersen}\ \emph {et~al.}(2013)\citenamefont
  {Petersen}, \citenamefont {Weyland}, \citenamefont {Paganin}, \citenamefont
  {Simula}, \citenamefont {Eastwood},\ and\ \citenamefont
  {Morgan}}]{Petersen2013}%
  \BibitemOpen
  \bibfield  {author} {\bibinfo {author} {\bibfnamefont {T.~C.}\ \bibnamefont
  {Petersen}}, \bibinfo {author} {\bibfnamefont {M.}~\bibnamefont {Weyland}},
  \bibinfo {author} {\bibfnamefont {D.~M.}\ \bibnamefont {Paganin}}, \bibinfo
  {author} {\bibfnamefont {T.~P.}\ \bibnamefont {Simula}}, \bibinfo {author}
  {\bibfnamefont {S.~A.}\ \bibnamefont {Eastwood}}, \ and\ \bibinfo {author}
  {\bibfnamefont {M.~J.}\ \bibnamefont {Morgan}},\ }\href@noop {} {\bibfield
  {journal} {\bibinfo  {journal} {Phys. Rev. Lett.}\ }\textbf {\bibinfo
  {volume} {110}},\ \bibinfo {pages} {033901} (\bibinfo {year}
  {2013})}\BibitemShut {NoStop}%
\bibitem [{\citenamefont {Onsager}(1949)}]{Onsager1949}%
  \BibitemOpen
  \bibfield  {author} {\bibinfo {author} {\bibfnamefont {L.}~\bibnamefont
  {Onsager}},\ }\href@noop {} {\bibfield  {journal} {\bibinfo  {journal} {Nuovo
  Cimento}\ }\textbf {\bibinfo {volume} {2}},\ \bibinfo {pages} {279} (\bibinfo
  {year} {1949})}\BibitemShut {NoStop}%
\bibitem [{\citenamefont {Sinha}\ \emph {et~al.}(1976)\citenamefont {Sinha},
  \citenamefont {Sivaram},\ and\ \citenamefont {Sudarshan}}]{Sinha1976}%
  \BibitemOpen
  \bibfield  {author} {\bibinfo {author} {\bibfnamefont {K.~P.}\ \bibnamefont
  {Sinha}}, \bibinfo {author} {\bibfnamefont {C.}~\bibnamefont {Sivaram}}, \
  and\ \bibinfo {author} {\bibfnamefont {E.~C.~G.}\ \bibnamefont {Sudarshan}},\
  }\href@noop {} {\bibfield  {journal} {\bibinfo  {journal} {Found. Phys.}\
  }\textbf {\bibinfo {volume} {6}},\ \bibinfo {pages} {65} (\bibinfo {year}
  {1976})}\BibitemShut {NoStop}%
\bibitem [{\citenamefont {Sinha}\ and\ \citenamefont
  {Sudarshan}(1978)}]{Sinha1978}%
  \BibitemOpen
  \bibfield  {author} {\bibinfo {author} {\bibfnamefont {K.~P.}\ \bibnamefont
  {Sinha}}\ and\ \bibinfo {author} {\bibfnamefont {E.~C.~G.}\ \bibnamefont
  {Sudarshan}},\ }\href@noop {} {\bibfield  {journal} {\bibinfo  {journal}
  {Found. Phys.}\ }\textbf {\bibinfo {volume} {8}},\ \bibinfo {pages} {823}
  (\bibinfo {year} {1978})}\BibitemShut {NoStop}%
\bibitem [{\citenamefont {Sethna}(2006)}]{SethnaBook}%
  \BibitemOpen
  \bibfield  {author} {\bibinfo {author} {\bibfnamefont {J.~P.}\ \bibnamefont
  {Sethna}},\ }\href@noop {} {\emph {\bibinfo {title} {Statistical Mechanics:
  Entropy, Order Parameters and Complexity}}}\ (\bibinfo  {publisher} {Oxford
  University Press},\ \bibinfo {address} {Oxford},\ \bibinfo {year}
  {2006})\BibitemShut {NoStop}%
\bibitem [{\citenamefont {Freund}(1994)}]{Freund1994}%
  \BibitemOpen
  \bibfield  {author} {\bibinfo {author} {\bibfnamefont {I.}~\bibnamefont
  {Freund}},\ }\href@noop {} {\bibfield  {journal} {\bibinfo  {journal} {J.
  Opt. Soc. Am. A}\ }\textbf {\bibinfo {volume} {11}},\ \bibinfo {pages} {1644}
  (\bibinfo {year} {1994})}\BibitemShut {NoStop}%
\bibitem [{\citenamefont {Rothschild}\ \emph {et~al.}(2012)\citenamefont
  {Rothschild}, \citenamefont {Kitchen}, \citenamefont {Faulkner},\ and\
  \citenamefont {Paganin}}]{Rothschild2012}%
  \BibitemOpen
  \bibfield  {author} {\bibinfo {author} {\bibfnamefont {F.}~\bibnamefont
  {Rothschild}}, \bibinfo {author} {\bibfnamefont {M.~J.}\ \bibnamefont
  {Kitchen}}, \bibinfo {author} {\bibfnamefont {H.~M.~L.}\ \bibnamefont
  {Faulkner}}, \ and\ \bibinfo {author} {\bibfnamefont {D.~M.}\ \bibnamefont
  {Paganin}},\ }\href@noop {} {\bibfield  {journal} {\bibinfo  {journal} {Opt.
  Commun.}\ }\textbf {\bibinfo {volume} {285}},\ \bibinfo {pages} {4141}
  (\bibinfo {year} {2012})}\BibitemShut {NoStop}%
\bibitem [{\citenamefont {Werdiger}\ \emph {et~al.}(2016)\citenamefont
  {Werdiger}, \citenamefont {Kitchen},\ and\ \citenamefont
  {Paganin}}]{Werdiger2016}%
  \BibitemOpen
  \bibfield  {author} {\bibinfo {author} {\bibfnamefont {F.}~\bibnamefont
  {Werdiger}}, \bibinfo {author} {\bibfnamefont {M.~J.}\ \bibnamefont
  {Kitchen}}, \ and\ \bibinfo {author} {\bibfnamefont {D.~M.}\ \bibnamefont
  {Paganin}},\ }\href@noop {} {\bibfield  {journal} {\bibinfo  {journal} {Opt.
  Express}\ }\textbf {\bibinfo {volume} {24}},\ \bibinfo {pages} {10620}
  (\bibinfo {year} {2016})}\BibitemShut {NoStop}%
\end{thebibliography}%

\end{document}